\documentclass{aastex62}

\usepackage{amsmath}
\usepackage{graphicx}
\usepackage{amsfonts}
\usepackage{natbib}
\usepackage{color}
\usepackage{epstopdf}
\usepackage{graphicx}
\usepackage{multirow}
\usepackage{longtable}
\usepackage{rotating}

\shorttitle{Analysis on XMM-Newton X-ray flares of Mrk 421}
\shortauthors{Yan et al.}
\begin{document}

\title{Statistical analysis on XMM-Newton X-ray flares of Mrk 421: distributions of peak flux and flaring time duration}

\correspondingauthor{Dahai Yan (DHY), Benzhong Dai (BZD)}
\email{yandahai@ynao.ac.cn (DHY), bzdai@ynu.edu.cn (BZD)}

%\author[0000-0002-0786-7307]{Greg J. Schwarz}
%\affil{American Astronomical Society \\
%2000 Florida Ave., NW, Suite 300 \\
%Washington, DC 20009-1231, USA}

%\author{August Muench}
%\affiliation{American Astronomical Society \\
%2000 Florida Ave., NW, Suite 300 \\
%Washington, DC 20009-1231, USA}
%\collaboration{(AAS Journals Data Scientists collaboration)}

\author{Dahai Yan}
\affil{Key Laboratory for the Structure and Evolution of Celestial Objects, Yunnan Observatory, Chinese Academy of Sciences, Kunming 650011, China}
\affil{Center for Astronomical Mega-Science, Chinese Academy of Sciences, 20A Datun Road, Chaoyang District, Beijing, 100012, China}

\author{Shenbang Yang}
\affil{Key Laboratory of Astroparticle Physics of Yunnan Province, Yunnan University, Kunming 650091, China}

\author{Pengfei Zhang}
\affil{Key Laboratory of Dark Matter and Space Astronomy, Purple Mountain Observatory, Chinese Academy of Sciences, Nanjing 210008, China}
\affil{Key Laboratory of Astroparticle Physics of Yunnan Province, Yunnan University, Kunming 650091, China}

\author{Benzhong Dai}
\affil{Key Laboratory of Astroparticle Physics of Yunnan Province, Yunnan University, Kunming 650091, China}

\author{Jiancheng Wang}
\affil{Key Laboratory for the Structure and Evolution of Celestial Objects, Yunnan Observatory, Chinese Academy of Sciences, Kunming 650011, China}
\affil{Center for Astronomical Mega-Science, Chinese Academy of Sciences, 20A Datun Road, Chaoyang District, Beijing, 100012, China}

\author{Li Zhang}
\affil{Key Laboratory of Astroparticle Physics of Yunnan Province, Yunnan University, Kunming 650091, China}



\begin{abstract}

The energy dissipation mechanism in blazar jet is unknown.
Blazar's flares could provide insights on this problem.
Here we report statistical results of XMM-Newton X-ray flares of Mrk 421.
We analyze all public XMM-Newton X-ray observations for Mrk 421, and construct the light curves.
Through fitting light curves, we obtain the parameters of flare-profiles, such as peak flux ($F_{\rm p}$) and flaring time duration ($T_{\rm fl}$).
It is found that both the distributions of $F_{\rm p}$ and $T_{\rm fl}$ obey a power-law form, 
with the same index of $\alpha_{\rm F}=\alpha_{\rm T}\approx1$.
The statistical properties are consistent with the predictions by a self-organized criticality (SOC)
system with energy dissipation in one-dimensional space. This is similar to solar flare, 
but with different space dimensions of the energy dissipation
domain. This suggests that X-ray flaers of Mrk 421 are possibly driven by a magnetic reconnection mechanism. 
Moreover, in the analysis, we find that variability on timescale of $\sim1000\ $s frequently appears.
Such rapid variability indicates a magnetic field of $\geq 2.1\delta_{\rm D}^{-1/3}$ G ($\delta_{\rm D}$ is the Doppler factor) in emission region.
%which is much higher than that derived in spectral modeling.

\end{abstract}



\section{Introduction} \label{sec:intro}

Blazars are a rather extreme class
of radio-loud active galactic nuclei (AGNs), consisting of BL Lac objects (BL Lacs) and flat-spectrum radio
quasars (FSRQs).
Due to relativistic Doppler boosting, blazar emission is dominated by the non-thermal emission coming from its jet.
Its spectral energy distribution (SED) extends from radio to $\gamma$-rays, and shows two bumps.
The low-energy bump is believed to be the synchrotron radiation of relativistic electrons, while the origin of high-energy bump is under debate.
A variety of emission mechanisms are proposed to explain this origin: (1) inverse Compton (IC) scattering of low-energy photons by relativistic electrons, 
including synchrotron self-Compton model \citep[i.e., SSC; e.g.,][]{Maraschi1992} and external Compton model \citep[i.e., EC; e.g.,][]{Dermer93,Sikora1994}; 
(2) relativistic proton synchrotron radiation  \citep{mann92,Aharonian2000,mucke2003}; 
(3) synchrotron radiation of secondary particles produced in proton-photon interaction \citep{mucke2001,bottcher09,Yan15,2015MNRAS.448..910C} .

According to the peak frequency of synchrotron bump ($\nu_{\rm p}$), BL Lacs are divided into three classes \citep[e.g.,][]{Padovani}: 
high-synchrotron-frequency peaked BL Lacs (HBLs; $\nu_{\rm p}>10^{15}\ $Hz),  
intermediate-synchrotron-frequency peaked BL Lacs (IBLs; $10^{14}<\nu_{\rm p}<10^{15}\ $Hz), and 
low-synchrotron-frequency peaked BL Lacs (LBLs; $\nu_{\rm p}<10^{14}\ $Hz).
FSRQs usually have $\nu_{\rm p}<10^{14}\ $Hz.

Blazars show strong variability across entire electromagnetic emission on timescales from
minutes to years \citep[e.g.,][]{Aharonian2007,Dai,Ackermann,zhu}.
Variability timescale can be used to constrain the size of emission region, the magnetic field in emission region, 
and the location of high energy emission region \citep[e.g.,][]{bottcher03, Yan17}.
The correlations of variabilities in different spectral bands could carry abundant information 
on the emission and acceleration mechanisms of particles in blazar jet \citep[e.g.,][]{Fossati,Chen}.
Although variability is important and useful for understanding blazar jet physics, 
the physical origin of blazar variability is not well understood.
Several scenarios have been proposed to understand the production of variability \citep[see][for a review]{Aharonian2017}, 
like magnetospheric gap model \citep[e.g.,][]{Neronov}, jet-star interaction model \citep[e.g.,][]{Barkov}, and jet-in-jet model \citep[e.g.,][]{Giannios}.
Magnetic reconnection may play an important role in the latter two models \citep[e.g.,][]{Aharonian2017}.
\citet{Sironi} argued that magnetic reconnection, rather than shock, powers blazar jet emission.

Statistical properties of flares could provide insights into the trigging mechanism of flares.
The flares trigged by magnetic reconnection is thought to form a self-organized criticality (SOC) system, such as solar flares \citep[e.g.,][]{Lu,Aschwanden}.
The SOC flare system expects that event parameters, e.g., the flux and the flaring time duration, should follow power-law distributions.
The indices of these power laws are related to the effective geometric dimension of the system \citep[e.g.,][]{Aschwanden12}.
Such a statistical approach has been used to investigate whether the SOC model can explain the X-ray flares of $\gamma$-ray bursts \citep[GRBs;][]{Wang13,Yi16,Yi17}, M87 \citep{Wang15}, and Sgr A$^*$ \citep{Wang15,Li,Yuan}.
Here, we apply this approach to X-ray flares of Mrk 421.

Mrk 421 is a HBL.
It is the brightest blazar in the X-ray sky.
Its X-ray emission is believed to be the synchrotron radiation of relativistic electrons, showing strong and rapid variabilities \citep[e.g.,][]{Cui,Fossati,Paliya2,Kapanadze16,Kapanadze18}.
The main goal of this paper is to investigate whether the statistical properties of Mrk 421 XMM-Newton X-ray flares are consistent with the expectations of a SOC system.

\section{Data reduction} \label{sec:obser}
We search for the archived data of Mrk 421 in XMM-Newton Science Archive\footnote{\url{http://nxsa.esac.esa.int/nxsa-web/}}, and collect a total of 50 observations containing EPIC exposures. All the observational information are collected in Table \ref{sec:intro}. Following the standard procedures of the SAS threads, we reprocess the raw data files to obtain calibrated and concatenated event lists, using the SAS version 15.0.0. 
We then check flaring high background periods; and create clear event lists which are free of high background (namely the data is screened by a new good time interval), 
using the background count rate threshold of \texttt{"RATE$<=$0.4"} for PN data and \texttt{"RATE$<=$0.35"} for MOS data. 

Before selecting the source regions, we check whether there are pile-up effects on the observations. 
It is found that most of the observations are affected by pile-up effects, as listed in Table \ref{tab:phafit} (column 10). 
To reduce the pile-up effects, we extract the light curves and spectra from core-excised regions. 
For image mode, the region is an annular area; for timing mode, each region is comprised of two columns. 
The selected source regions are listed in Table \ref{tab:phafit} (column 11). 
The background regions are extracted from the places near the source regions in the same event maps. 
Before the light curves and spectra extraction, we set the extracted event pattern and flag to be \texttt{"FLAG==0 \&\& PATTERN$<=$4"} for all PN data, 
while \texttt{"PATTERN$<=$12"} for image mode MOS data  and  \texttt{"FLAG ==0 \&\& PATTERN==0"} for timing mode MOS data. 
We create response matrix files and ancillary response files for the extracted spectra using the \texttt{rmfgen} and \texttt{arfgen} tasks. 
The spectra prepared for analysis are grouped in order to make sure that there are at least 20 counts for each spectral channel.   

XSPEC (version 12.9) is used for spectral analysis. 
We use $\chi^2$ statistic in spectral fittings. 
We first separately apply a power-law, a broken power-law and a log-parabola models to data,  to determine the best-fit model. 
We find that all the three models are failed to fit the spectra, with the reduced $\chi^2_r \gtrsim 2.0$. 
The log-parabola (LP) model is better than the others. 
Therefore, we determine the LP model as the basic component of a new model, 
and then try to add other components to reduce $\chi^2_r$. 
Ultimately, we find that a model containing a log-parabola and a black body (i.e., LP+BB) is the best-fit model for most of the spectra. 
In the fittings, we notice that there are some unknown line-shape humps in some spectra around 0.5 keV, resulting in a large $\chi^2_r$. 
We speculate that those phenomena are caused by out-of-time events\footnote{\url{https://xmm-tools.cosmos.esa.int/external/xmm_user_support/documentation/sas_usg/USG/epicOoT.html}}. 
To eliminate this effect, we add Lorenz line (Lor) components in the models. 
%Although some spectra are fitted by the best models we think, the fittings just get better, not enough to be called ``good fits". 
%The reason causing ``bad" fits is probably because of the timing variations during the relatively long observations. 
All best-fitting parameters are given in Table \ref{tab:phafit}. 
In all spectral fittings, the Galactic hydrogen absorption is considered. 
The Galactic hydrogen column density (N$_H$) is fixed to, by default, 1.92$\times$10$^{20}$ cm$^{-2}$ which is produced by  Leiden/Argentine/Bonn (LAB) Survey \citep{2005AA...440..775K}, except for two cases in which N$_H$ is set to be free, in order to obtain a successful fit.

%%%tabel 1%%%
\startlongtable
\begin{deluxetable*}{lllllllc}
\tablecaption{The XMM-Newton observational information for Mrk 421. (1) The observation ID,  (2) the observation instrument, (3) the observation mode, (4) the exposure ID, 
(5) the used filter, (6-7) the start time and stop time of the observation, (8) the total exposure time. }
\label{tab:info}
\tablehead{\colhead{Obs.ID} & \colhead{Instrument} & \colhead{Mode} & \colhead{Expo.ID} & \colhead{Filter} & \colhead{StartTime} & \colhead{StopTime} & \colhead{Exposure} \\ 
\colhead{} & \colhead{} & \colhead{} & \colhead{} & \colhead{} & \colhead{(MJD)} & \colhead{(MJD)} & \colhead{(s)} } 
%\decimals
\colnumbers
\startdata
0099280101 & PN   & Timing & S008 & Thick  & 51689.1624 & 51689.4251 & 16399.63 \\
0099280101 & PN   & Image  & S010 & Thick  & 51689.4414 & 51689.8176 & 12325.25 \\
0099280201 & PN   & Image  & S010 & Thick  & 51850.006  & 51850.4342 & 24242.19 \\
0099280301 & PN   & Image  & S010 & Thick  & 51861.9324 & 51862.4729 & 25639.59 \\
0136540101 & PN   & Image  & S008 & Thin1  & 52037.3989 & 52037.8352 & 25726.63 \\
0136540301 & MOS1 & Timing & S003 & Thin1  & 52582.0308 & 52582.3028 & 22839.90 \\
0136540401 & MOS1 & Timing & S003 & Thin1  & 52582.3202 & 52582.5922 & 22934.73 \\
0136540801 & PN   & Image  & S008 & Thick  & 52592.8741 & 52592.984  & 5489.68  \\
0136541001 & PN   & Timing & S008 & Medium & 52609.9727 & 52610.7817 & 56761.88 \\
0136541101 & PN   & Image  & S008 & Medium & 52610.8351 & 52610.9451 & 7241.35  \\
0136541201 & PN   & Image  & S008 & Medium & 52611.0031 & 52611.113  & 7107.85  \\
0150498701 & PN   & Timing & S003 & Thin1  & 52957.6897 & 52958.2417 & 18988.06 \\
0153950601 & MOS1 & Timing & S003 & Thin1  & 52398.6795 & 52399.1297 & 38360.06 \\
0153950701 & PN   & Image  & S005 & Thick  & 52399.1911 & 52399.389  & 15850.26 \\
0153951201 & PN   & Timing & S005 & Thin1  & 53681.8447 & 53681.9465 & 3781.63  \\
0153951301 & PN   & Timing & S005 & Medium & 53681.7058 & 53681.8041 & 8331.73  \\
0158970101 & MOS1 & Timing & U002 & Medium & 52791.557  & 52792.0269 & 39851.21 \\
0158970201 & PN   & Image  & S009 & Thick  & 52792.0603 & 52792.2721 & 14606.30 \\
0158970701 & MOS1 & Timing & S010 & Thick  & 52797.897  & 52798.4607 & 48071.07 \\
0158971201 & PN   & Timing & S003 & Medium & 53131.1251 & 53131.8762 & 12838.59 \\
0158971301 & PN   & Timing & S003 & Thick  & 53683.7759 & 53684.4553 & 30751.74 \\
0162960101 & PN   & Image  & S007 & Medium & 52983.8975 & 52984.2459 & 13430.16 \\
0302180101 & MOS2 & Timing & S002 & Thin1  & 53854.8676 & 53855.3479 & 39774.04 \\
0411080301 & PN   & Image  & S003 & Medium & 53883.0932 & 53883.8849 & 29605.57 \\
0411080701 & PN   & Timing & S003 & Medium & 54074.5064 & 54074.7113 & 17400.58 \\
0411081301 & PN   & Image  & S003 & Medium & 54230.1689 & 54230.3668 & 9475.17  \\
0411081401 & PN   & Image  & S003 & Medium & 54230.4143 & 54230.4964 & 4783.49  \\
0411081501 & PN   & Image  & S003 & Medium & 54230.5439 & 54230.6261 & 5754.45  \\
0411081601 & PN   & Image  & S003 & Medium & 54230.6735 & 54230.7557 & 2878.52  \\
0411081901 & MOS1 & Image  & S001 & Medium & 54423.5489 & 54423.7653 & 18328.05 \\
0411082701 & PN   & Image  & U002 & Thick  & 54617.1091 & 54617.2098 & 6267.40  \\
0411083201 & PN   & Image  & S600 & Thick  & 55151.7552 & 55151.8501 & 6518.67  \\
0502030101 & PN   & Timing & S003 & Thin1  & 54593.0812 & 54593.5673 & 27711.17 \\
0510610101 & PN   & Timing & S003 & Medium & 54228.6283 & 54228.9061 & 11045.06 \\
0510610201 & PN   & Timing & S003 & Medium & 54228.3529 & 54228.6017 & 16714.07 \\
0560980101 & PN   & Image  & S600 & Thick  & 54792.6095 & 54792.716  & 8467.44  \\
0560983301 & PN   & Image  & S600 & Thick  & 54976.1719 & 54976.2783 & 8456.58  \\
0656380101 & PN   & Image  & S600 & Thick  & 55319.3278 & 55319.4112 & 6356.08  \\
0656380801 & PN   & Image  & S600 & Thick  & 55512.8889 & 55512.985  & 7622.14  \\
0656381301 & PN   & Image  & S600 & Thick  & 55514.8844 & 55514.9805 & 7628.62  \\
0658800101 & PN   & Image  & S600 & Thick  & 55698.4452 & 55698.5574 & 4854.50  \\
0658800801 & PN   & Timing & S600 & Thick  & 55894.0036 & 55894.1181 & 8627.35  \\
0658801301 & PN   & Image  & S003 & Thick  & 57179.0079 & 57179.3261 & 19270.32 \\
0658801801 & PN   & Image  & S003 & Thick  & 57334.6077 & 57334.9584 & 21199.86 \\
0658802301 & PN   & Image  & S003 & Thick  & 57514.17   & 57514.4929 & 19543.77 \\
0670920301 & PN   & Timing & S003 & Thin1  & 56776.1859 & 56776.3363 & 8592.70  \\
0670920401 & PN   & Timing & S003 & Thin1  & 56778.1597 & 56778.331  & 13485.07 \\
0670920501 & PN   & Timing & S003 & Thin1  & 56780.1518 & 56780.3231 & 11295.34 \\
0791780101 & PN   & Image  & S001 & Thick  & 57695.5677 & 57695.7529 & 11215.14 \\
0791780601 & PN   & Image  & S001 & Thick  & 57877.186  & 57877.3134 & 7709.57 \\
\enddata
%\tablecomments{The observational information of Mrk 421 in XMM-Newton}
%\tablerefs{tab:obsinfo}
\end{deluxetable*}
%%%%%%%%%%%%%   
%%%tabel 2%%%
\begin{longrotatetable}
\begin{deluxetable*}{lcccccccccl}
\tabletypesize{\tiny}
\tablewidth{0pt}
\tablenum{2}
\tablecaption{The spectral fitting results. (1) The start time of the observation,  (2) the observation ID.,  (3) the best-fitting model, black body (BB), log-parabola  (LP), Lorenz line (Lor), 
(4) Galactic Hydrogen column density, (5) the black body temperature,  (6-7) the parameters of LP model, (8) reduced chi-square, (9) the integrated flux between 0.3-10 keV, 
in unit of 10$^{-11}$\rm  ergs/cm$^2$/s, (10-11) the pile-up effect and the corresponding source region selection. 
RAWX is the physical X axis coordinate of the event map, and $r$ is the radius from the centre of the source in units of pixel. 
The numbers in parentheses represent the second region.}
\label{tab:phafit}
\tablehead{\colhead{StartTime} & \colhead{Obs.ID} & \colhead{Model} & \colhead{N$_{\rm{H}}$} & \colhead{kT} & \colhead{$\alpha$} & \colhead{$\beta$} & \colhead{$\chi^{2}_{r}$} & \colhead{F$_{0.3-10}$} & \colhead{Pile-up} & \colhead{Region} \\ 
\colhead{(MJD)} & \colhead{} & \colhead{} & \colhead{($10^{22}\rm{cm^{-2}}$)} & \colhead{(keV)} & \colhead{}  & \colhead{} & \colhead{} & \colhead{} & \colhead{} & \colhead{} } 
\colnumbers
\startdata
51689.1624 & 0099280101 & BB+LP     & 0.0192 (fixed) & 0.128$\pm$0.004     & 2.19$\pm$0.01     & 0.05$\pm$0.01    & 1.9159 (325.70/170)  & 63.611$\pm$0.075   & yes    & 23.00 (39.00)$<$RAWX$<$37.00 (53.00)     \\
51689.4414 & 0099280101 & BB+LP     & 0.0192 (fixed) & 0.117$\pm$0.003     & 2.16$\pm$0.02     & 0.19$\pm$0.02    & 1.4076 (240.71/171)  & 77.742$\pm$0.11    & yes    & 200.00$<$r$<$800.00                                        \\
51850.006  & 0099280201 & BB+LP     & 0.0192 (fixed) & 0.101$\pm$0.004     & 2.40$\pm$0.01     & 0.18$\pm$0.02    & 1.5107 (258.32/171)  & 28.645$\pm$0.025   & yes    & 100.00$<$r$<$800.00                                        \\
51861.9324 & 0099280301 & BB+LP     & 0.0192 (fixed) & 0.109$\pm$0.002     & 2.13$\pm$0.01     & 0.34$\pm$0.01    & 2.3361 (401.81/172)  & 99.991$\pm$0.095   & yes    & 250.00$<$r$<$800.00                                        \\
52037.3989 & 0136540101 & BB+LP     & 0.0192 (fixed) & 0.100$\pm$0.002     & 2.13$\pm$0.01     & 0.29$\pm$0.02    & 1.9353 (330.94/171)  & 74.695$\pm$0.08    & yes    & 300.00$<$r$<$1000.00                                       \\
52582.0308 & 0136540301 & LP+Lor    & 0.0192 (fixed) & -      & 2.34$\pm$0.00     & 0.34$\pm$0.01    & 2.5112 (424.39/169)  & 55.339$\pm$0.52    & yes    & 290.65 (314.65)$<$RAWX$<$ 306.65 (330.65) \\
52582.3202 & 0136540401 & BB+LP+Lor & 0.0192 (fixed) & 0.151$\pm$0.007     & 2.23$\pm$0.01     & 0.22$\pm$0.01    & 2.7944 (497.40/178)  & 85.644$\pm$0.68    & no     & 284.39$<$RAWX$<$334.39                                     \\
52592.8741 & 0136540801 & BB+LP     & 0.0192 (fixed) & 0.108$\pm$0.003     & 1.84$\pm$0.04     & 0.49$\pm$0.04    & 1.6255 (263.34/162)  & 127.37$\pm$0.35    & yes    & 500.00$<$r$<$1200.00                                       \\
52609.9727 & 0136541001 & BB+LP     & 0.0037$\pm$0.0021 & 0.058$\pm$0.005     & 2.26$\pm$0.01     & 0.19$\pm$0.01    & 3.4342  (580.39/169)  & 51.353$\pm$0.115   & no     & 16.00$<$RAWX$<$56.00                                       \\
52610.8351 & 0136541101 & BB+LP     & 0.0192 (fixed) & 0.091$\pm$0.004     & 2.16$\pm$0.04     & 0.40$\pm$0.05    & 1.4890 (226.32/152)  & 59.842$\pm$0.195   & yes    & 500.00$<$r$<$1200.00                                       \\
52611.0031 & 0136541201 & BB+LP     & 0.0192 (fixed) & 0.094$\pm$0.003     & 2.02$\pm$0.03     & 0.33$\pm$0.03    & 1.8367 (295.70/161)  & 62.533$\pm$0.17    & yes    & 300.00$<$r$<$1200.00                                       \\
52957.6897 & 0150498701 & BB+LP     & 0.0192 (fixed) & 0.096$\pm$0.003     & 2.13$\pm$0.02     & 0.38$\pm$0.03    & 1.7040 (287.97/169)  & 117.67$\pm$0.25    & yes    & 17.01 (41.01)$<$RAWX$<$33.01 (57.01)     \\
52398.6795 & 0153950601 & BB+LP+Lor & 0.0192 (fixed) & 0.030$\pm$0.002     & 2.64$\pm$0.01     & 0.18$\pm$0.02    & 2.7949 (480.72/172)  & 31.898$\pm$0.095   & no     & 284.68$<$RAWX$<$324.68                                     \\
52399.1911 & 0153950701 & BB+LP     & 0.0192 (fixed) & 0.098$\pm$0.005     & 2.61$\pm$0.04     & 0.37$\pm$0.06    & 1.0920 (150.70/138)  & 18.775$\pm$0.06    & yes    & 350.00$<$r$<$1200.00                                       \\
53681.8447 & 0153951201 & BB+LP     & 0.0192 (fixed) & 0.094$\pm$0.007     & 2.16$\pm$0.04     & 0.34$\pm$0.06    & 1.1953 (188.86/158)  & 119.23$\pm$0.45    & yes    & 17.51 (41.51)$<$RAWX$<$33.51 (57.51)     \\
53681.7058 & 0153951301 & LP        & 0.0192 (fixed) & -      & 2.34$\pm$0.04     & -0.0$\pm$0.11    & 0.82803(119.24/144) & 1.2443$\pm$0.0235  & no     & 16.55$<$RAWX$<$46.55                                       \\
52791.557  & 0158970101 & BB+LP+Lor & 0.0192 (fixed) & 0.143$\pm$0.004     & 2.41$\pm$0.01     & 0.35$\pm$0.02    & 3.0534 (546.55/179)  & 47.302$\pm$0.045   & no     & 290.79$<$RAWX$<$330.79                                     \\
52792.0603 & 0158970201 & BB+LP     & 0.0192 (fixed) & 0.090$\pm$0.006     & 2.35$\pm$0.06     & 0.64$\pm$0.08    & 1.0577 (138.56/131)  & 24.23 $\pm$0.115   & no     & r$<$1600.00                                                      \\
52797.897  & 0158970701 & BB+LP+Lor & 0.0192 (fixed) & 0.153$\pm$0.007     & 2.54$\pm$0.01     & 0.35$\pm$0.02    & 1.8045 (315.79/175)  & 26.594$\pm$0.045   & no     & 292.71$<$RAWX$<$332.71                                     \\
53131.1251 & 0158971201 & BB+LP     & 0.0192 (fixed) & 1.126$\pm$0.064     & 2.10$\pm$0.01     & 0.06$\pm$0.02    & 1.6384 (278.53/170)  & 133.81$\pm$0.25    & yes    & 17.54 (40.54)$<$RAWX$<$34.54 (57.54)    \\
53683.7759 & 0158971301 & BB+LP     & 0.0192 (fixed) & 0.102$\pm$0.003     & 2.22$\pm$0.02     & 0.41$\pm$0.03    & 1.9563 (330.61/169)  & 112.43$\pm$0.2     & yes    & 16.48 (40.48)$<$RAWX$<$32.48 (56.48)     \\
52983.8975 & 0162960101 & BB+LP     & 0.0192 (fixed) & 0.090$\pm$0.003     & 2.15$\pm$0.02     & 0.32$\pm$0.02    & 1.4069 (237.77/169)  & 63.088$\pm$0.085   & yes    & 250.00$<$r$<$1000.00                                       \\
53854.8676 & 0302180101 & BB+LP+Lor & 0.0192 (fixed) & 0.145$\pm$0.003     & 2.02$\pm$0.01     & 0.21$\pm$0.01    & 4.7147 (881.66/187)  & 93.475$\pm$0.095   & no     & 283.08$<$RAWX$<$333.08                                     \\
53883.0932 & 0411080301 & BB+LP+Lor & 0.0192 (fixed) & 0.097$\pm$0.001     & 1.87$\pm$0.01     & 0.25$\pm$0.01    & 4.7907 (828.80/173)  & 198.76$\pm$0.15    & yes    & 300.00$<$r$<$1000.00                                       \\
54074.5064 & 0411080701 & BB+LP     & 0.0192 (fixed) & 0.074$\pm$0.004     & 2.46$\pm$0.02     & 0.30$\pm$0.03    & 1.1580 (192.23/166)  & 36.226$\pm$0.12    & yes    & 15.99 (37.99)$<$RAWX$<$33.99 (55.99)  \\
54230.1689 & 0411081301 & BB+LP     & 0.0192 (fixed) & 0.081$\pm$0.005     & 2.48$\pm$0.04     & 0.46$\pm$0.05    & 1.1083 (156.27/141)  & 41.803$\pm$0.14    & yes    & 500.00$<$r$<$1200.00                                       \\
54230.4143 & 0411081401 & BB+LP     & 0.0192 (fixed) & 0.070$\pm$0.009     & 2.48$\pm$0.04     & 0.55$\pm$0.07    & 1.0065 (126.82/126)  & 47.031$\pm$0.205   & yes    & 500.00$<$r$<$1200.00                                       \\
54230.5439 & 0411081501 & BB+LP     & 0.0192 (fixed) & 0.078$\pm$0.007     & 2.45$\pm$0.04     & 0.45$\pm$0.05    & 1.1346 (154.30/136)  & 47.145$\pm$0.165   & yes    & 450.00$<$r$<$1200.00                                       \\
54230.6735 & 0411081601 & BB+LP     & 0.0192 (fixed) & 0.082$\pm$0.012     & 2.46$\pm$0.06     & 0.48$\pm$0.10    & 1.0281 (119.26/116)  & 45.171$\pm$0.26    & yes    & 500.00$<$r$<$1200.00                                       \\
54423.5489 & 0411081901 & BB+LP     & 0.0192 (fixed) & 0.051$\pm$0.013     & 2.24$\pm$0.02     & 0.27$\pm$0.03    & 1.2642 (202.28/160)  & 71.484$\pm$0.285   & yes    & 400.00$<$r$<$1200.00                                       \\
54617.1091 & 0411082701 & BB+LP     & 0.0192 (fixed) & 0.095$\pm$0.004     & 2.07$\pm$0.03     & 0.53$\pm$0.04    & 1.2858 (205.73/160)  & 255.16$\pm$0.6     & yes    & 600.00$<$r$<$1200.00                                       \\
55151.7552 & 0411083201 & BB+LP     & 0.0192 (fixed) & 0.101$\pm$0.003     & 1.93$\pm$0.02     & 0.38$\pm$0.03    & 1.3597 (228.44/168)  & 140.58$\pm$0.3     & yes    & 400.00$<$r$<$1200.00                                       \\
54593.0812 & 0502030101 & BB+LP     & 0.0192 (fixed) & 0.090$\pm$0.003     & 2.33$\pm$0.01     & 0.21$\pm$0.02    & 1.2511 (211.43/169)  & 90.554$\pm$0.14    & yes    & 18.43 (41.43)$<$RAWX$<$35.43 (58.43)     \\
54228.6283 & 0510610101 & BB+LP     & 0.0192 (fixed) & 0.039$\pm$0.032     & 2.65$\pm$0.03     & 0.15$\pm$0.06    & 1.2484 (202.24/162)  & 33.974$\pm$0.7     & yes    & 17.52 (40.52)$<$RAWX$<$34.52 (57.52)     \\
54228.3529 & 0510610201 & BB+LP     & 0.0192 (fixed) & 0.038$\pm$0.018     & 2.63$\pm$0.01     & 0.11$\pm$0.03    & 1.2361 (203.96/165)  & 36.129$\pm$0.555   & yes    & 17.44 (40.44)$<$RAWX$<$34.44 (57.44)     \\
54792.6095 & 0560980101 & BB+LP     & 0.0192 (fixed) & 0.098$\pm$0.004     & 2.32$\pm$0.03     & 0.54$\pm$0.04    & 1.1647 (180.53/155)  & 71.664$\pm$0.15    & yes    & 400.00$<$r$<$1200.00                                       \\
54976.1719 & 0560983301 & BB+LP     & 0.0192 (fixed) & 0.096$\pm$0.004     & 2.50$\pm$0.03     & 0.50$\pm$0.05    & 1.1404 (172.20/151)  & 63.154$\pm$0.135   & yes    & 400.00$<$r$<$1200.00                                       \\
55319.3278 & 0656380101 & BB+LP     & 0.0192 (fixed) & 0.102$\pm$0.003     & 2.05$\pm$0.03     & 0.44$\pm$0.04    & 1.2368 (201.61/163)  & 114.38$\pm$0.25    & yes    & 500.00$<$r$<$1200.00                                       \\
55512.8889 & 0656380801 & BB+LP     & 0.0192 (fixed) & 0.104$\pm$0.004     & 2.17$\pm$0.03     & 0.38$\pm$0.04    & 1.2674 (204.05/161)  & 74.897$\pm$0.155   & yes    & 400.00$<$r$<$1200.00                                       \\
55514.8844 & 0656381301 & BB+LP     & 0.0192 (fixed) & 0.091$\pm$0.005     & 2.44$\pm$0.03     & 0.38$\pm$0.04    & 1.1864 (181.51/153)  & 53.691$\pm$0.13    & yes    & 350.00$<$r$<$1200.00                                       \\
55698.4452 & 0658800101 & BB+LP     & 0.0192 (fixed) & 0.095$\pm$0.006     & 2.53$\pm$0.05     & 0.40$\pm$0.07    & 1.3032 (179.84/138)  & 36.678$\pm$0.13    & yes    & 300.00$<$r$<$1200.00                                       \\
55894.0036 & 0658800801 & BB+LP     & 0.0192 (fixed) & 0.119$\pm$0.012     & 2.80$\pm$0.03     & 0.15$\pm$0.04    & 1.2851 (205.62/160)  & 17.917$\pm$0.055   & no     & 16.55$<$RAWX$<$56.55                                       \\
57179.0079 & 0658801301 & BB+LP     & 0.0192 (fixed) & 0.105$\pm$0.002     & 2.31$\pm$0.01     & 0.30$\pm$0.02    & 2.6138 (449.57/172)  & 83.575$\pm$0.075   & yes    & 200.00$<$r$<$1000.00                                       \\
57334.6077 & 0658801801 & BB+LP     & 0.0192 (fixed) & 0.116$\pm$0.005     & 2.65$\pm$0.01     & 0.28$\pm$0.02    & 1.7979 (298.45/166)  & 47.076$\pm$0.055   & yes    & 200.00$<$r$<$1000.00                                       \\
57514.17   & 0658802301 & BB+LP     & 0.0192 (fixed) & 0.121$\pm$0.007     & 2.65$\pm$0.01     & 0.25$\pm$0.02    & 1.4298 (237.34/166)  & 43.549$\pm$0.055   & yes    & 200.00$<$r$<$1000.00                                       \\
56776.1859 & 0670920301 & BB+LP     & 0.0192 (fixed) & 0.099$\pm$0.007     & 2.49$\pm$0.04     & 0.20$\pm$0.05    & 1.2370 (197.92/160)  & 91.905$\pm$0.355   & yes    & 18.12 (42.12)$<$RAWX$<$34.12 (58.12)     \\
56778.1597 & 0670920401 & BB+LP     & 0.0192 (fixed) & 0.104$\pm$0.014     & 2.68$\pm$0.03     & 0.15$\pm$0.04    & 0.97964 (159.68/163) & 56.571$\pm$0.17    & yes    & 18.00 (41.00)$<$RAWX$<$35.00 (58.00)     \\
56780.1518 & 0670920501 & BB+LP     & 0.0192 (fixed) & 0.099$\pm$0.006     & 2.38$\pm$0.02     & 0.09$\pm$0.03    & 1.1384 (188.97/166)  & 82.456$\pm$0.215   & yes    & 17.74 (40.74)$<$RAWX$<$34.74 (57.74)    \\
57695.5677 & 0791780101 & BB+LP     & 0.0192 (fixed) & 0.128$\pm$0.011     & 2.62$\pm$0.02     & 0.25$\pm$0.03    & 0.94886 (148.97/157) & 16.435$\pm$0.035   & yes    & 100.00$<$r$<$1000.00                                       \\
57877.186  & 0791780601 & BB+LP     & 0.0102$\pm$0.0034 & 0.126$\pm$0.007     & 2.06$\pm$0.03     & 0.28$\pm$0.03    & 1.8218 (309.71/170)  & 106.97$\pm$0.15    & yes    & 200.00$<$r$<$1000.00                                                           \\
\enddata
\end{deluxetable*}
\end{longrotatetable}
%%%%%%%%%%%%%

\section{Results}

After performing the procedures given in Section \ref{sec:obser}, 
we extract the 100-second sampled XMM-Newton X-ray light curves of Mrk 421.
All fifty light curves are shown in Appendix where we also give the normalized power spectrum density (NPSD) and the fraction root mean square (rms) variability amplitude for each light curve.

\subsection{Light curve fitting}

We aim to determine the parameters of each flare-profile, such as rise and decay times.
Therefore, we select these light curves, which have one complete flare-profile at least. 
17 out of 50 light curves are accordingly selected.

We fit each flare-profile in the 17 light curves using the function given by \citet{2010ApJ...722..520A},
\begin{equation}\label{fun:lcfit}
F(t) = F_c+F_0\left( e^{\frac{t_0-t}{T_r}}+e^{\frac{t-t_0}{T_d}}\right)^{-1}\ , 
\end{equation}
where $F_c$ is a constant, $ F_0 $ the amplitude of the peak, $ t_0 $ the time of the peak when the peak is symmetric,  and $ T_r $ and $T_d $ respectively represent the rise and decay time. 
For multi-peak light curves, we consider one light curve as a whole, and fit the light curve using a model comprising of several components of equation (\ref{fun:lcfit}). 

We first identify the clear flares in each light curve by our eyes. We then use a model with as many components as
the flares' number to fit each light curve. In order to examine this procedure, we calculate the distribution of the residuals
(i.e., the ratio of the difference between the observed count and the modeled one and the count error) for each fitting.
If our identification for the flares is correct, the distribution of the residuals should be compatible with a constant level.
If there are peaks in the distribution of the residuals, we accordingly increase the flares' number and re-fit the light curve. 
If the parameters of the added flares are well constrained, the added flares are significant. 
If the parameters of the added flares are poorly constrained, the added flares are not significant, and we will not adopt the results of the re-fitting.

The best-fitting results for the light curves are show in Fig.~\ref{fig:lcfit}, and the best-fitting parameters are given in Table \ref{tab:lcfit}. 
We also show the distribution of the residuals obtained in each fitting.
Each distribution of the residuals is fitted by a constant.
The best-fitting results are given in Fig.~\ref{fig:lcfit}.
One can see that except for some random events, our procedure works satisfactorily.

Here, we convert peak count rate into peak flux using Portable Interactive Multi-Mission Simulator (PIMMS)\footnote{\url{https://heasarc.gsfc.nasa.gov/docs/software/tools/pimms.html}} . 
We use a single power-law model with $N_{H}=1.92\times10^{20}\ \rm{cm}^{2}$, 
and set the photon index as the $\alpha$ given in Table \ref{tab:phafit}. 
The peak fluxes are given in Table \ref{tab:lcfit}.

We obtain the rise and decay times as well as the peak flux for 48 flares.
However, one can see that in 22 flares the rise or decay time is poorly constrained.

\begin{figure}
\centering
\includegraphics[scale=0.50]{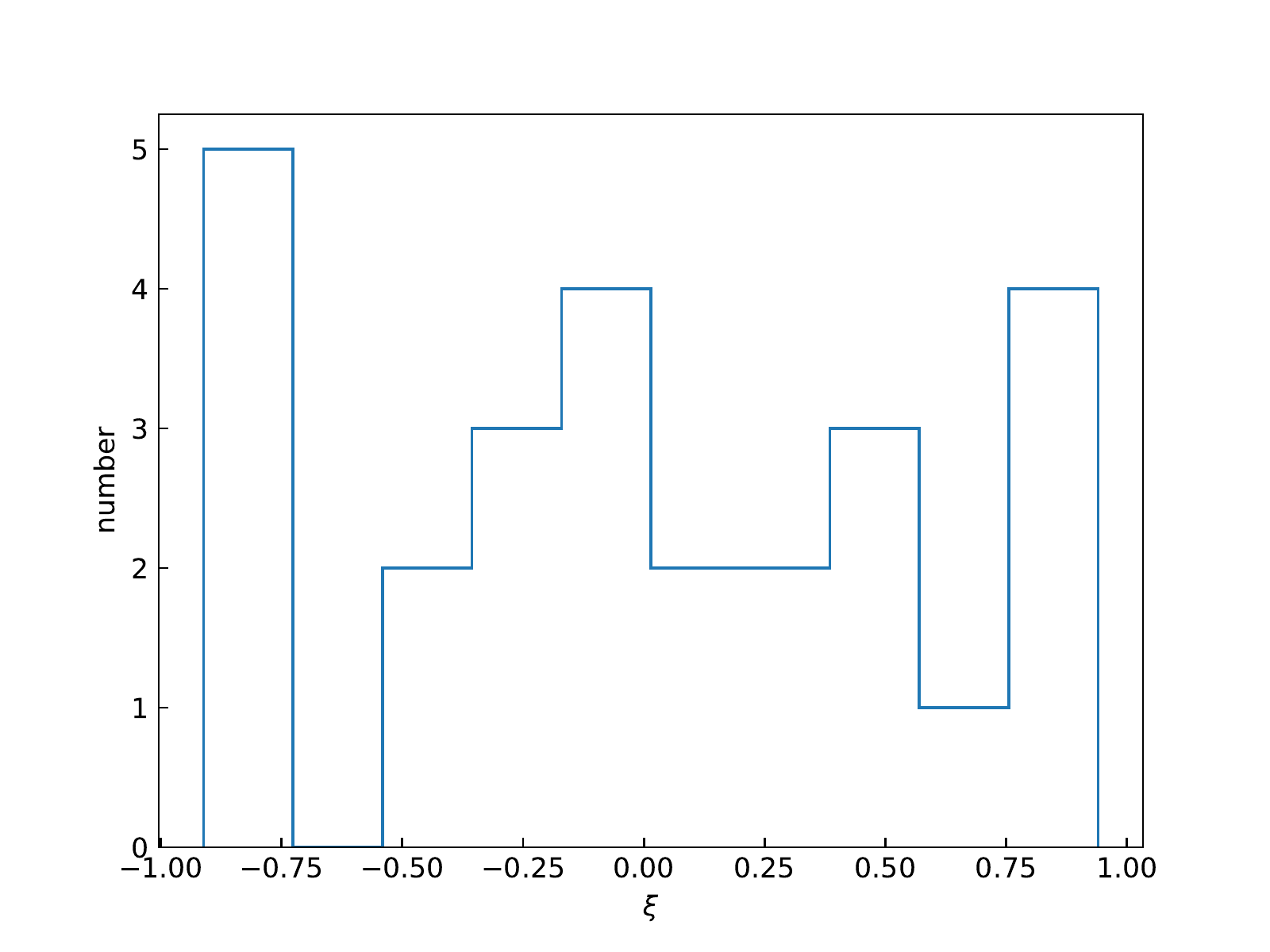} 
\caption{Distribution of flare symmetry parameter $\xi$. \label{fig:xi}}
\end{figure}

The following parameter can be defined to describe the symmetry of the flares:
\begin{equation}\label{td}
\xi=\frac{T_{d}-T_{r}}{T_{d}+T_{r}}\ , 
\end{equation}
which is in the range of -1 to 1, indicating completely right (-1) and left (1)
asymmetric flares, respectively. The distribution of is shown in Fig.~\ref{fig:xi}.
No tendency is found in this distribution.

\subsection{Distributions of peak flux and flaring time duration}

The flaring time duration is defined as \citep{2010ApJ...722..520A}
\begin{equation}\label{td}
T_{fl}\simeq2(T_{r}+T_{d})\ , 
\end{equation}
which are listed in Table~\ref{tab:lcfit}. 

\begin{figure}[!h]
%\figurenum{1}
\plottwo{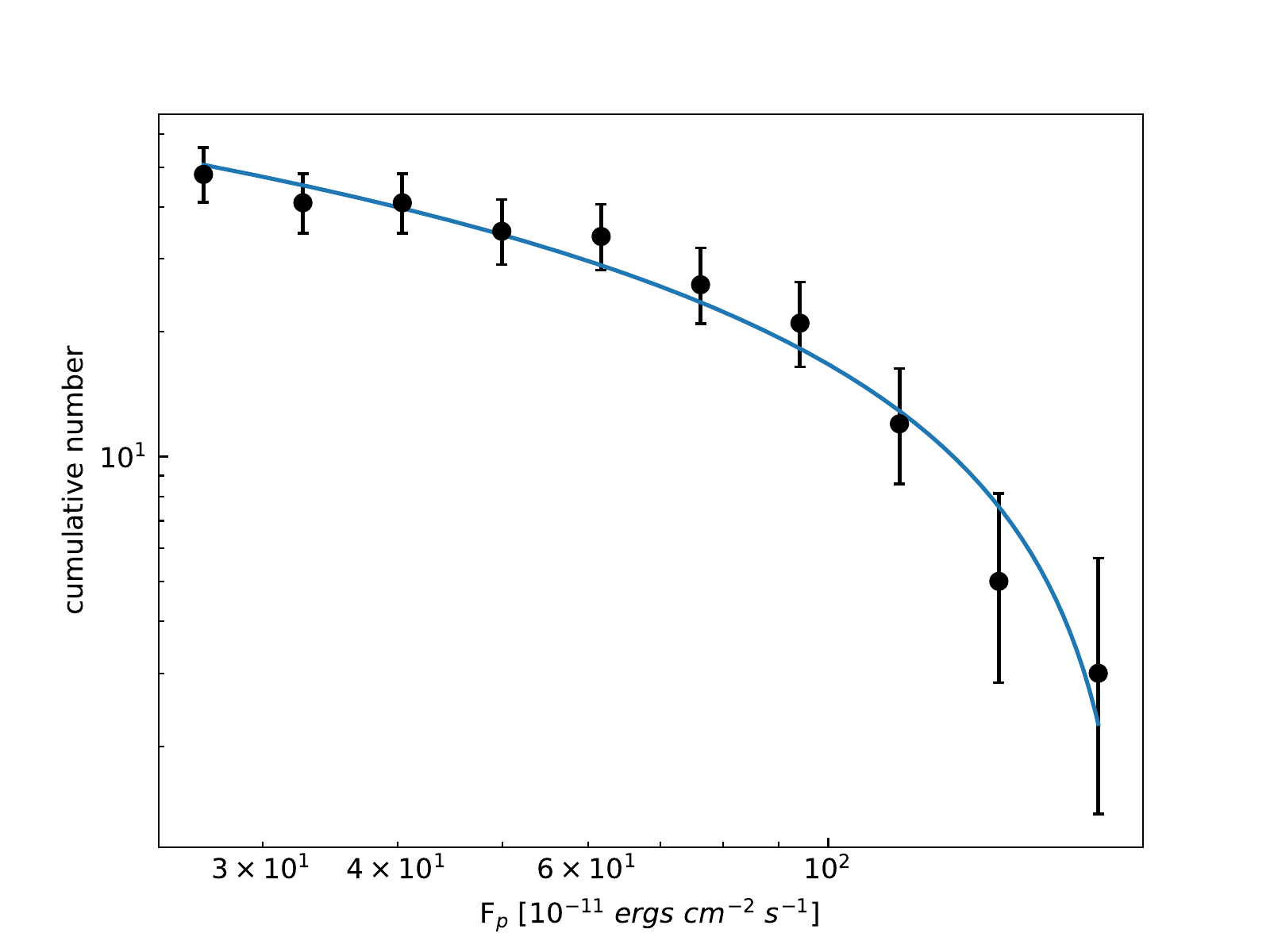}{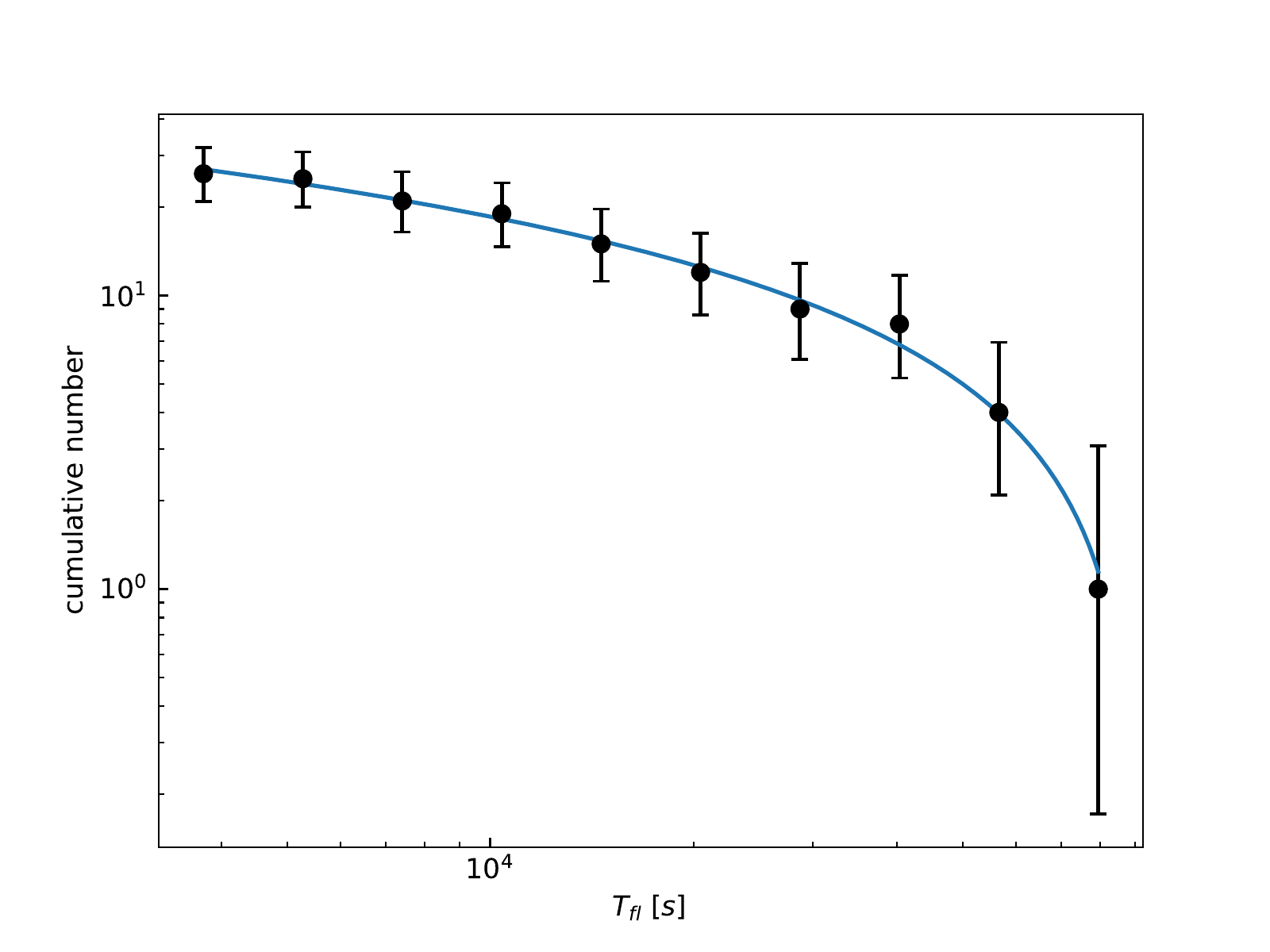}
\caption{Cumulative distributions of peak flux (left) and flaring time duration (right). The solid lines are the best-fitting results to the distributions with the equation~(\ref{N}). \label{fig:dis}}
\end{figure}

The binned cumulative distributions of peak flux and flaring time duration are shown in Fig.~\ref{fig:dis}.
The errors are estimated by the method given in \citet{Gehrels}.
Note that the 22 flares with poorly constrained rise or decay time are excluded in the cumulative distribution of flaring time duration .
The two distributions are fitted with the following function through a $\chi^2$-minimization procedure,
\begin{equation}\label{N}
N(>x)=a+b(x^{1-\alpha}-x_{\rm max}^{1-\alpha})\ , 
\end{equation}
where $\alpha$ is the index of the distribution of $N(x)$, i.e., $N(x)\propto x^{-\alpha}$; $x_{\rm max}$ is a cutoff parameter; 
$a$ and $b$ is two parameters. We obtain $\alpha_{\rm F}=1.02\pm0.25$ for the peak flux distribution with the reduced $\chi^2_{\rm r}=0.38$, 
and $\alpha_{\rm T}=1.01\pm0.08$ for the flaring time duration distribution with the reduced $\chi^2_{\rm r}=0.04$.

Alternatively, we also use an unbinned maximum likelihood method to obtain the indices for the distributions of peak flux and flaring time duration.
For a power-law distribution, $P(x)=A\cdot x^{-\alpha}$,  the logarithmic likelihood function of $P(x)$ is 
\begin{equation}
{\rm ln}\ L=-\alpha\sum_{i=1}^N{\rm log}(x_{\rm i})+N\cdot\rm{log}(A)\ .
\label{like}
\end{equation}
We use the Markov Chain
Monte Carlo (MCMC) technique \citep[e.g.,][]{Yan13} to maximize the right side of equation (\ref{like}) and constrain the index $\alpha$.
We obtain $\alpha_{\rm F}=0.99^{+0.24}_{-0.23}$ for the peak flux distribution with $-{\rm ln}\ L=239$, 
and $\alpha_{\rm T}=1.05^{+0.21}_{-0.20}$ for the flaring time duration distribution with $-{\rm ln}\ L=104$. 
The two methods obtain the consistent results.

No correlation is found between $T_{\rm fl}$, $\xi$ and $F_{\rm peak}$.

\subsection{Magnetic field strength constrained by X-ray variability timescale}

Variability of synchrotron emission can be used to constrain the magnetic field strength in the emission region \citep[e.g.,][]{bottcher03}.
If electron cooling is dominated by synchrotron cooling, 
the cooling timescale of electron in comoving frame is written as~\citep[e.g.,][]{Tavecchio98}
\begin{equation}
 t'_{\rm cooling} 
          = \frac{6\pi m_ec}{\sigma_T\gamma B'^2}\ ,
\label{tcool}
\end{equation}
where $ B'$ is the magnetic field in comoving frame; 
$m_e$ is the electron rest mass; $\sigma_T$ is the cross section of Thomson scattering; 
$\gamma$ is the Lorentz factor of electron.  The observational synchrotron photon energy is
\begin{equation}
 E_{\rm syn} \approx1.5\times10^{-11} \gamma^2 \frac{B'}{1\ \rm G} \frac{\delta_{\rm D}}{1+z}\ \rm keV\  ,
\label{nu}
\end{equation}
where $z$ is redshift and $\delta_{\rm D}$ is Doppler factor.

The observational variability timescale  $t_{\rm var}$ should be longer than (or equal to at least) the  cooling timescale in observer frame $t_{\rm cooling}=t'_{\rm cooling}(1+z)/\delta_{\rm D}$, i.e., $t_{\rm var} \geq t_{\rm cooling}$ \citep[e.g.,][]{bottcher03,Paliya2}. We therefor derive the lower limit for magnetic field strength from equation~(\ref{tcool}) and~(\ref{nu}), i.e.,
\begin{equation}
 B' \geq 2.1 (\frac{t_{\rm var}}{1000\ \rm s})^{-2/3} (\frac{E_{\rm syn}}{1\ \rm keV})^{-1/3} \delta_{\rm D}^{-1/3} (1+z)^{1/3}\ \rm G\ .
\label{estimatemag}
\end{equation}

In Table~\ref{tab:lcfit}, one can see that the decay or rise time can be short as $\sim$400 s.
In many flares, the decay or rise time can be short as $\sim$1000 s.
Considering that the X-ray emission of Mrk 421 is believed to the synchrotron radiation of electrons,
taking $t_{\rm var}=1000\ $s and $E_{\rm syn}=1\ $keV, we derive $B' \geq 2.1\delta_{\rm D}^{-1/3}$ G.

\begin{figure}[!h]
%\figurenum{1}
\plottwo{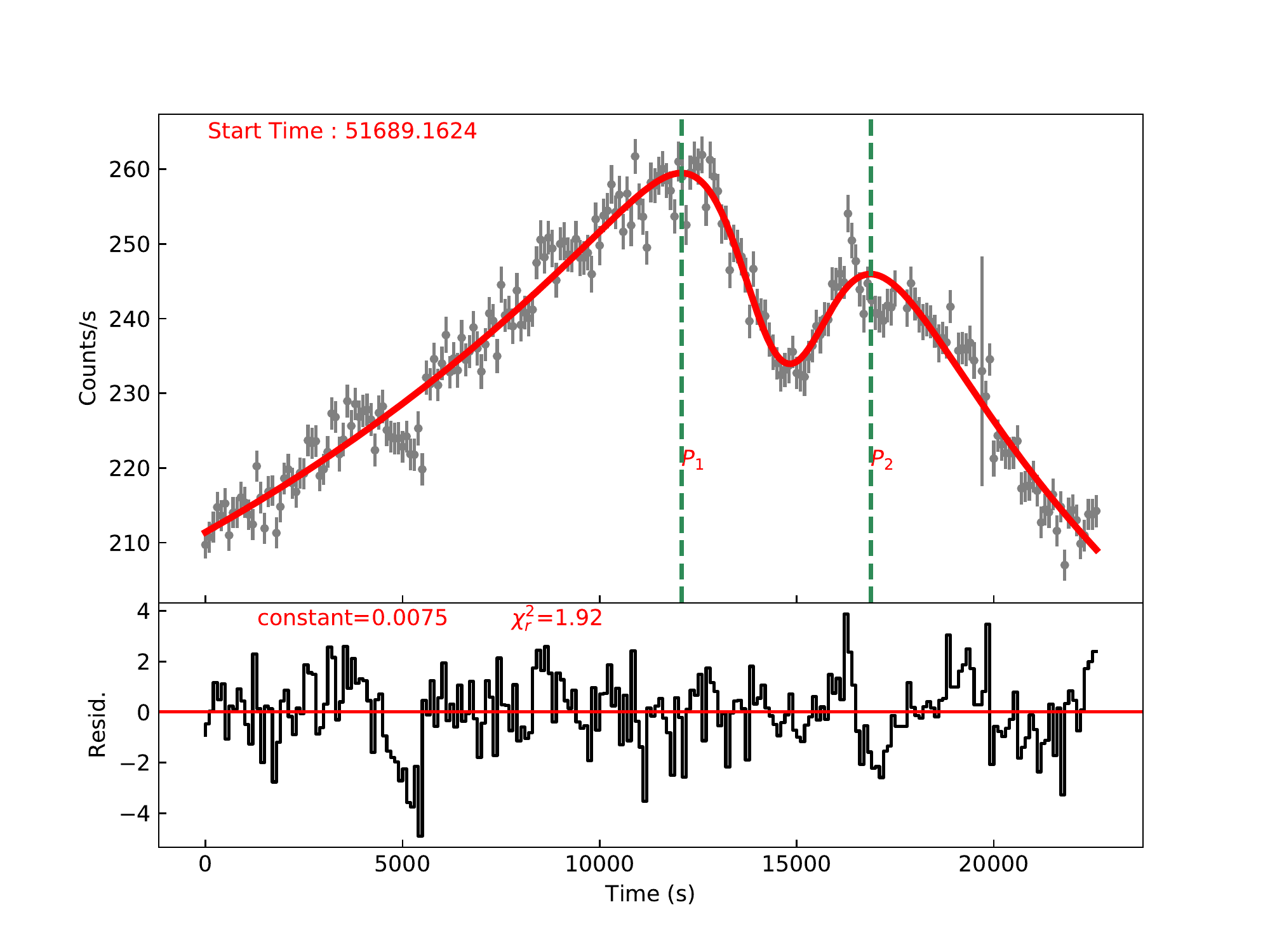}{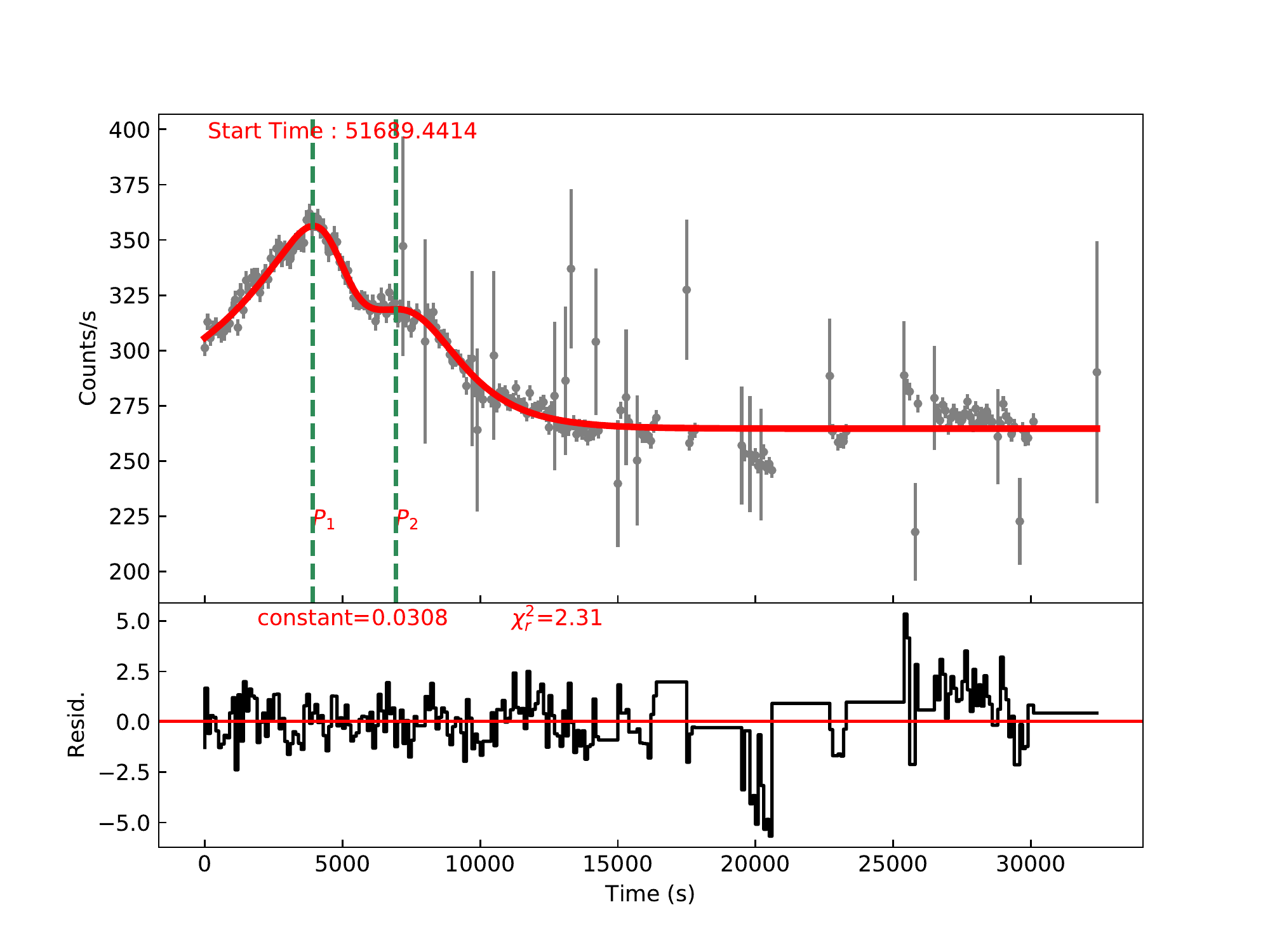}
\plottwo{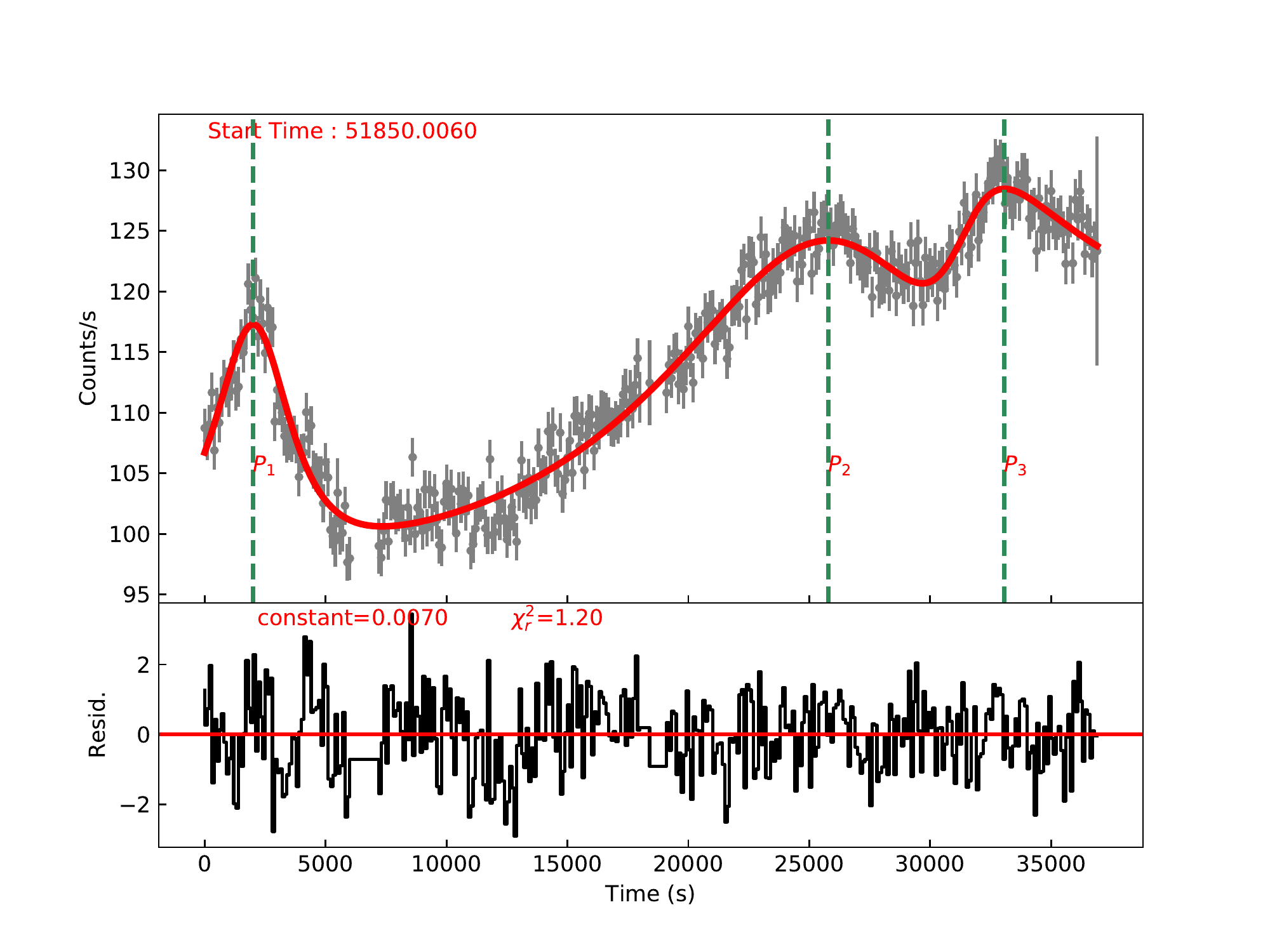}{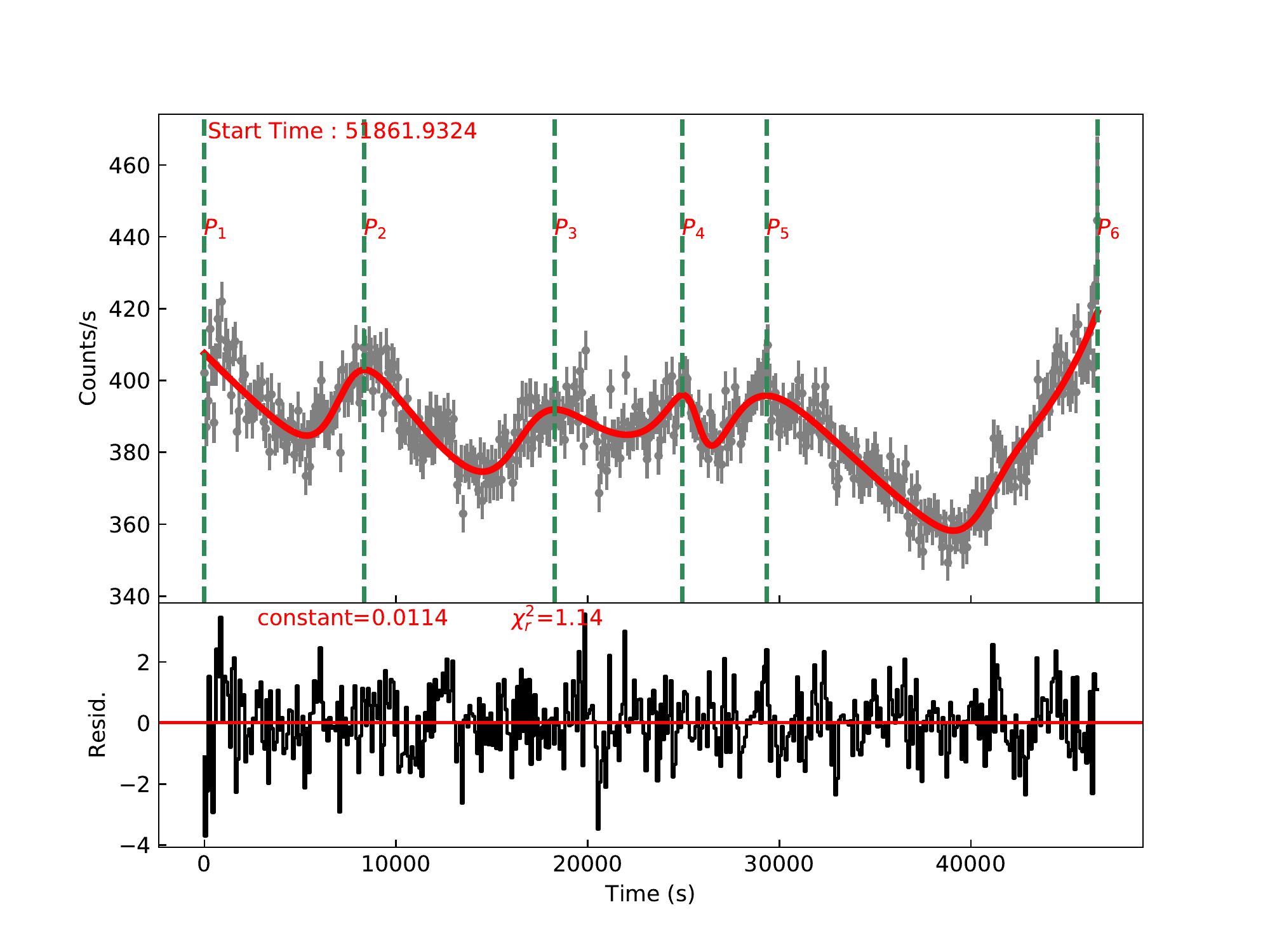}
\plottwo{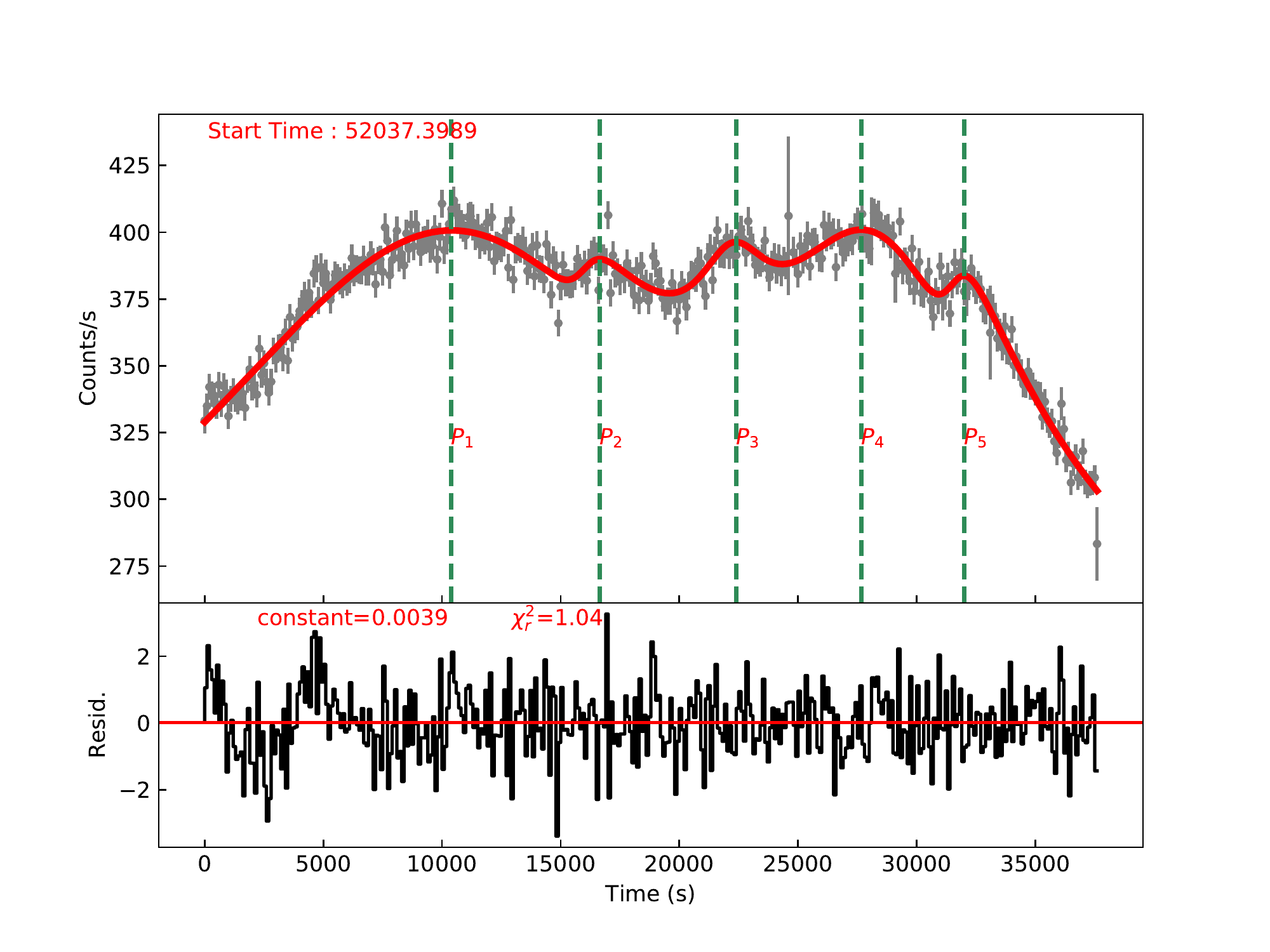}{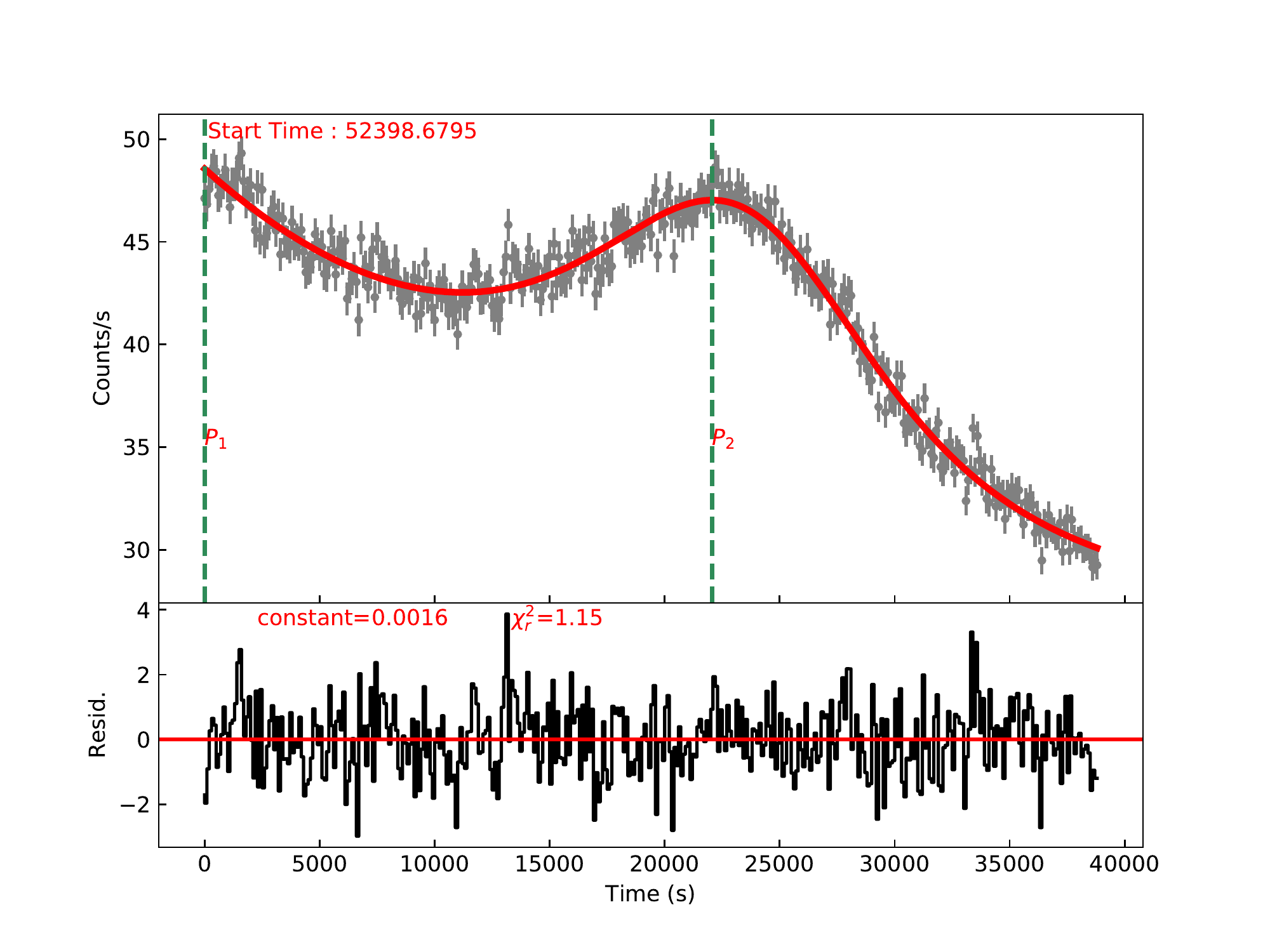}
\plottwo{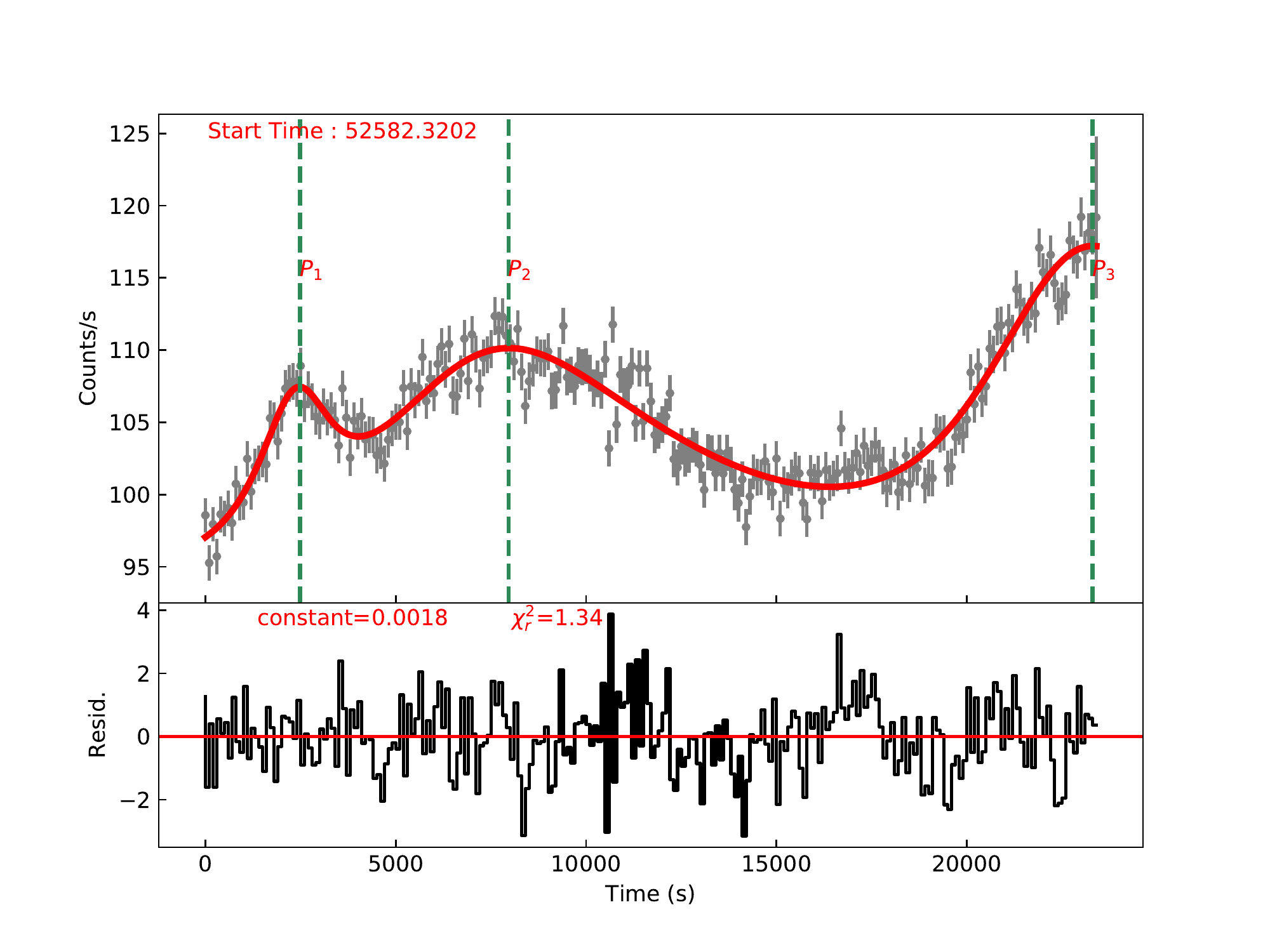}{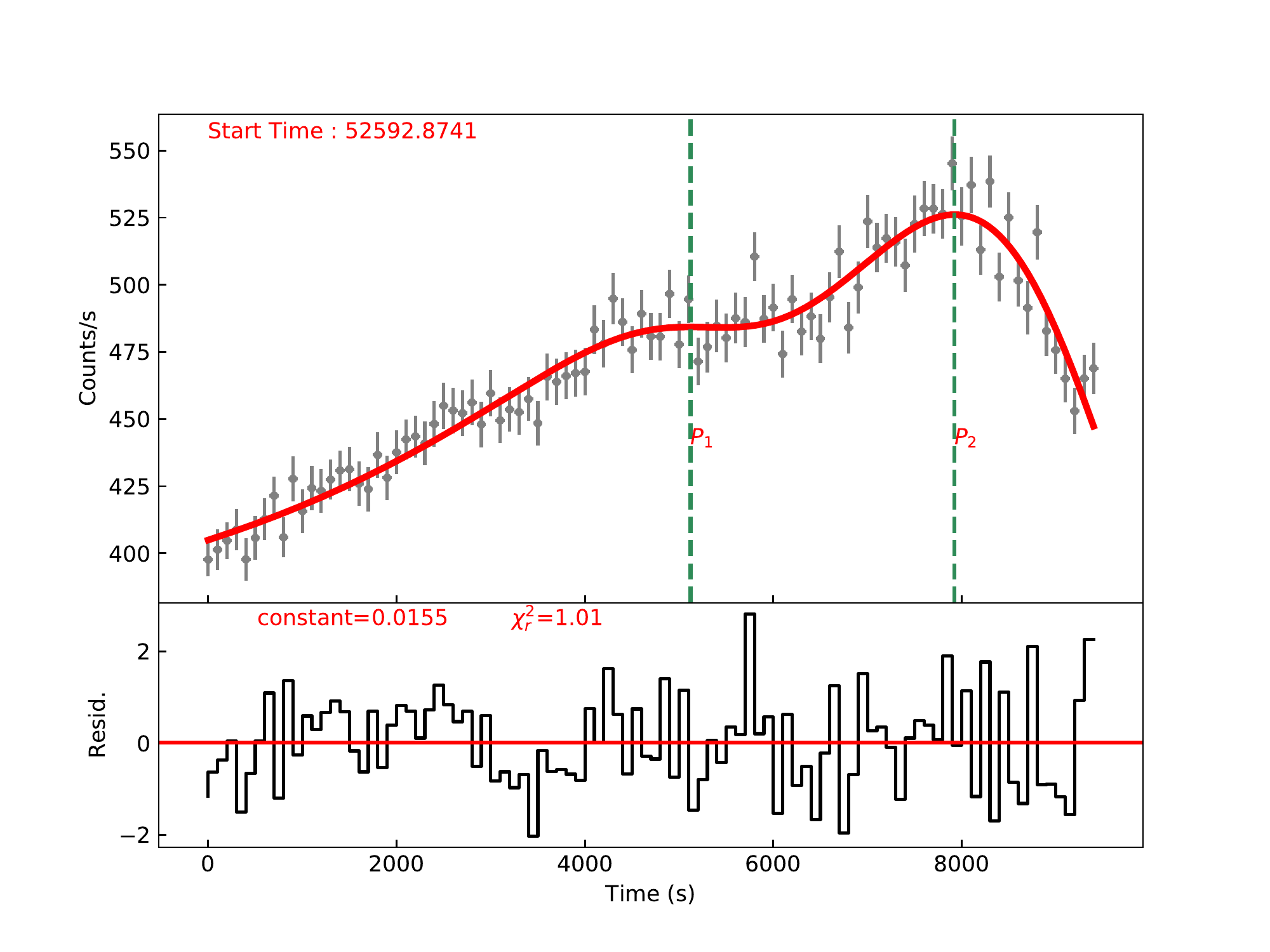}
\caption{The best-fitting results for light curves. In each panel: the gray points are the observation data; the thick red line is the best-fitting model; 
the green vertical line indicates the light curves peaking at these time, and every peak is marked by a red word. 
The distribution of the residuals obtained in each fitting is also showed (the bottom in each panel), which is fitted by a constant (thin red line). \label{fig:lcfit}}
\end{figure}

\begin{figure}[!h]
\figurenum{1}
\plottwo{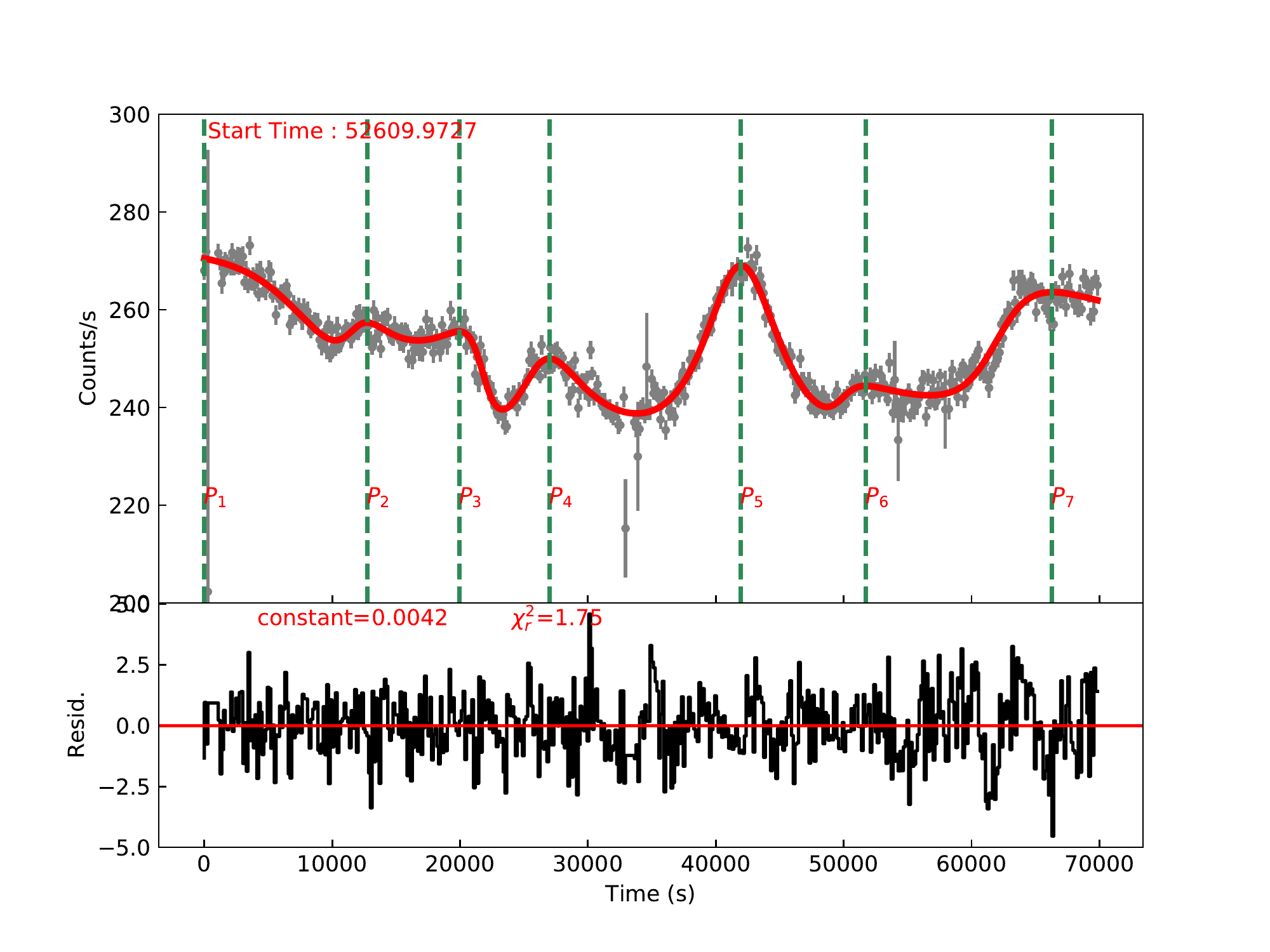}{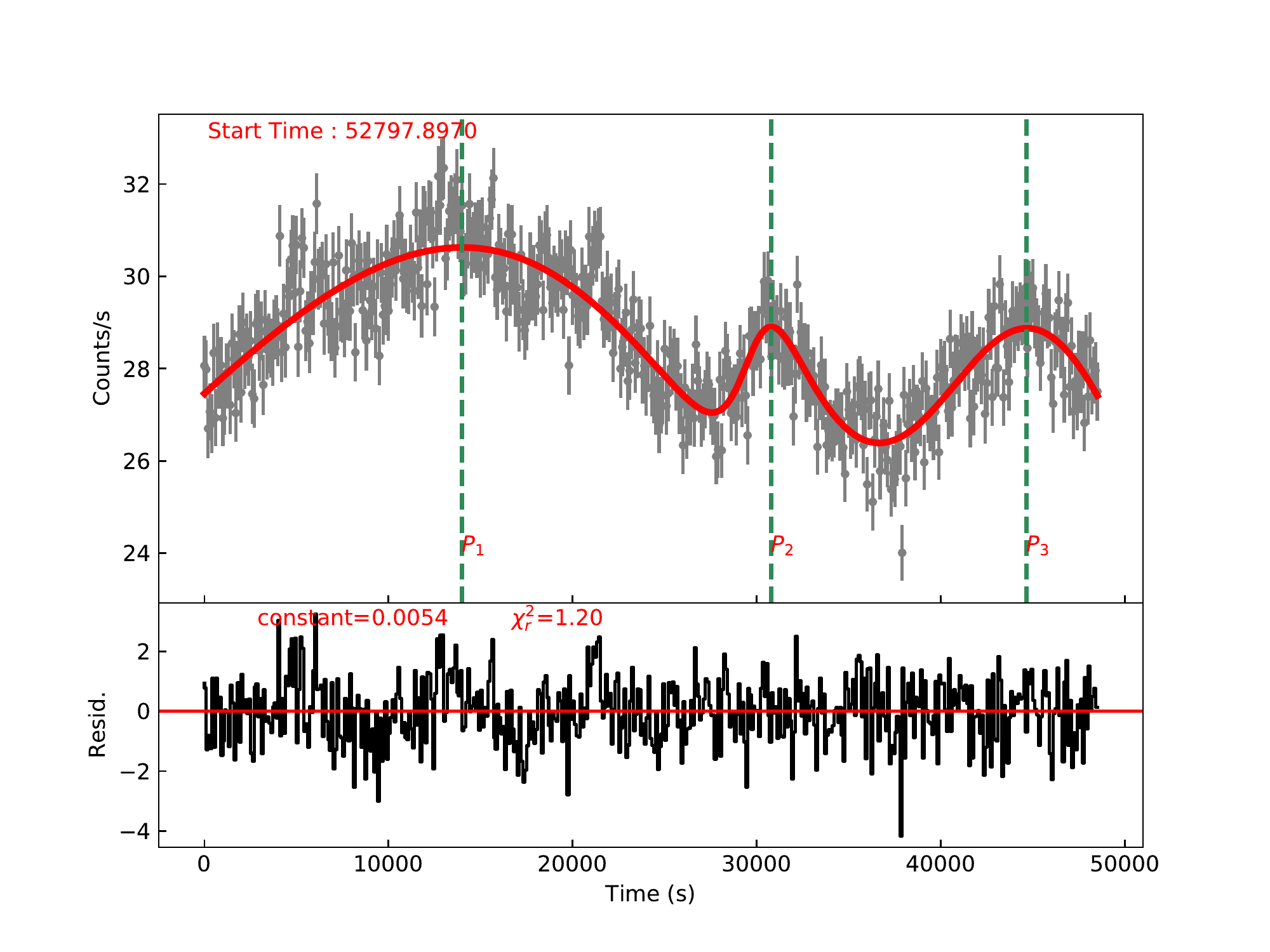}
\plottwo{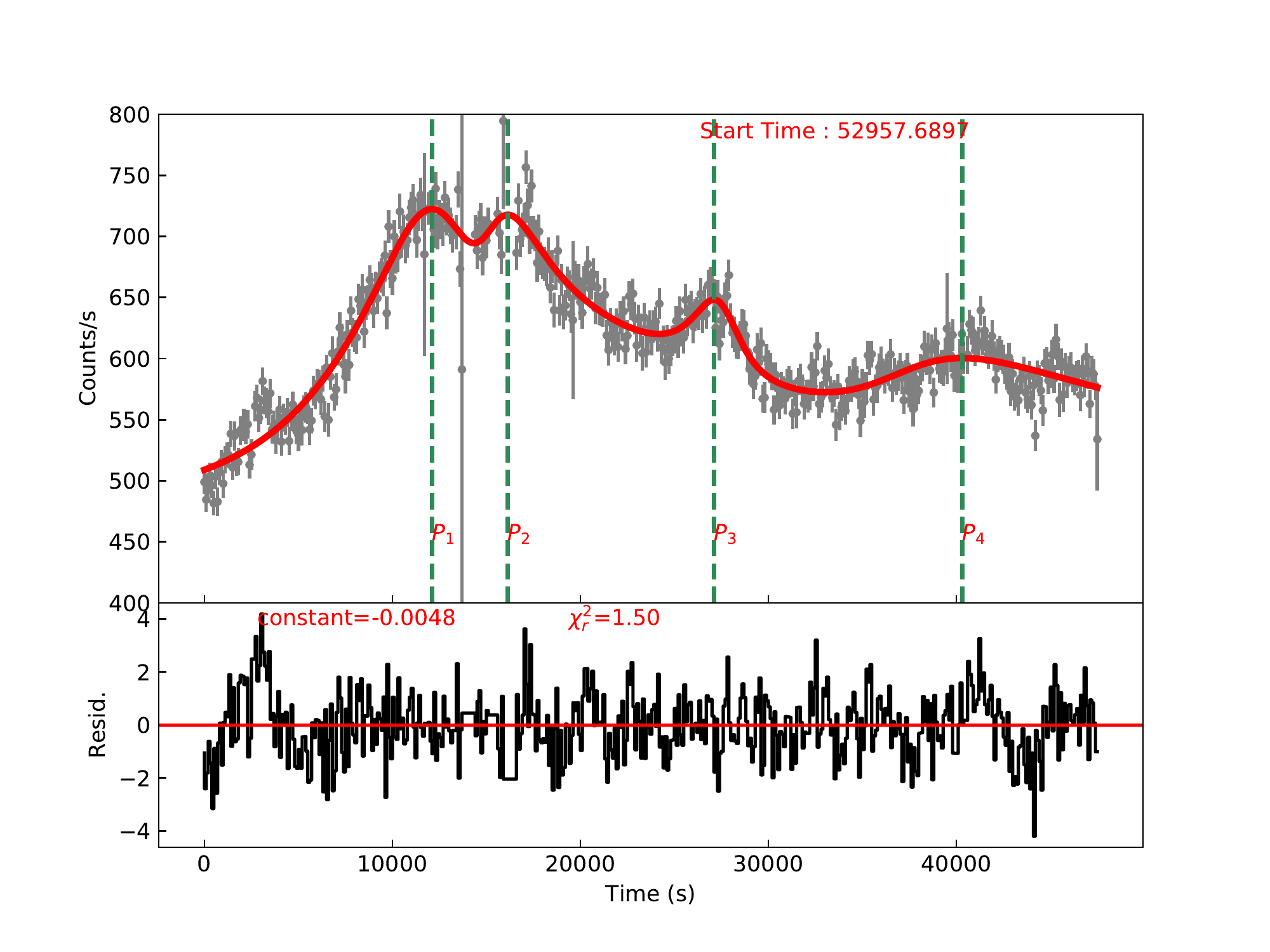}{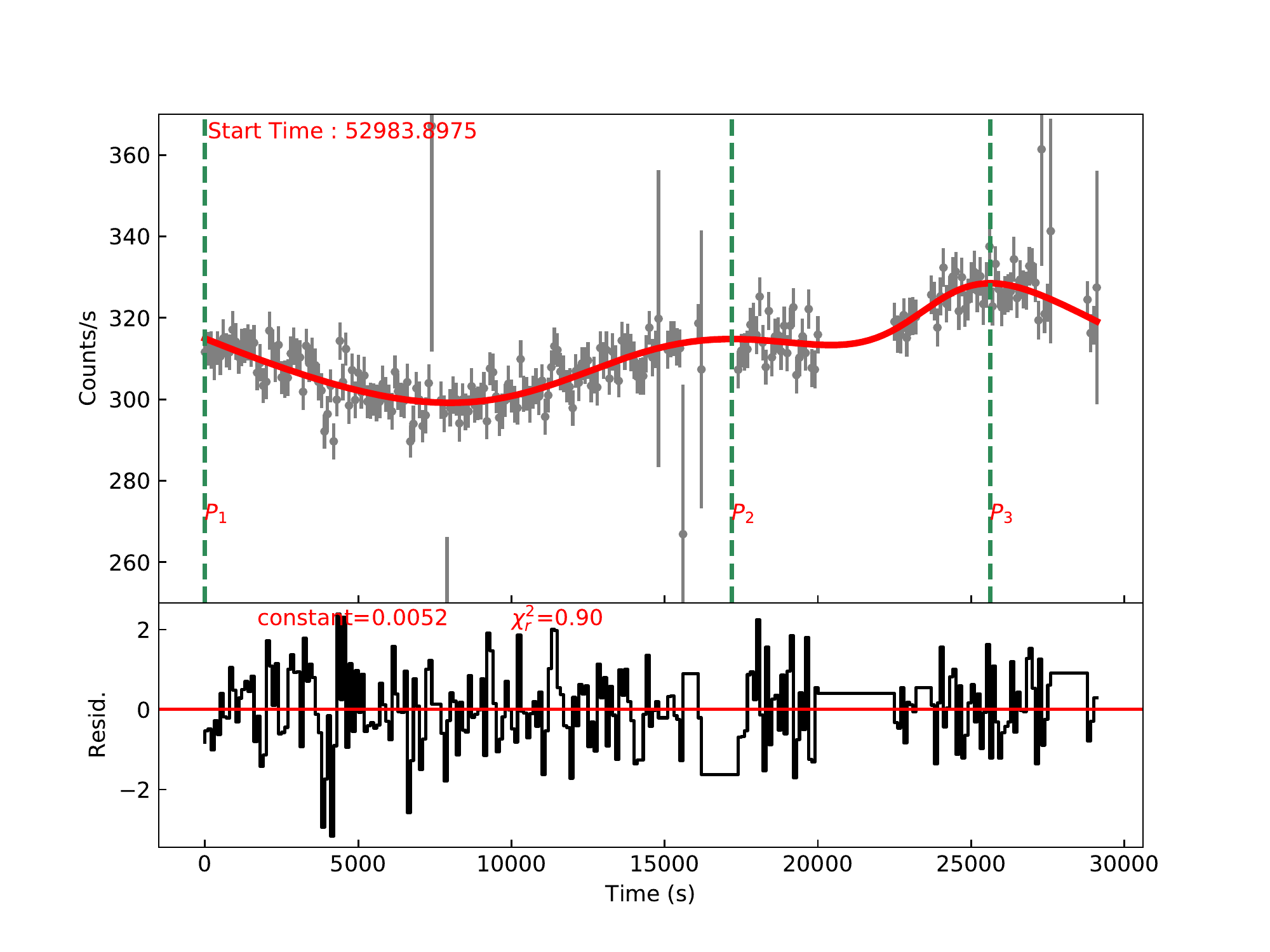}
\plottwo{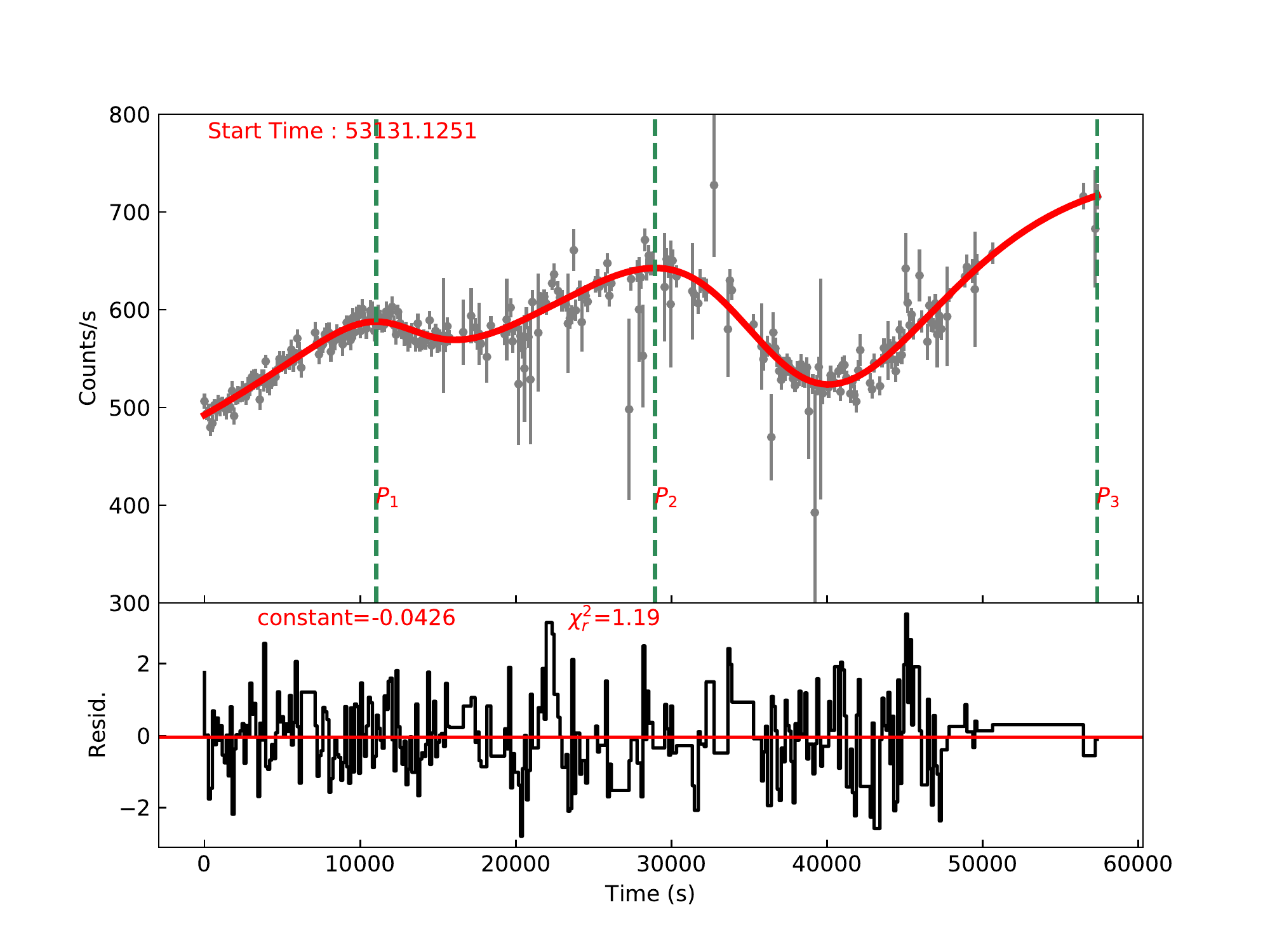}{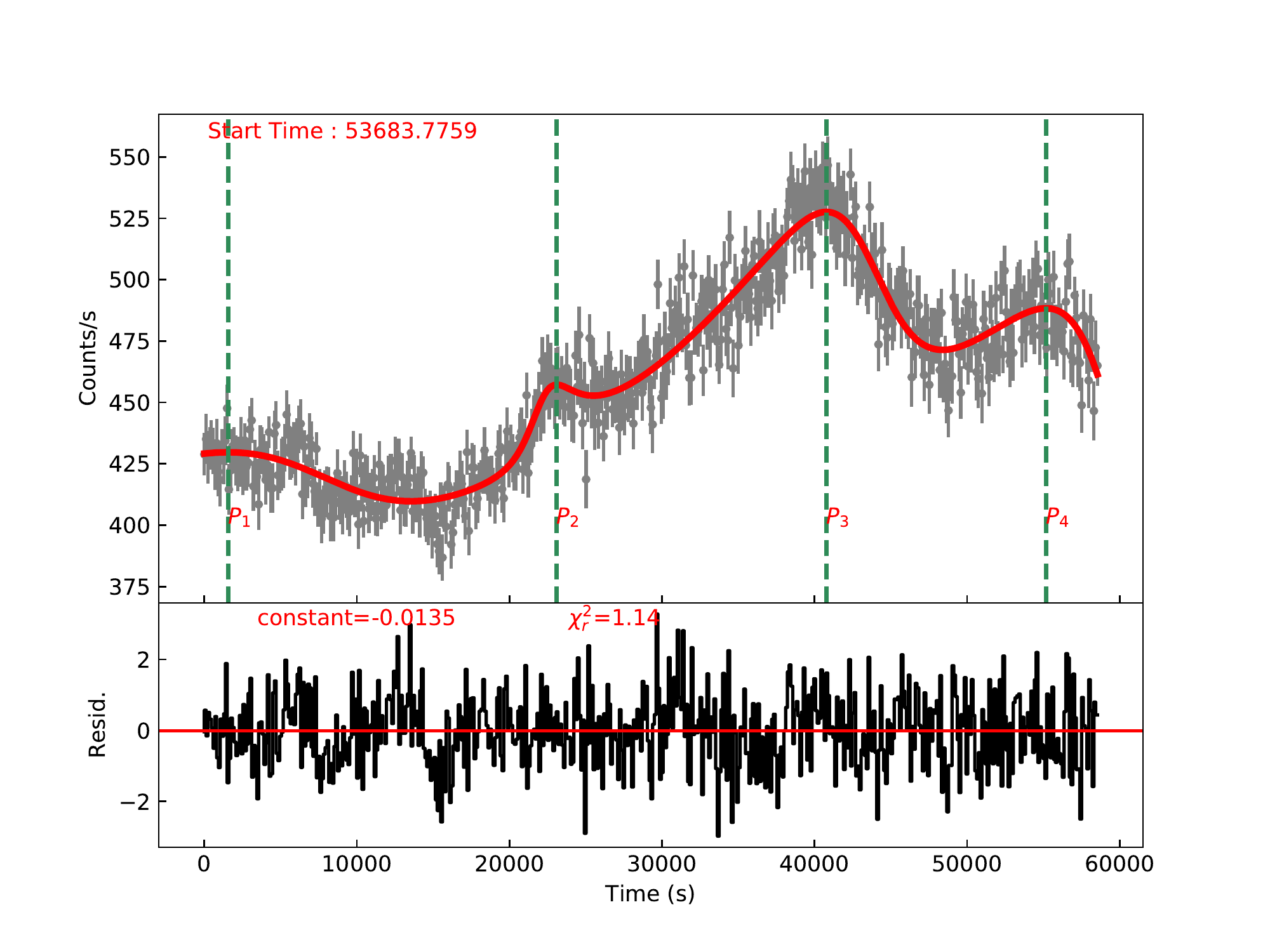}
\plottwo{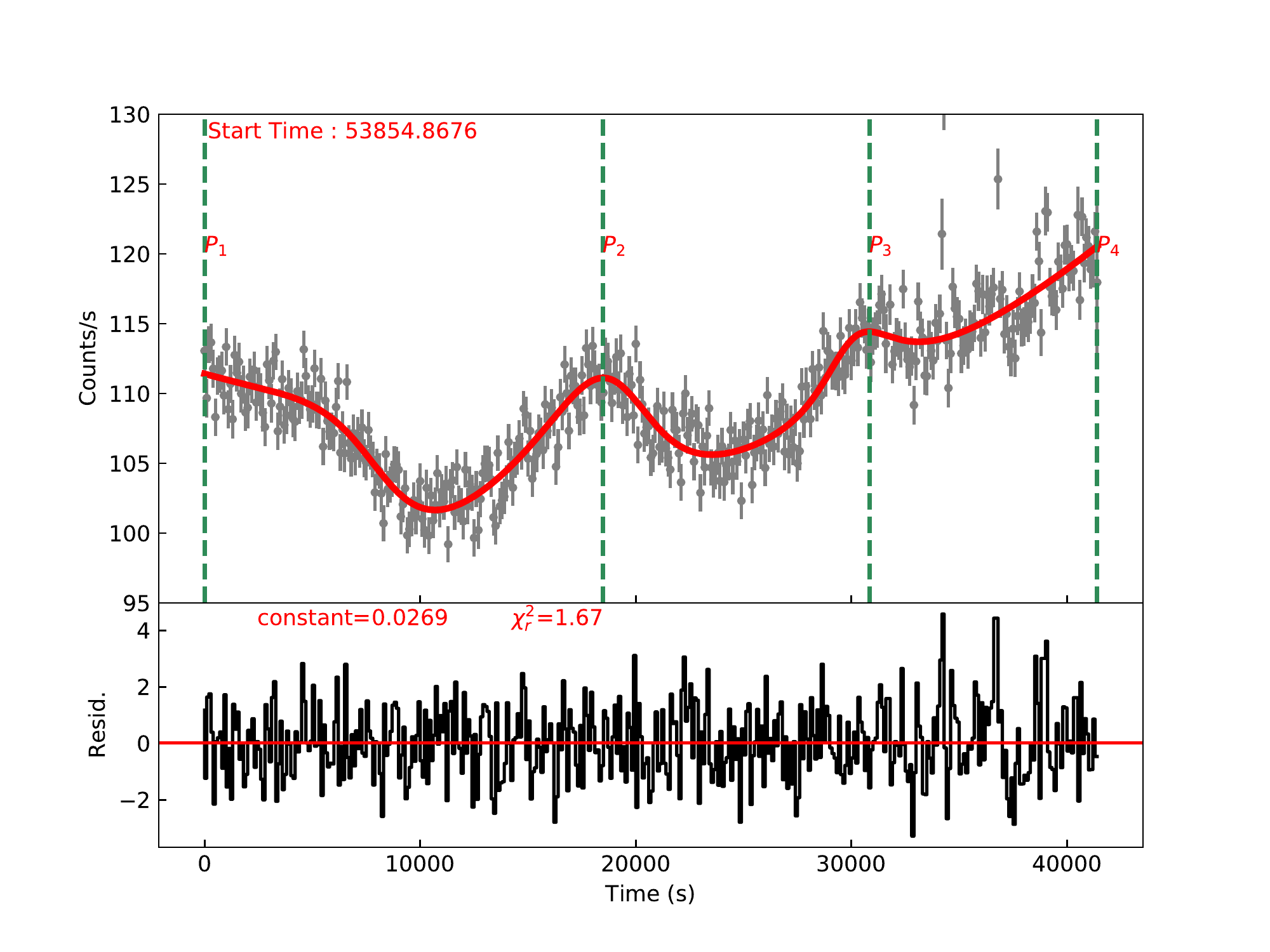}{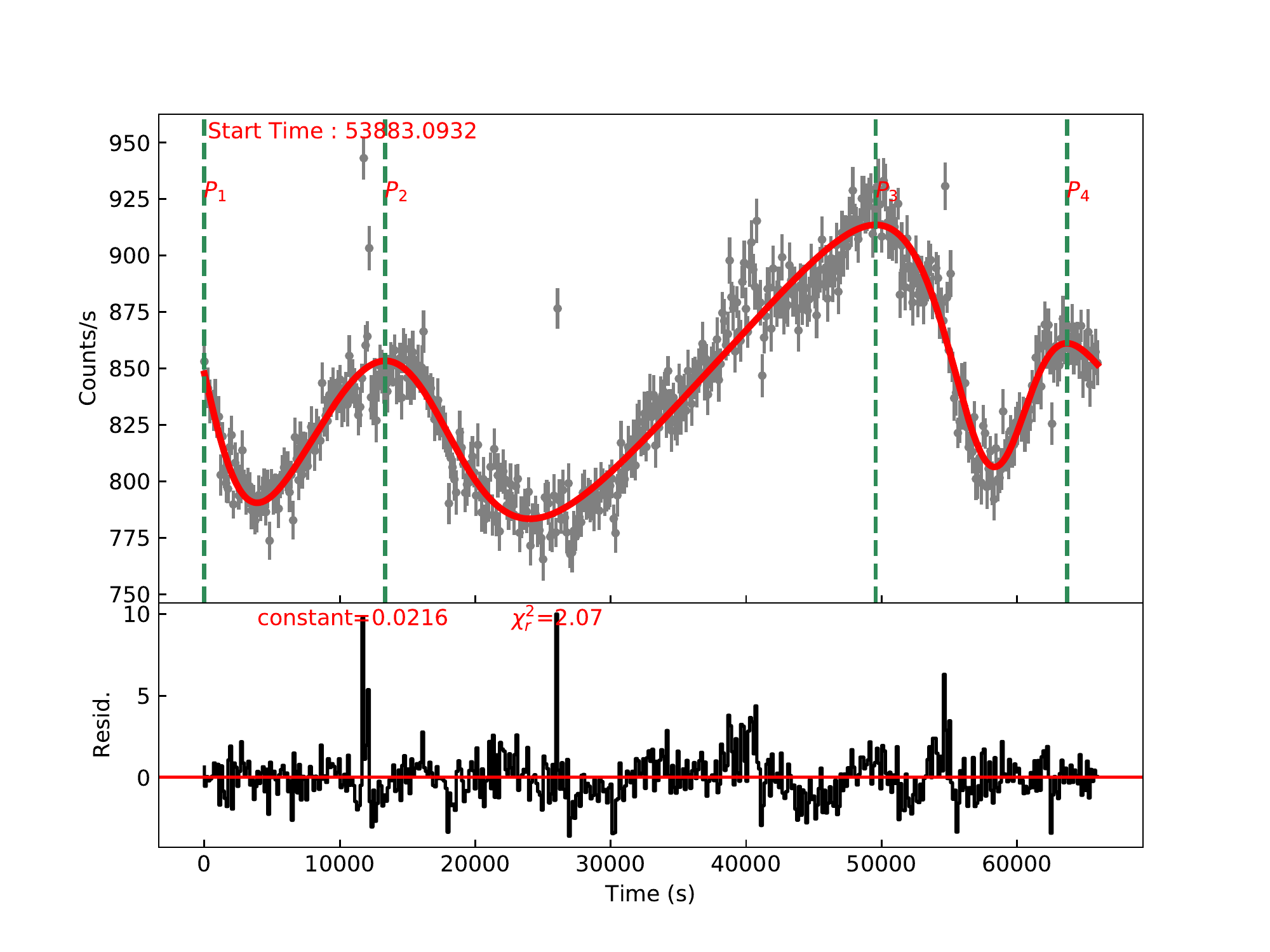}
\caption{\em{Continued.}}
\end{figure}

\begin{figure}[!h]
\figurenum{1}
\includegraphics[scale=0.40]{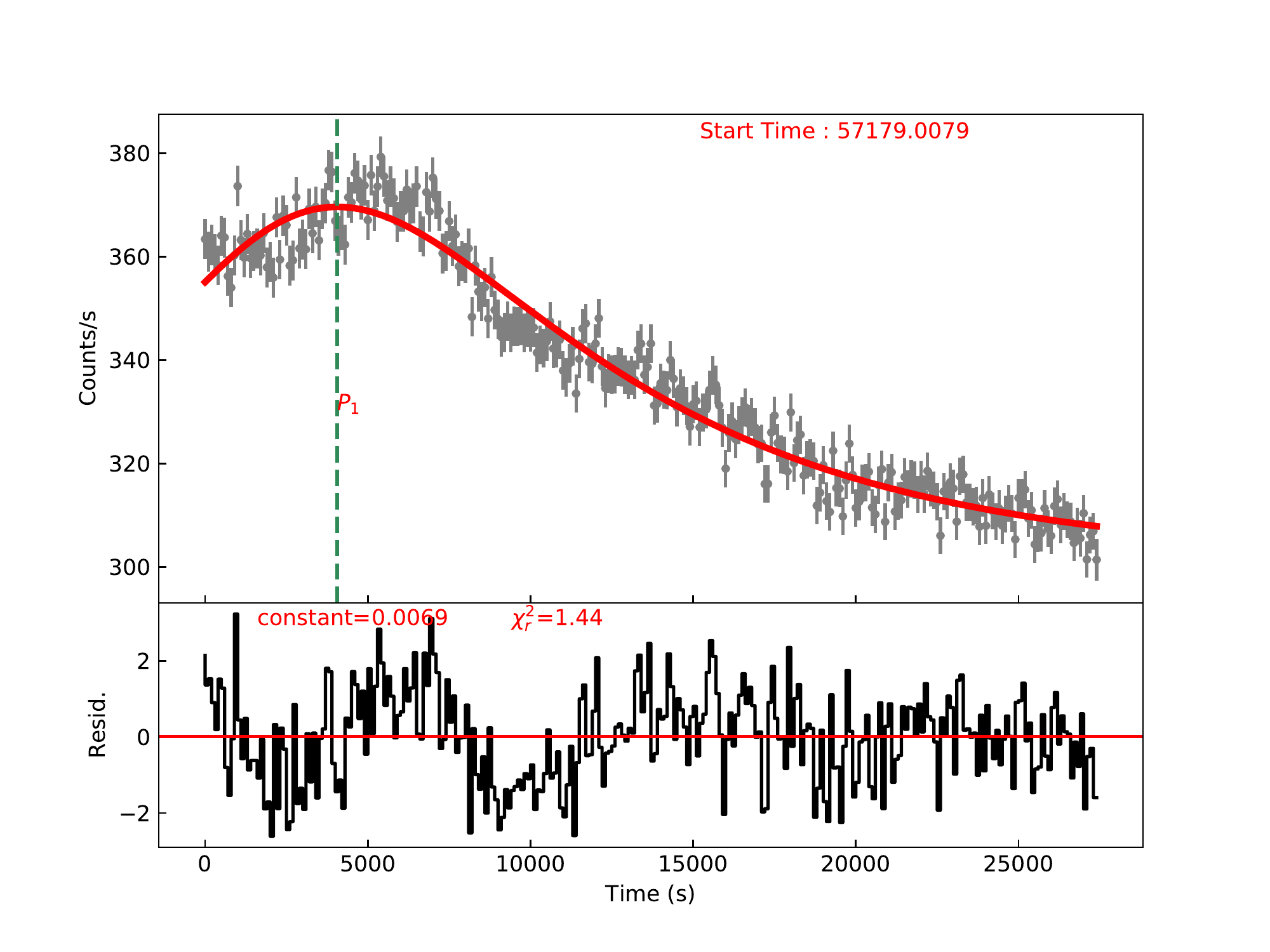} 
\caption{\em{Continued.}}
\end{figure}

\clearpage

%%%%%table 4 %%%%%%%%%
\startlongtable
\begin{deluxetable*}{cccccccccccc}
\tabletypesize{\tiny}
\tablenum{3}
\tablecaption{The results of light-curve fittings. (1) The start time, (2) the observation ID,  (3) the observation instrument, (4) the observation filter, (5) the peak number, (6-8) the parameters in equation~(\ref{fun:lcfit}), (9) the flaring time duration,  (10) the symmetry  parameter $\xi$, (11) the peak flux, in units of $10^{-11}$ ergs/cm$^2$/s, (12) the reduced chi-square of the fitting for each light curve.}
\label{tab:lcfit}
\tablehead{\colhead{StartTime} & \colhead{Obs.ID} & \colhead{Instru.} & \colhead{Filter} & \colhead{Peak NO.} & \colhead{$t_0$}& \colhead{$T_r$}& \colhead{$T_d$}& \colhead{$\rm{T_{fl}}$}& \colhead{$\xi$}& \colhead{$\rm{F_{peak}}$}& \colhead{$\chi^2_r$} \\ 
\colhead{(MJD)} & \colhead{} & \colhead{} & \colhead{} & \colhead{} & \colhead{(ks)} & \colhead{(ks)}& \colhead{(ks)}& \colhead{(ks)}& \colhead{}& \colhead{}& \colhead{} } 
%\decimals
\colnumbers
\startdata
51689.1624&		0099280101&	 EPN    &	 Thick  &	 p1   &	14.0	$\pm$	0.2&	17.0	$\pm$	4.2&	0.8	     $\pm$	0.1&	35.5	$\pm$	5.9&	-0.91	$\pm$	0.32&	61.84&	 2        \\ 
&		&	     &	   &	 p2   &	15.3	$\pm$	0.3&	0.9		$\pm$	0.1&	8.4	     $\pm$	2.2&	18.6	$\pm$	3.2&	0.80  $\pm$	0.31&	58.62&	         \\
51689.4414&		0099280101&	 EPN    &	 Thick  &	 p1   &	4.8     $\pm$	0.1&	4.6		$\pm$	0.8&	0.6	     $\pm$	0.2&	10.5	$\pm$	1.2&	-0.77	$\pm$	0.20&	86.39&	 2.41     \\
&		&	     &	   &	 p2   &	7.5     $\pm$	0.6&	2.0		$\pm$	1.6&	1.6	     $\pm$	0.3&	7.4     $\pm$	2.3&	-0.11	$\pm$	0.44&	77.27&	     \\
51850.006&	    0099280201&	 EPN    &	 Thick  &	 p1   &	2.1     $\pm$	0.3&	1.5		$\pm$	0.2&	1.2	     $\pm$	0.2&	5.4     $\pm$	0.4&	-0.09	$\pm$	0.11&	25.04&	 1.25     \\
&	    &	     &	   &	 p2   &	26.6	$\pm$	1.3&	6.6		$\pm$	1.2&	4.9	     $\pm$	1.5&	22.9	$\pm$	2.8&	-0.15	$\pm$	0.17&	26.52&	     \\
&	    &	     &	   &	 p3   &	31.3	$\pm$	0.2&	0.8		$\pm$	0.2&	3565.1   $\pm$	316037.0&-     &       -		&	27.42&	    \\
51861.9324&		0099280301&	 EPN    &	 Thick  &	 p2   &	7.2     $\pm$	0.4&	0.7		$\pm$	0.2&	9.6	     $\pm$	12.3&   -        &     -  		&	99.41&	 1.22     \\
&		&	     &	   &	 p3   &	16.6	$\pm$	0.6&	1.0		$\pm$	0.4&	13.1     $\pm$	21.3&      -     &       -		&	96.68&	     \\
&		&	     &	   &	 p4   &	25.7	$\pm$	0.7&	5.5		$\pm$	40.7&	0.4	     $\pm$	0.3&      -      &       	-	&	97.68&	      \\
&		&	     &	   &	 p5   &	26.6	$\pm$	3.7&	-44.0	$\pm$	197.8&	-1.3     $\pm$	1.2&    -        &       	-	&	97.62&	      \\
52037.3989&		0136540101&	 EPN    &	 Thin1  &	 p1   &	12.2	$\pm$	5.5&	11.7	$\pm$	9.9&	7.6      $\pm$	4.0&	38.7    $\pm$	15.1&	-0.21	$\pm$	0.56&	64.11&	 1.12     \\
&		&	     &	   &	 p2   &	16.0	$\pm$	0.3&	0.4		$\pm$	0.2&	29.2     $\pm$	256.7&     -	    &	      -	&	62.39&	      \\
&		&	     &	   &	 p3   &	22.2	$\pm$	0.8&	1.2		$\pm$	0.7&	1.3      $\pm$	1.0&	4.9     $\pm$	1.7&	0.04  $\pm$	0.50&	63.42&	      \\
&		&	     &	   &	 p4   &	27.9	$\pm$	2.6&	3.5		$\pm$	3.6&	5.1      $\pm$	2.6&        -	    &	    -	&	64.15&	     \\
&		&	     &	   &	 p5   &	31.5	$\pm$	0.3&	0.4		$\pm$	0.1&	31.8     $\pm$	384.2&      -	    &	   -	&	61.39&	      \\
52582.3202&		0136540401&	 EMOS1  &	 Thin1  &	 p1   &	2.4     $\pm$	0.4&	0.8		$\pm$	0.4&	0.7      $\pm$	0.3&	3.2     $\pm$	0.7&	-0.06	$\pm$	0.31&	71.97&	 1.43     \\
&		&	   &	   &	 p2   &	6.7     $\pm$	0.6&	2.3		$\pm$	0.6&	5.2      $\pm$	1.0&	14.8    $\pm$	1.7&	0.39  $\pm$	0.17&	73.76&	     \\
52592.8741&		0136540801&	 EPN    &	 Thick  &	 p1   &	5.4     $\pm$	0.6&	71.4	$\pm$	1667.5&	-0.6     $\pm$	0.5&     -        &       -		&	143.9&	 1.12     \\
&		&	     &	   &	 p2   &	9.1     $\pm$	0.8&	3.8		$\pm$	4.3&	1.0      $\pm$	1.0&    -        &       -		&	156.4&	      \\
52609.9727&		0136541001&	 EPN    &	 Medium &	 p2   &	11.8	$\pm$	0.9&	0.9		$\pm$	0.6&	4.2      $\pm$	7.3&    -        &       -		&	40.5&	 1.88     \\
&		&	     &	  &	 p3   &	21.7	$\pm$	0.3&	-0.6	$\pm$	0.1&	-9.9     $\pm$	11.8&   -        &       -		&	40.22&	      \\
&		&	     &	  &	 p4   &	26.0	$\pm$	0.6&	1.4		$\pm$	0.5&	3.7      $\pm$	1.2&	10.3    $\pm$	1.8&	0.44  $\pm$	0.27&	39.34&	     \\
&		&	     &	  &	 p5   &	41.8	$\pm$	0.2&	2.4		$\pm$	0.2&	2.7      $\pm$	0.3&	10.2    $\pm$	0.5&	0.07  $\pm$	0.07&	42.34&	     \\
&		&	     &	  &	 p6   &	49.9	$\pm$	0.3&	-29.5	$\pm$	29.8&	-0.7     $\pm$	0.3&	-	    &	    -	&	38.46&	     \\
&		&	     &	  &	 p7   &	62.1	$\pm$	0.3&	1.4		$\pm$	0.3&	45.6     $\pm$	35.8&	94.1    $\pm$	50.6&	0.94  $\pm$	1.04&	41.48&	      \\
52957.6897&		0150498701&	 EPN    &	 Thin1  &	 p1   &	13.3	$\pm$	0.4&	4.8		$\pm$	0.7&	1.9      $\pm$	0.4&	13.4    $\pm$	1.2&	-0.44	$\pm$	0.14&	115.9&	 1.56     \\
&		&	     &	   &	 p2   &	15.2	$\pm$	0.2&	0.7		$\pm$	0.1&	19.4     $\pm$	4.5&	40.2    $\pm$	6.4&	0.93  $\pm$	0.31&	115.2&	     \\
&		&	     &	   &	 p3   &	27.4	$\pm$	0.5&	1.3		$\pm$	0.4&	0.9      $\pm$	0.3&	4.5     $\pm$	0.8&	-0.19	$\pm$	0.25&	104&	     \\
&		&	     &	   &	 p4   &	37.2	$\pm$	0.9&	-25.4	$\pm$	16.9&	-2.0     $\pm$	0.8&	54.8    $\pm$	23.9&	0.85  $\pm$	0.81&	96.35&	      \\
52398.6795&		0153950601&	 EMOS1  &	 Thin1  &	 p2   &	24.1	$\pm$	0.5&	7.1		$\pm$	0.8&	4.8      $\pm$	0.3&	23.8    $\pm$	1.2&	-0.20	$\pm$	0.07&	26.06&	 1.17     \\
52797.897&	    0158970701&	 EMOS1  &	 Thick  &	 p1   &	26.1	$\pm$	18.3&	44.2	$\pm$	67.7&	12.3     $\pm$	4.9&	   -	    &	 -	&	25.24&	 1.23     \\
&	    &	   &	   &	 p2   &	29.8	$\pm$	0.5&	0.9		$\pm$	0.2&	5.5      $\pm$	3.6&	12.9    $\pm$	5.0&	0.72  $\pm$	0.68&	23.83&	     \\
&	    &	   &	   &	 p3   &	42.7	$\pm$	3.4&	4.5		$\pm$	2.5&	16.7     $\pm$	28.5&    -  &      -	&	23.8&	     \\
53131.1251&		0158971201&	 EPN    &	 Medium &	 p1   &	12.6	$\pm$	1.4&	13.9	$\pm$	21.1&	2.1      $\pm$	0.4&     -      &    -	&	104.8&	 1.26     \\
&		&	     &	  &	 p2   &	35.1	$\pm$	5.9&	51.9	$\pm$	588.8&	3.5      $\pm$	0.4&     -      &    -	&	114.6&	     \\
53683.7759&		0158971301&	 EPN    &	 Thick  &	 p1   &	7.8     $\pm$	5.6&	96.6	$\pm$	823.1&	3.5      $\pm$	2.9&     -    &    -	&	100.5&	 1.18     \\
&		&	     &	   &	 p2   &	22.2	$\pm$	0.6&	0.8		$\pm$	0.4&	2.2      $\pm$	0.9&	6.2     $\pm$	1.4&	0.46  $\pm$	0.36&	107&	      \\
&		&	     &	   &	 p3   &	43.7	$\pm$	0.4&	19.7	$\pm$	8.0&	1.7      $\pm$	0.3&	42.9	$\pm$	11.3&	-0.84	$\pm$	0.49&	123.5&	     \\
&		&	     &	   &	 p4   &	60.0	$\pm$	1.4&	30.2	$\pm$	27.5&	1.8      $\pm$	0.8&	63.9	$\pm$	38.9&	-0.89	$\pm$	1.15&	114.3&	      \\
52983.8975&		0162960101&	 EPN    &	 Medium &	 p2   &	12.8	$\pm$	7.0&	2.7		$\pm$	3.9&	76.6     $\pm$	1914.1& -  &	   -	&	53.65&	 0.95     \\
&		&	     &	  &	 p3   &	23.7	$\pm$	3.2&	1.2		$\pm$	1.4&	21.5     $\pm$	282.3&   -	    &	 -	&	55.98&	     \\
53854.8676&		0302180101&	 EMOS2  &	 Thin1  &	 p2   &	19.3	$\pm$	0.6&	4.3		$\pm$	1.8&	1.5      $\pm$	0.3&	11.7	$\pm$	2.6&	-0.48	$\pm$	0.36&	84.02&	 1.76     \\
&		&	   &	   &	 p3   &	29.9	$\pm$	0.7&	1.1		$\pm$	0.5&	2.0      $\pm$	0.9&	6.3     $\pm$	1.4&	0.29  $\pm$	0.33&	86.52&	      \\
53883.0932&		0411080301&	 EPN    &	 Medium &	 p2   &	16.8	$\pm$	0.7&	24.1	$\pm$	10.8&	3.2      $\pm$	0.4&	54.6	$\pm$	15.2&	-0.77	$\pm$	0.50&	184.5&	 2.15     \\
&		&	     &	  &	 p3   &	59.3	$\pm$	1.5&	4843.5	$\pm$325805.3&	2.2      $\pm$	0.7&	 -	    &	  -	&	197.5&	      \\
&		&	     &	  &	 p4   &	59.3	$\pm$	2.3&	2.2		$\pm$	0.4&	9298.6   $\pm$	597976.9 &	-	    &	 -	&	186.2&	      \\
57179.0079&		0658801301&	 EPN    &	 Thick  &	 p1   &	2.1     $\pm$	0.8&	4.4		$\pm$	0.7&	8.5      $\pm$	1.0&	25.7	$\pm$	1.7&	0.32  $\pm$	0.10&	82.69&	1.46\\
\enddata
\end{deluxetable*}
%%%%%%%%%%%%%%%%%%%%5%%

\section{Conclusions and discussion}

We analyzed 50 XMM-Newton X-ray observations for Mrk 421, and constructed the X-ray light curves.
The basic information, such as $F_{\rm var}$ and NPSD can be found in Appendix.
We fitted the 17 light curves which have one complete flare-profile at least, in order to determine the flare rise and decay times as well as peak fluxes (see Fig.~\ref{fig:lcfit}).
We obtained these parameters for 48 flares.
Among them, 26 flares have well constrained parameters (see Table~\ref{tab:lcfit}).
The flaring time duration is evaluated by using the flare rise and decay times.
It is found that both the peak flux and flaring time duration distributions follow a power-law form 
with the same index of $\alpha_{\rm F}=\alpha_{\rm T}\approx1$.

It has been argued that the behavior that event parameters obey a power-law distribution is a characteristic of SOC system \citep[e.g.,][]{Aschwanden12}.
The SOC model has been used to explain astrophysical X-ray flares, such as GRB X-ray flares \citep{Wang13,Yi16,Yi17} and 
super massive black hole (SMBH) X-ray flares \citep{Wang15,Li}.
They found that the statistical properties of the GRB and SMBH X-ray flares can be explained by a fractal-diffusive SOC model, but with different spatial
dimensions $S$ ($S=1$ for GRB X-ray flares and $S=3$ for SMBH X-ray flares). However, \citet{Yuan} argued that a simple SOC model failed to explain the X-ray flares in 
SMBH Sgr A$^*$.

The SOC model expects the index of peak flux distribution $\alpha_{\rm F}=1+(S-1)/D_S$ ($D_S$ is the fractal
Hausdorff dimension, spanning from 1 to $S$), and the index of flaring time duration distribution $\alpha_{\rm T}=1+(S-1)\beta/S$ where $\beta$ is a diffusion parameter \citep[e.g.,][]{Aschwanden12}. The statistical results of Mrk 421 X-ray flares are consistent with the expectation of a SOC mode with $S=1$.
This indicates that the X-ray flares of Mrk 421 are possibly driven by magnetic reconnection.
Our results support the finding that magnetic reconnection is a promising process for energy dissipation in blazar jet \citep{Sironi}.

The $S$ of Mrk 421 is similar to that of GRB X-ray flares, but different from that of X-ray flares in M87 and Sgr A$^*$ (i.e., the SMBH X-ray flares).
M87 is a radio galaxy. It is thought that the main difference between radio galaxy and blazar is 
the angle of jet direction with respect to the line-of-sight.
 We notice that the flaring time duration for M87 in \cite{Wang15} spans from 0.2 years to several years.
 Such a long timescale indicates that the X-ray emissions come 
 from a large region which locates relatively far away from the central SMBH.
On the other hand, the rapid X-ray emission from blazar is believed to come from a region locating at sub-pc scale.
It is possible that the results of M87 and Mrk 421 reveal the situations about magnetic field in different regions along the jet in radio-loud AGN.
However, note that the uncertainties on the results of M87 are very large \citep{Wang15}.

Sub-hour variabilities are presented in our analysis.
Flares with $t_{\rm var}\sim t_{\rm d}\sim 1000\ $s frequently appear.
Such rapid flare indicates that the magnetic field in emission region is  $B' \geq 2.1\delta_{\rm D}^{-1/3}$ G.
This magnetic field is much higher than the value of $B'<0.1\ $G that derived in SSC modeling of SED \citep[e.g.,][]{Yan13,Zhu}.
Taking advantage of the radio core-shift-effect obtained in Very Long Baseline Array (VLBA) observations, \cite{zam} derived the 
magnetic field strength $B'_{\rm 1pc}$ at one pc along the jet for tens blazars, and they found $B'_{\rm 1pc}$ spanning from 0.2 G to 2 G.
This is consistent with the magnetic field strength expected by a magnetically powered jet \citep[e.g.,][]{Os}.
If Mrk 421 is not a outlier, the magnetic field of $B' \geq 2.1\delta_{\rm D}^{-1/3}$ G is also likely consistent with the prediction of a magnetically powered jet.
For a magnetically powered jet,  magnetic reconnection is a natural candidate
for energy dissipation in blazar jet \citep[e.g.,][]{Sironi}.

As a last remark, we note that \citet{Zub} proposed a model of tidal disruption of asteroids by the SMBH for the origin of Sgr A$^*$ X-ray flares.
This model probably also can explain the power-law distributions of the parameters of Sgr A$^*$ X-ray flares \citep{Zub}.
However, this scenario will not happen in blazar \citep[e.g.,][]{Komossa}, due to the SMBH with the mass of $\gtrsim10^8M_{\odot}$ in blazar, where $M_{\odot}$ is the solar mass.
As far as we know, the SOC model is the only model for explaining the power-law distributions of the parameters of blazar flares.

%%%%%%%%%%%%%%%%%%%%5%%
%Acknowledgements %
%%%%%%
\section*{Acknowledgements}
%\textbf{We thank the referees for helpful and valuable comments .}
We thank the reviewers for the constructive questions. 
We thank Dr. Zunli Yuan (YNAO) for helpful discussions.
We acknowledge financial supports from the National
Natural Science Foundation of China (NSFC-11573060, NSFC-11573026, NSFC-U1531131, NSFC-U1738124 and
NSFC-11661161010, NSFC-11803081), and the Key Laboratory of Astroparticle
Physics of Yunnan Province (No. 2016DG006).
The work of D. H. Yan is also supported by the CAS
``Light of West China'' Program.
\software{XSPEC \citep[version 12.9;][]{Arnaud}, PIMMS \citep{Mukai}}, SAS.
%%%%%%%%%%%%%%%%%%%%%%%%
\bibliography{mrk421}

\appendix
\section{Fifty XMM-Newton X-ray light curves of Mrk 421}

In Fig.~\ref{fig:NPSD}, we show all fifty XMM-Newton X-ray light curves obtained in our analysis.
To characterize the variability amplitude for each light curve, 
we use the fraction root mean square (rms) variability amplitude \citep{2003MNRAS.345.1271V},
\begin{equation}
F_{\rm{var}} = \sqrt{\frac{S^2-\overline{\sigma^2_{\rm{err}}}}{\overline{x}^2}}\ ,
\end{equation}
where $S^2$ is the variance of the light curve; $\overline{\sigma^2_{\rm{err}}}$ is the mean square error of the light curve, and $\overline{x}^2$ is the arithmetic mean of the data. 
The error of the $F_{\rm{var}}$ is calculated using the Equation B2 in \citep{2003MNRAS.345.1271V}. 
The results of the calculations are listed in Table \ref{tab:NPSDfvar}. 

The general timing analysis technique for characterizing the light curve is power spectrum density (PSD). 
However, because of the orbital effects of satellites and some CCD background problems, 
the X-ray light curves extracted from the clear events are usually uneven. To getting correct PSD calculations, we adopt the method called normalized power spectrum density \citep[NPSD; e.g., ][]{1998ApJ...500..642H,2001AIPC..558..660K}. The calculated frequency range is from 1/T to 1/(2$\times$bin) where T is the timing coverage and ``bin'' is the bin size of the light curve. 
If the light curve contains $n$ time gaps larger than twice the data bin size, each gap is considered as a break so that the light curve is separated into $n+1$ segments. 
Then the total power of the light curve is the average of the powers of these segments. 
Note that we do not subtract the white noise power in each calculation.

\begin{figure}
\figurenum{5}
\centering
\includegraphics[scale=0.60]{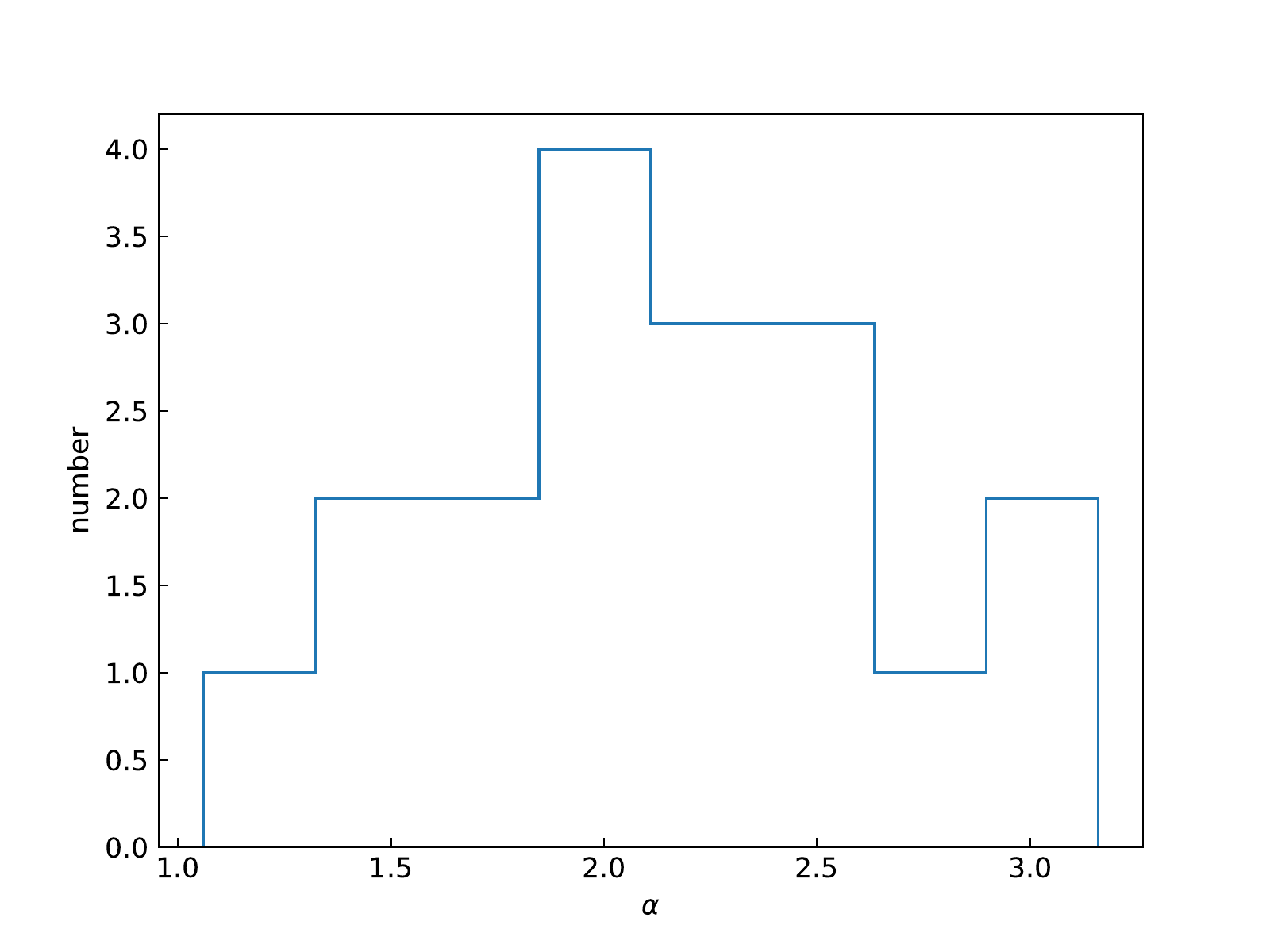} 
\caption{Distribution of the index of NPSD $\alpha$. \label{fig:alpha}}
\end{figure}

We then fit the NPSDs using a power-law form function,
\begin{equation}
P = af^{-\alpha}+b\, 
\end{equation}
where $P$ is the NPSD power, $f$ the frequency, $\alpha$ the power-law index, $b$ the white noise power, 
and $a$ a constant. The NPSD for each light curve and its best-fitting result is shown in Figure \ref{fig:NPSD}.
The best-fitting values of $\alpha$ and $b$ are listed in Table~\ref{tab:NPSDfvar}.
The distribution of $\alpha$ is shown in Fig.~\ref{fig:alpha}.
The poorly constrained $\alpha$ are excluded in Fig.~\ref{fig:alpha}.
One can see that $\alpha$ spans between 1 and 3, and most of them cluster around 2.

%%%%figure 1%%%%%%%%%%%%%%
\begin{figure}[!h]
	\figurenum{4}
	\includegraphics[scale=0.35]{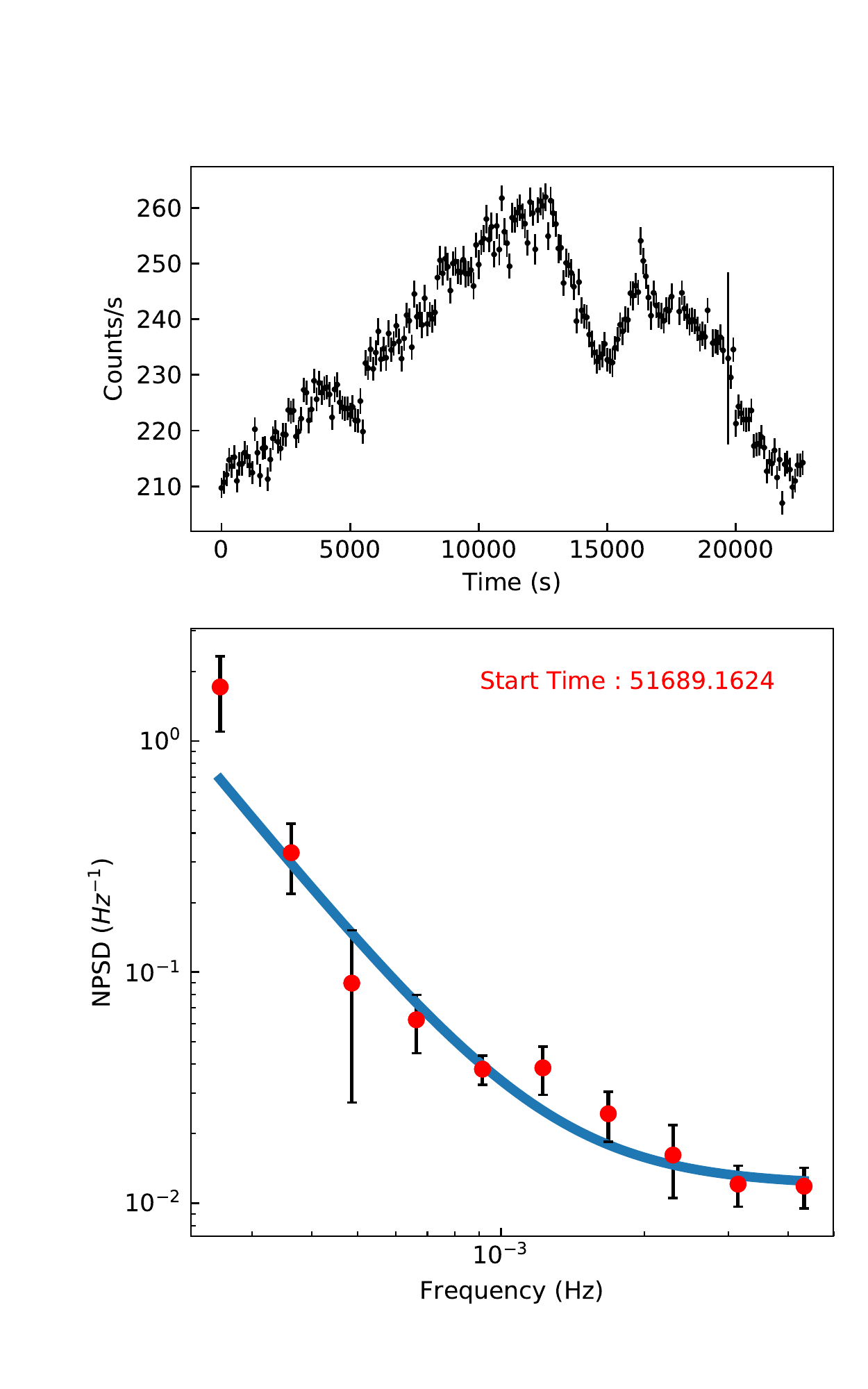}
	\includegraphics[scale=0.35]{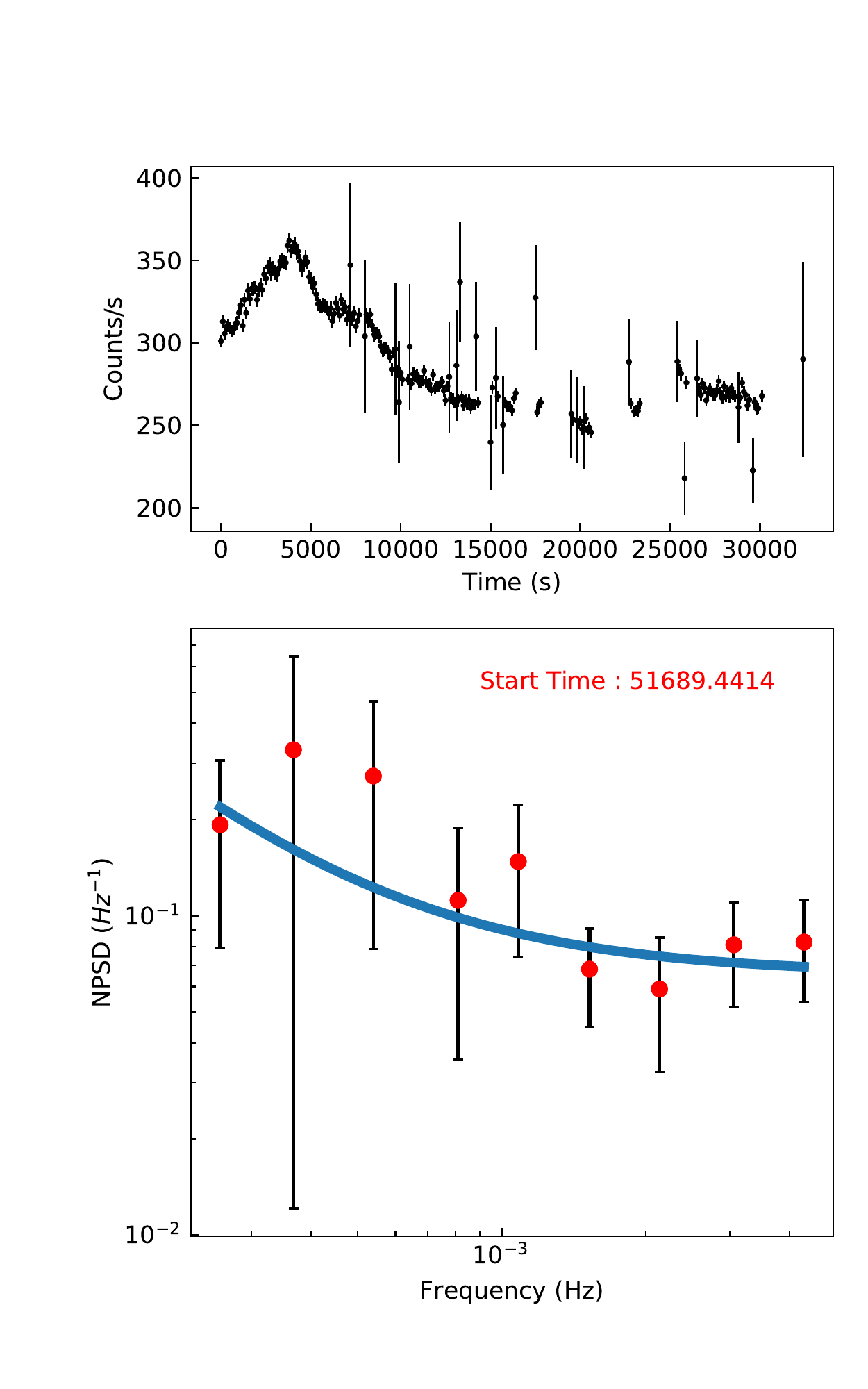}
	\includegraphics[scale=0.35]{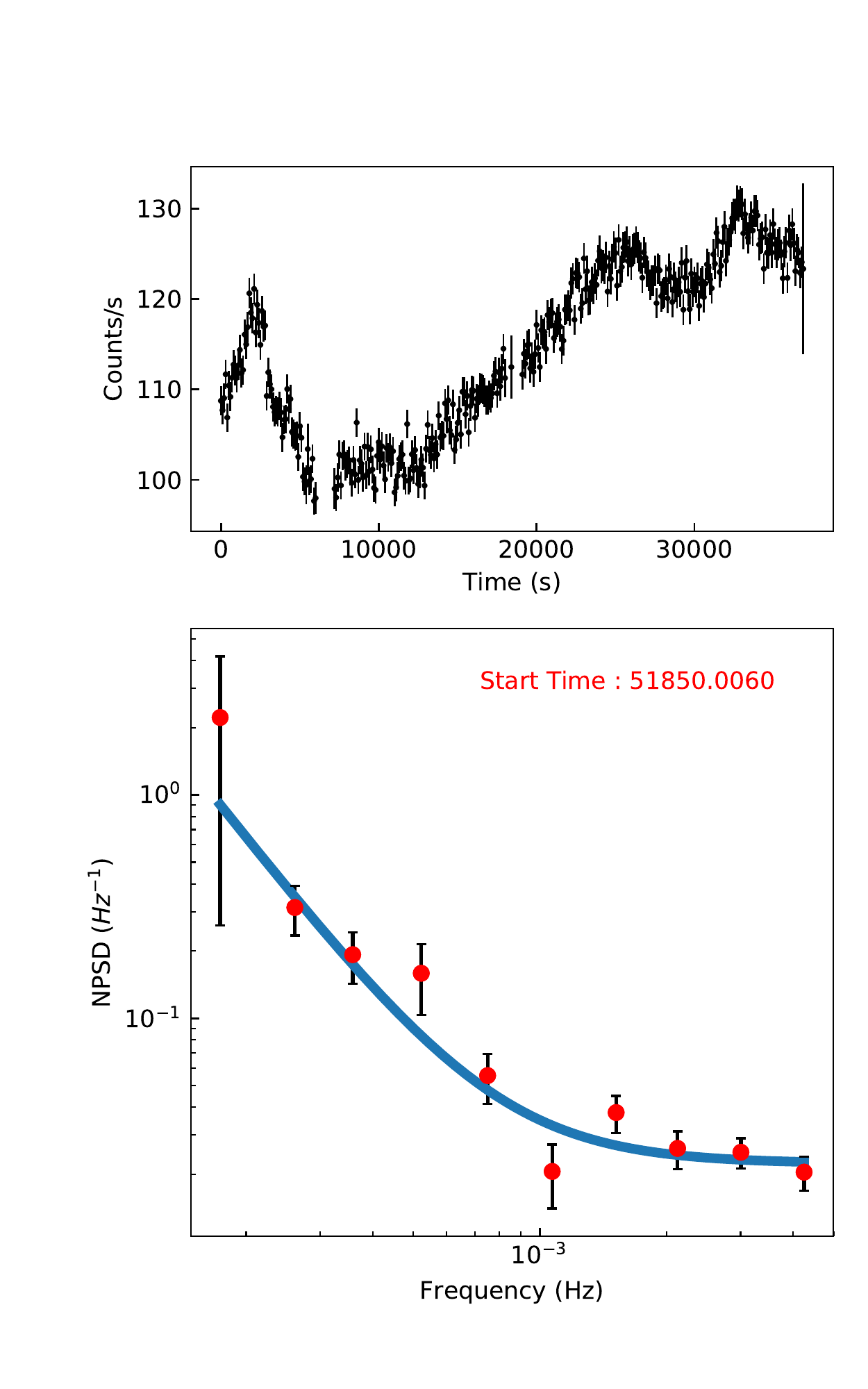}
	\includegraphics[scale=0.35]{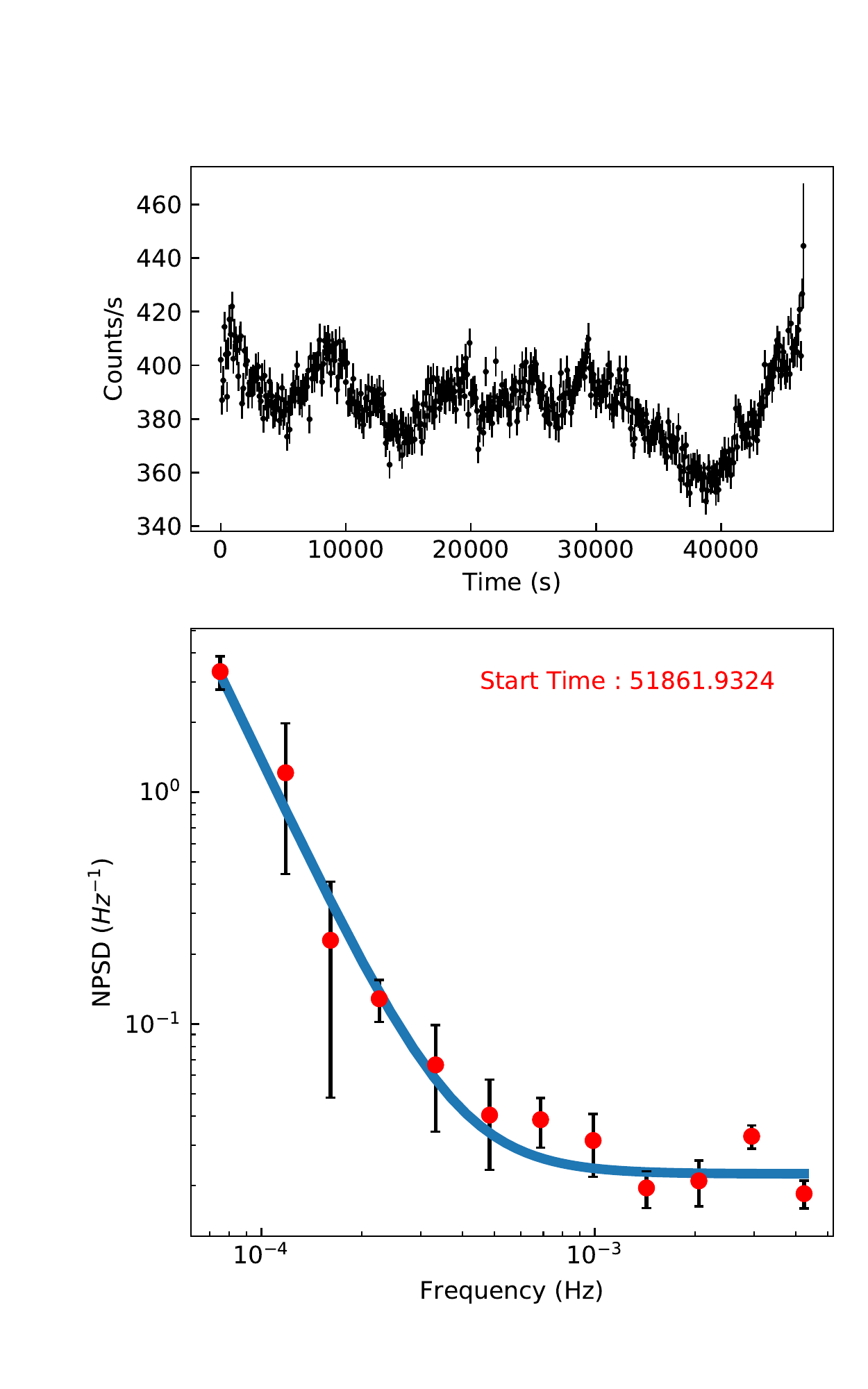}
	\includegraphics[scale=0.35]{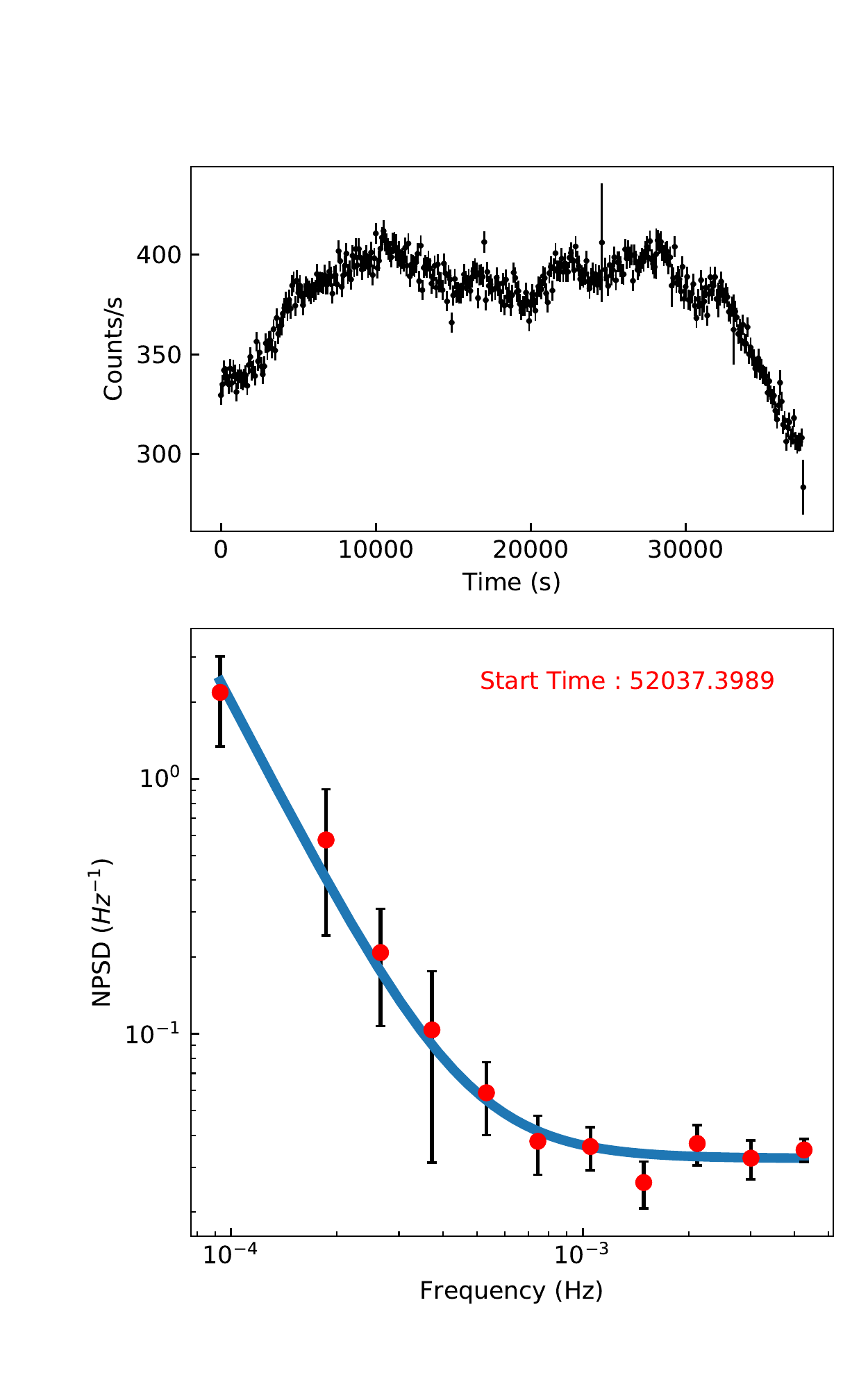}
	\includegraphics[scale=0.35]{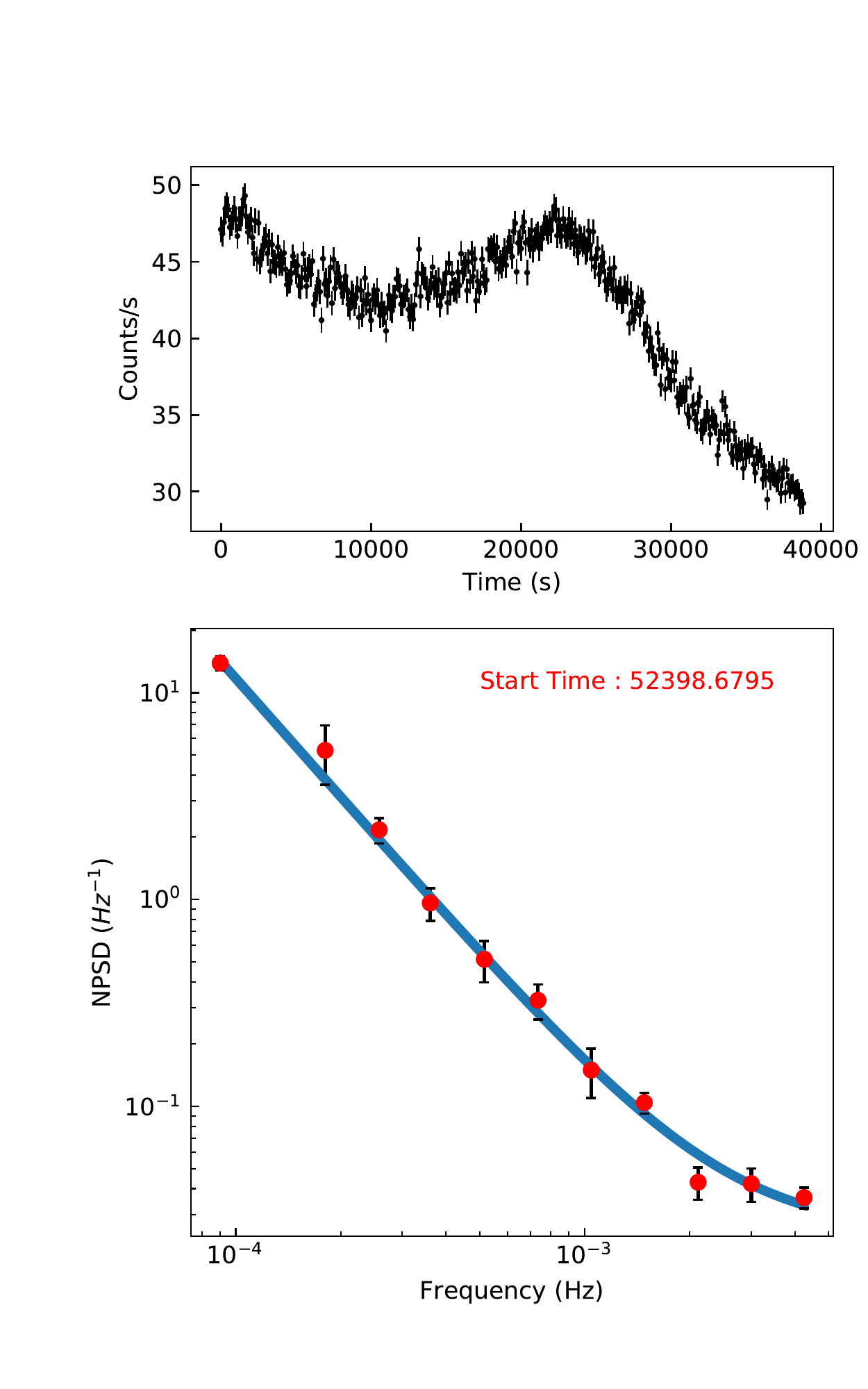}
	\includegraphics[scale=0.35]{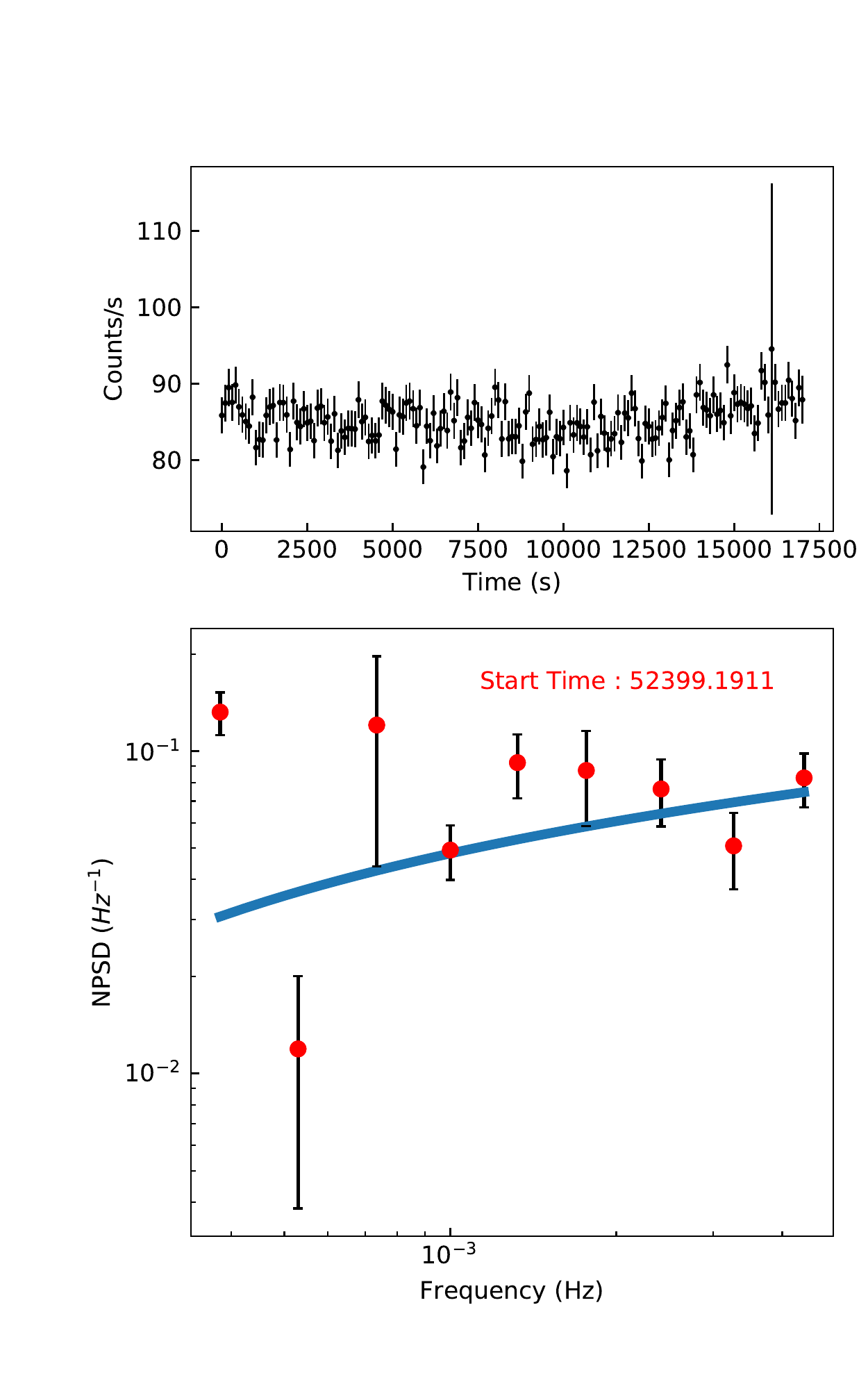}
	\includegraphics[scale=0.35]{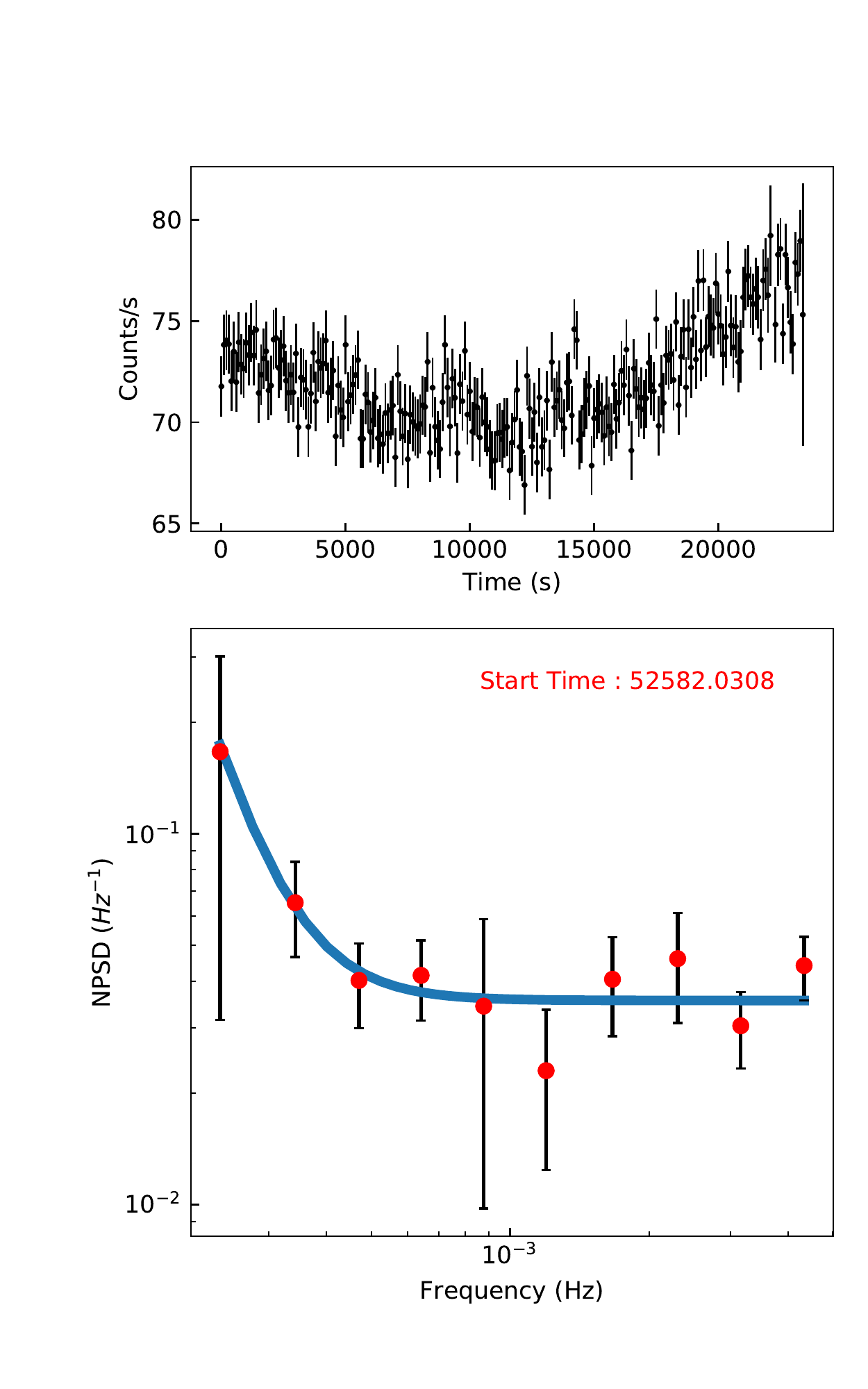}
	\includegraphics[scale=0.35]{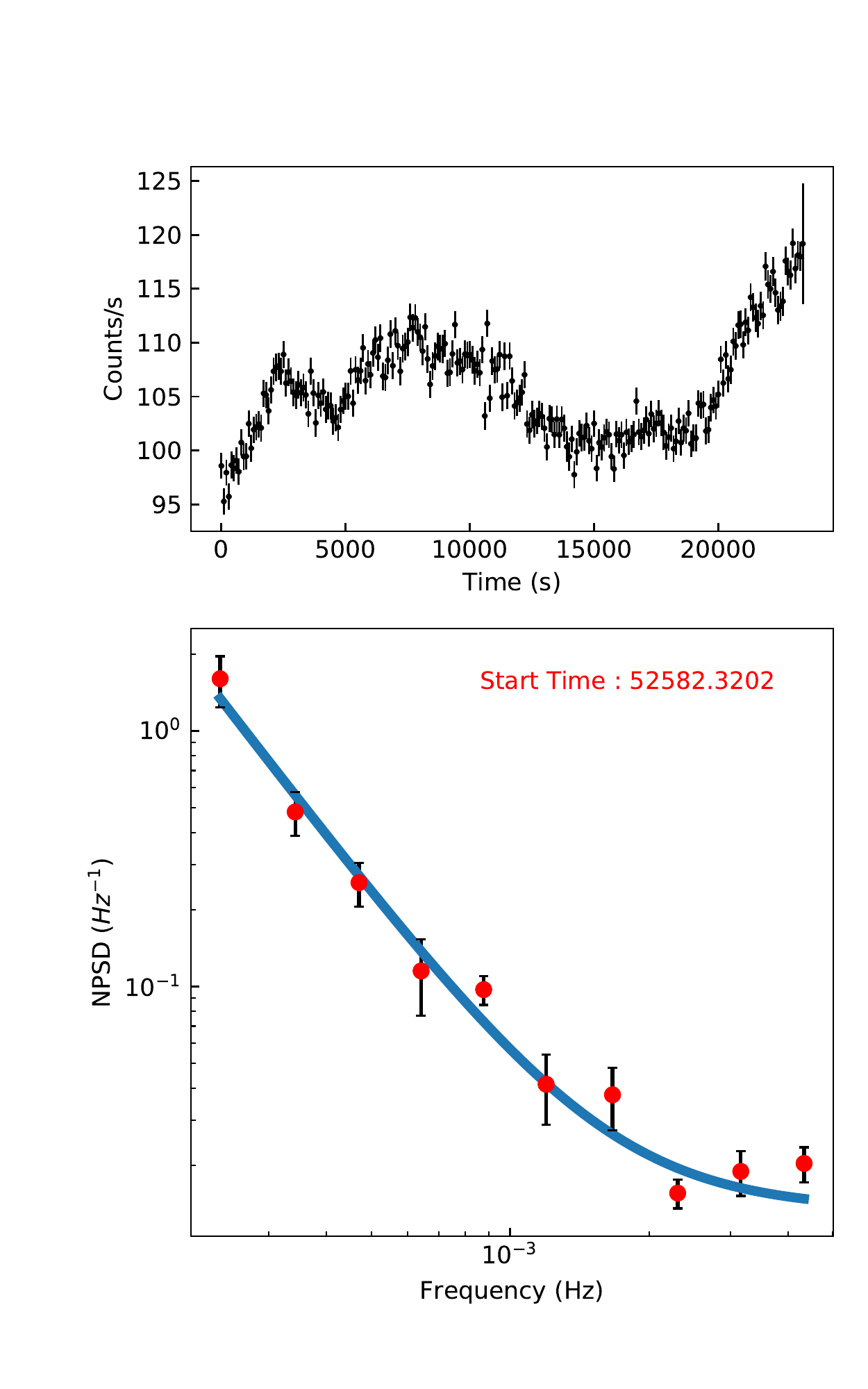}
	\includegraphics[scale=0.35]{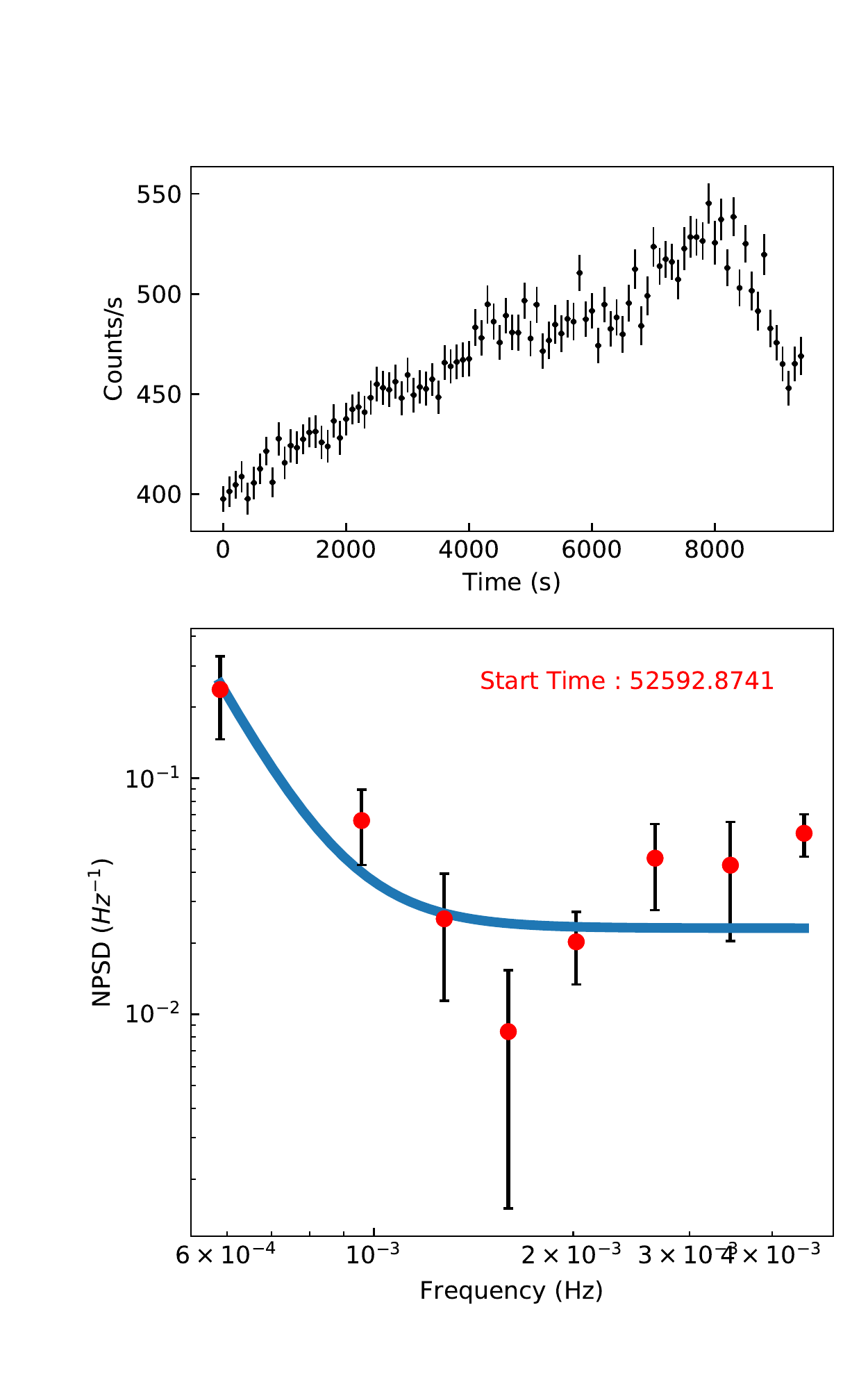}
	\includegraphics[scale=0.35]{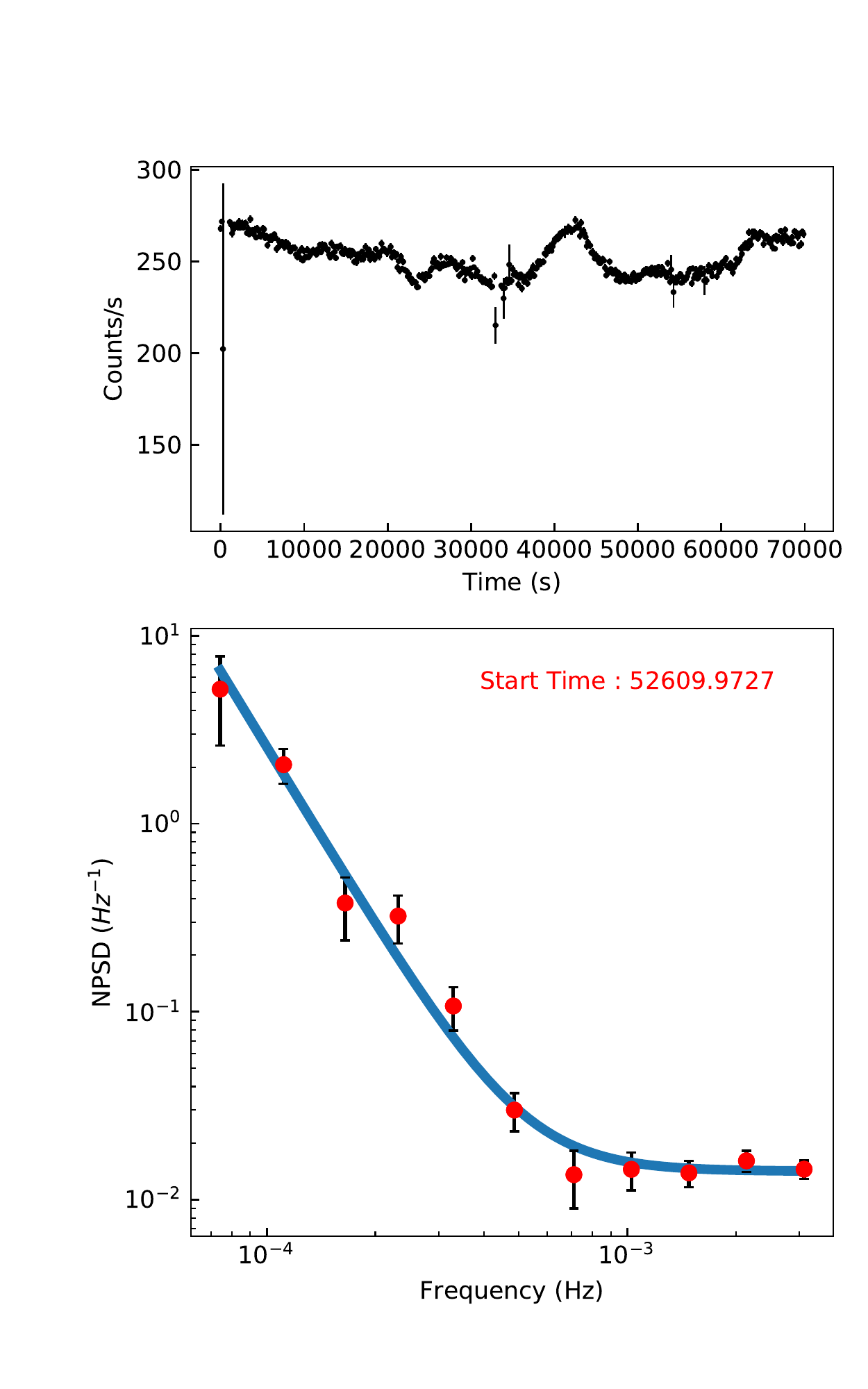}
	\includegraphics[scale=0.35]{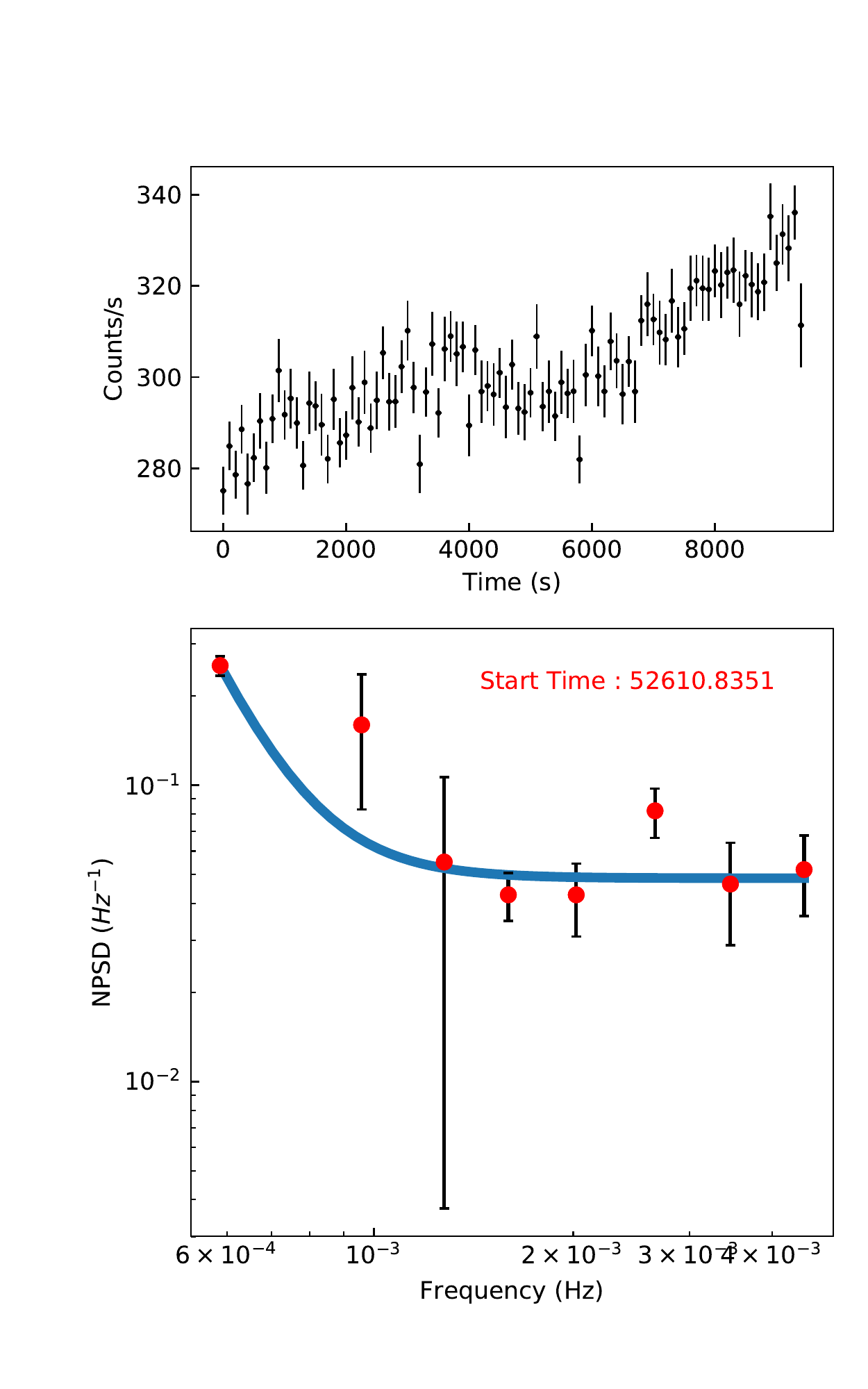}
	\caption{Fifty light curves and the corresponding NPSD. The upper panel is the light curve, and the lower panel is the NPSD. The red points are the calculated NPSD, and the steel blue line is the best fitting. The start time of each light curve is marked using the MJD in the lower panel. \label{fig:NPSD}}
\end{figure}
\begin{figure}[!h]
	\figurenum{4}
	\includegraphics[scale=0.35]{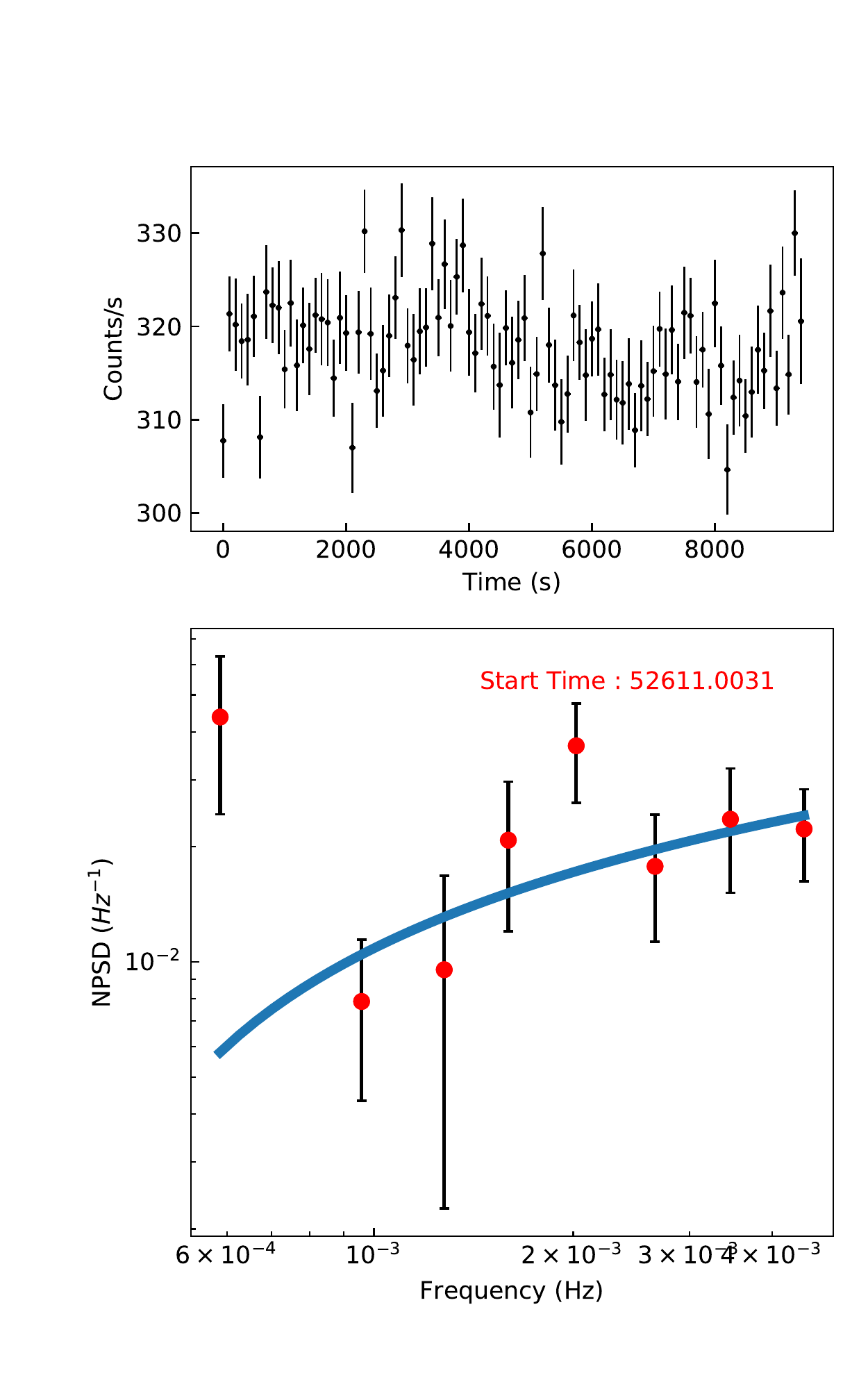}
	\includegraphics[scale=0.35]{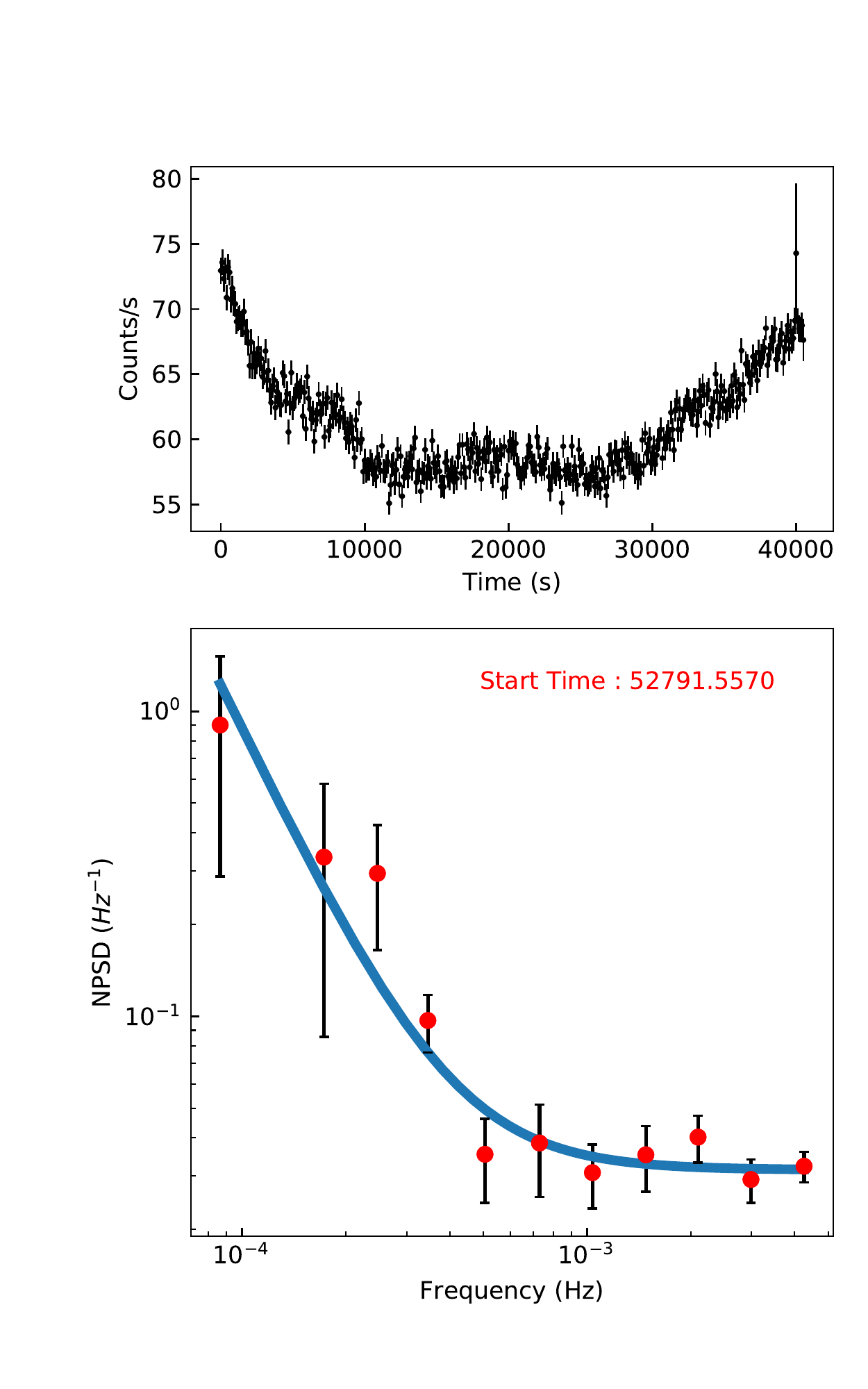}
	\includegraphics[scale=0.35]{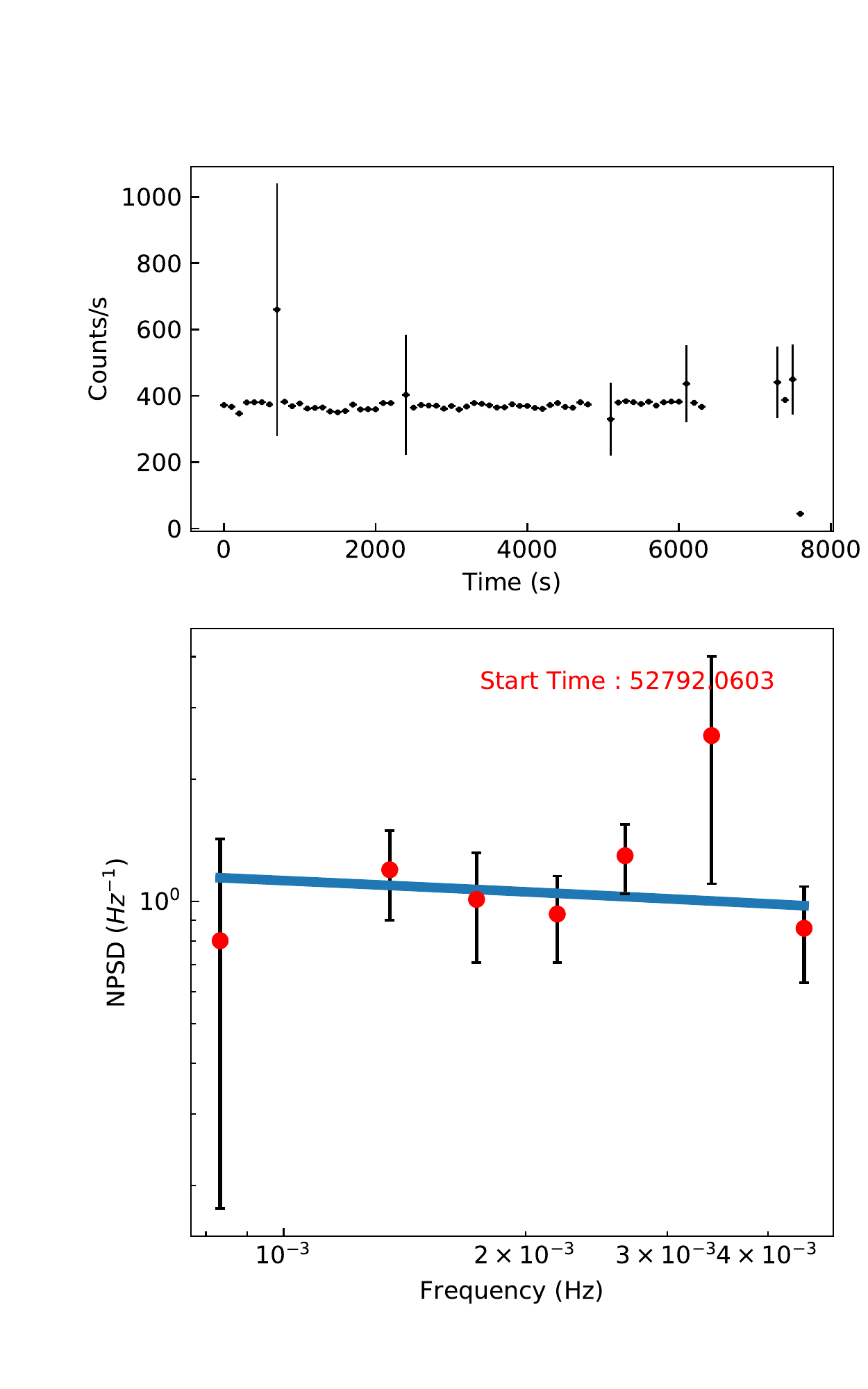}
	\includegraphics[scale=0.35]{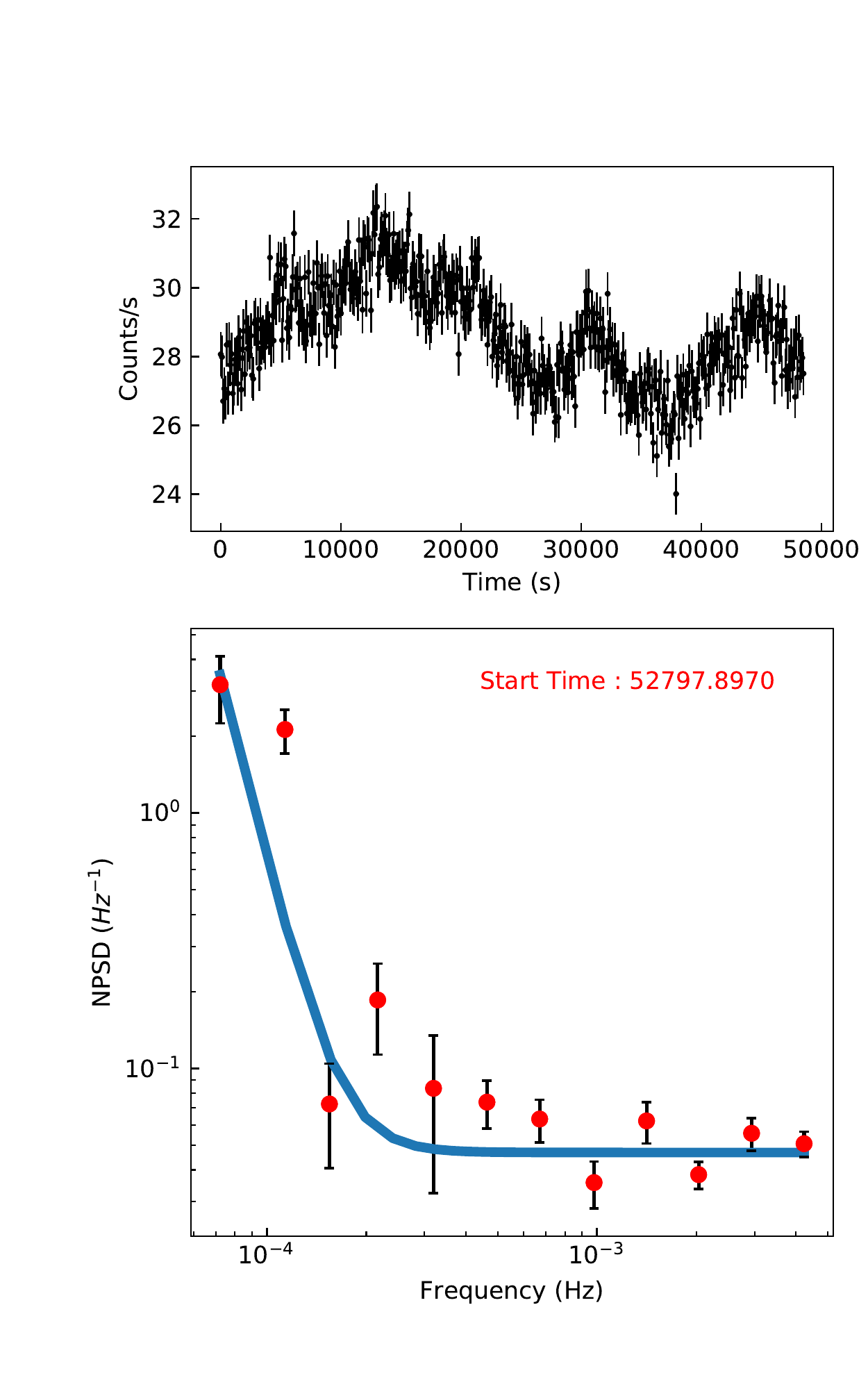}
	\includegraphics[scale=0.35]{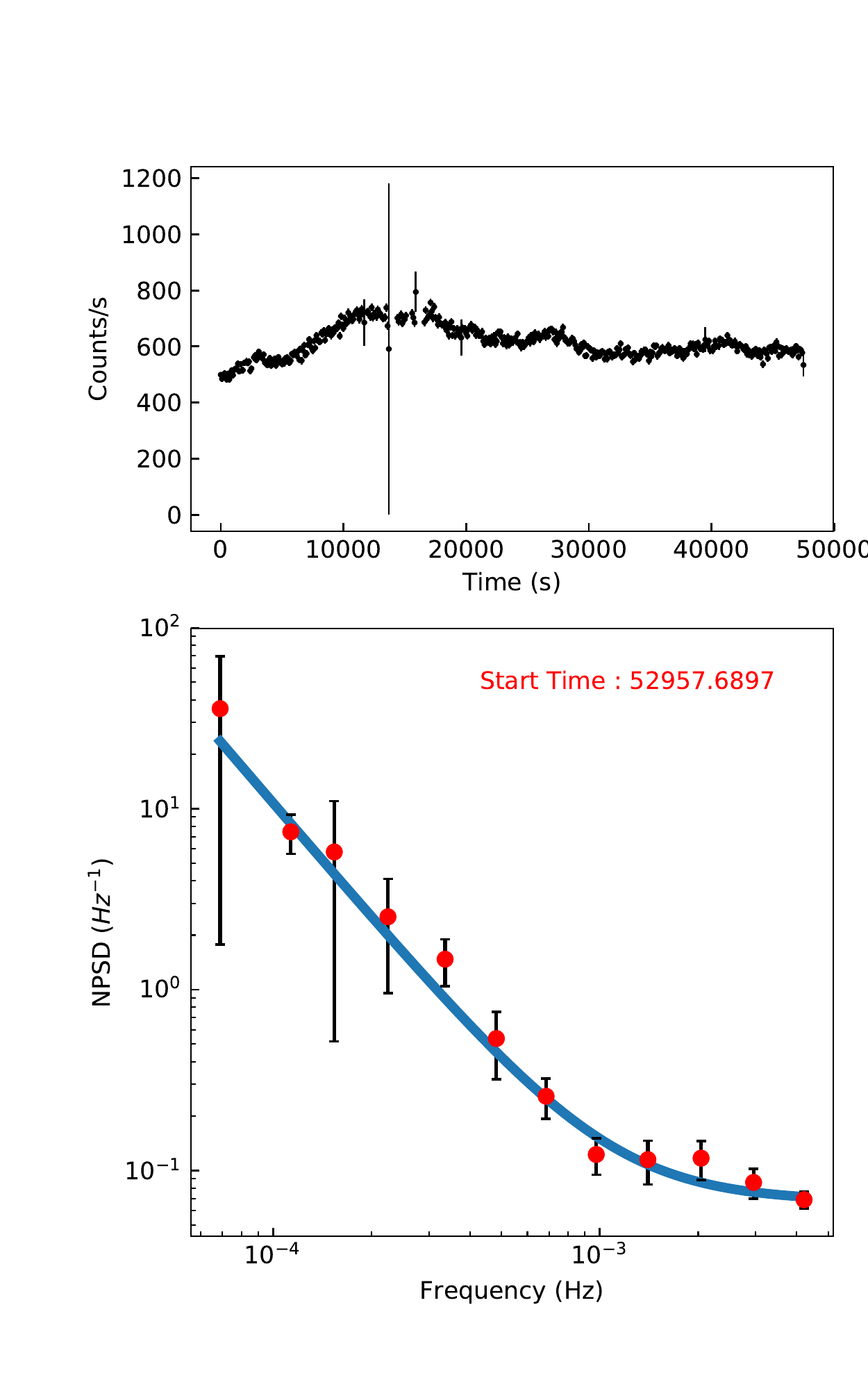}
	\includegraphics[scale=0.35]{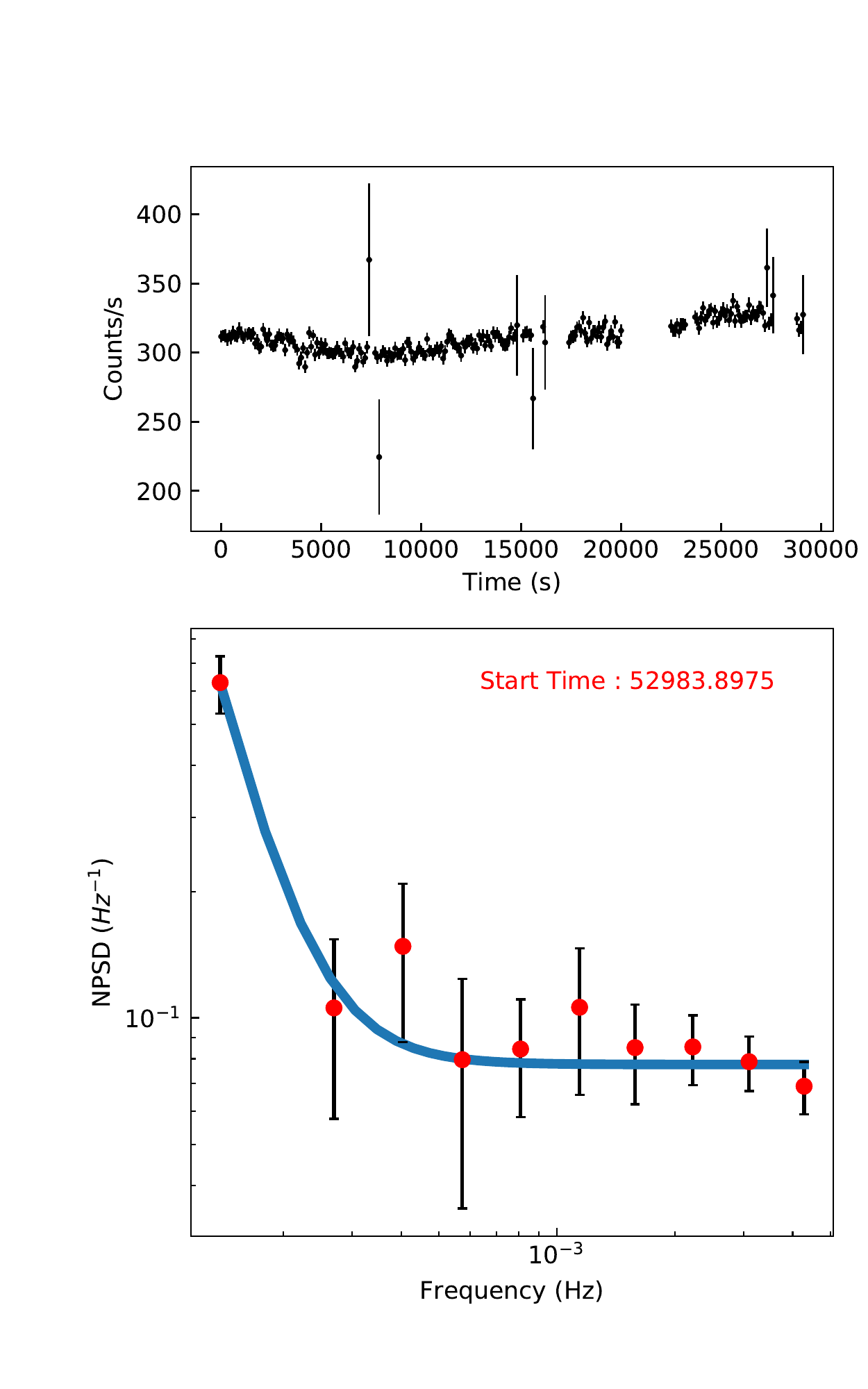}
	\includegraphics[scale=0.35]{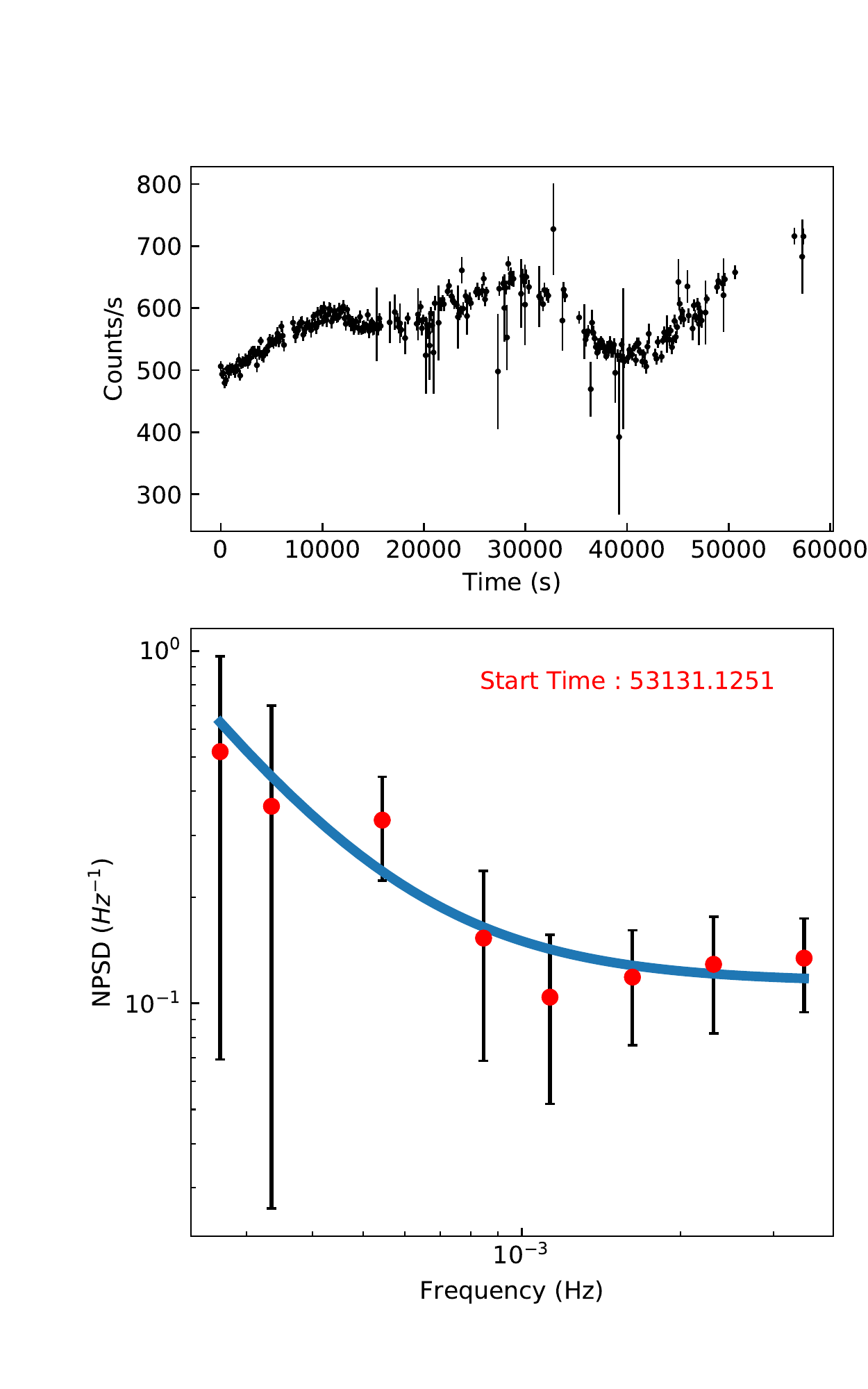}
	\includegraphics[scale=0.35]{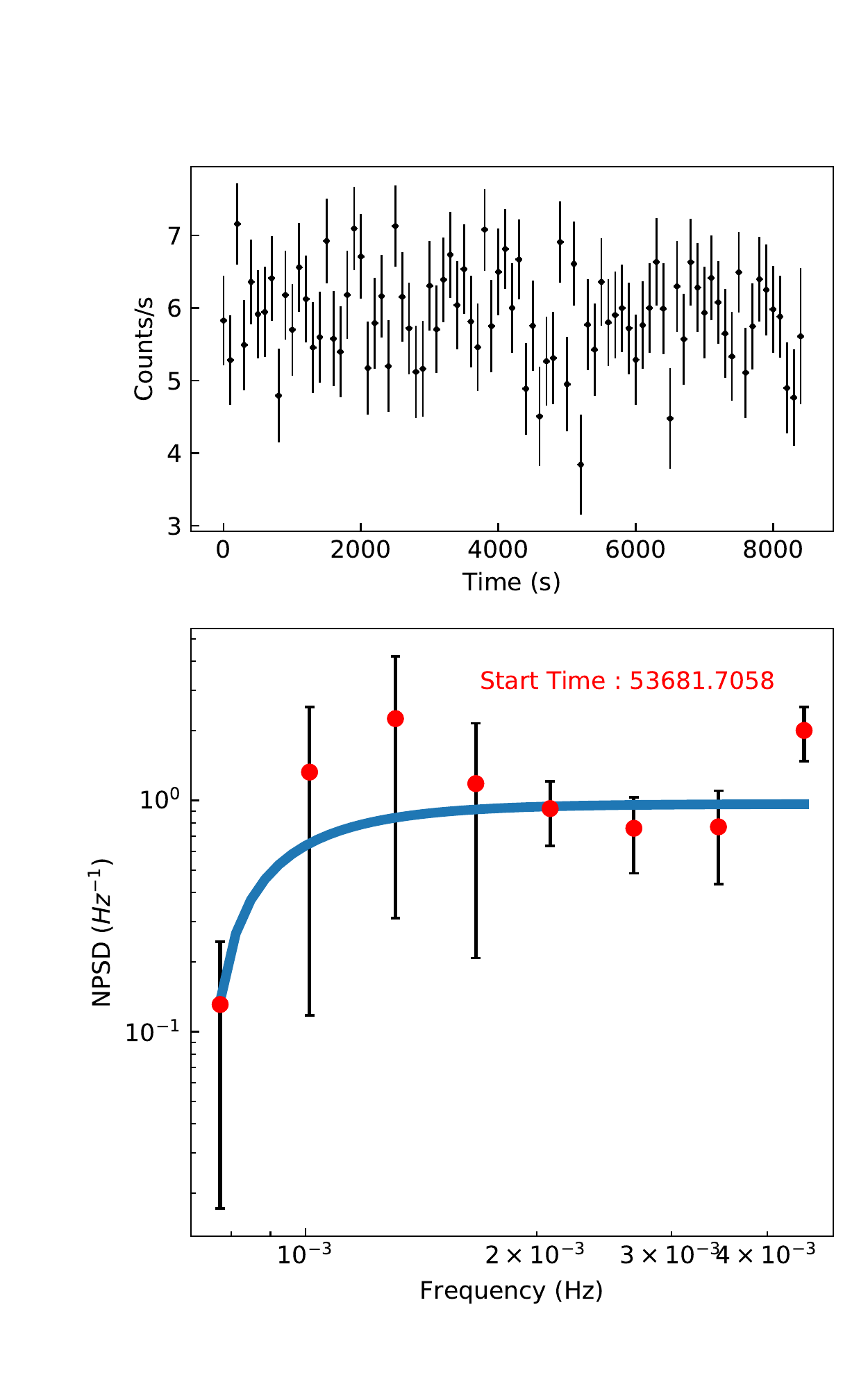}
	\includegraphics[scale=0.35]{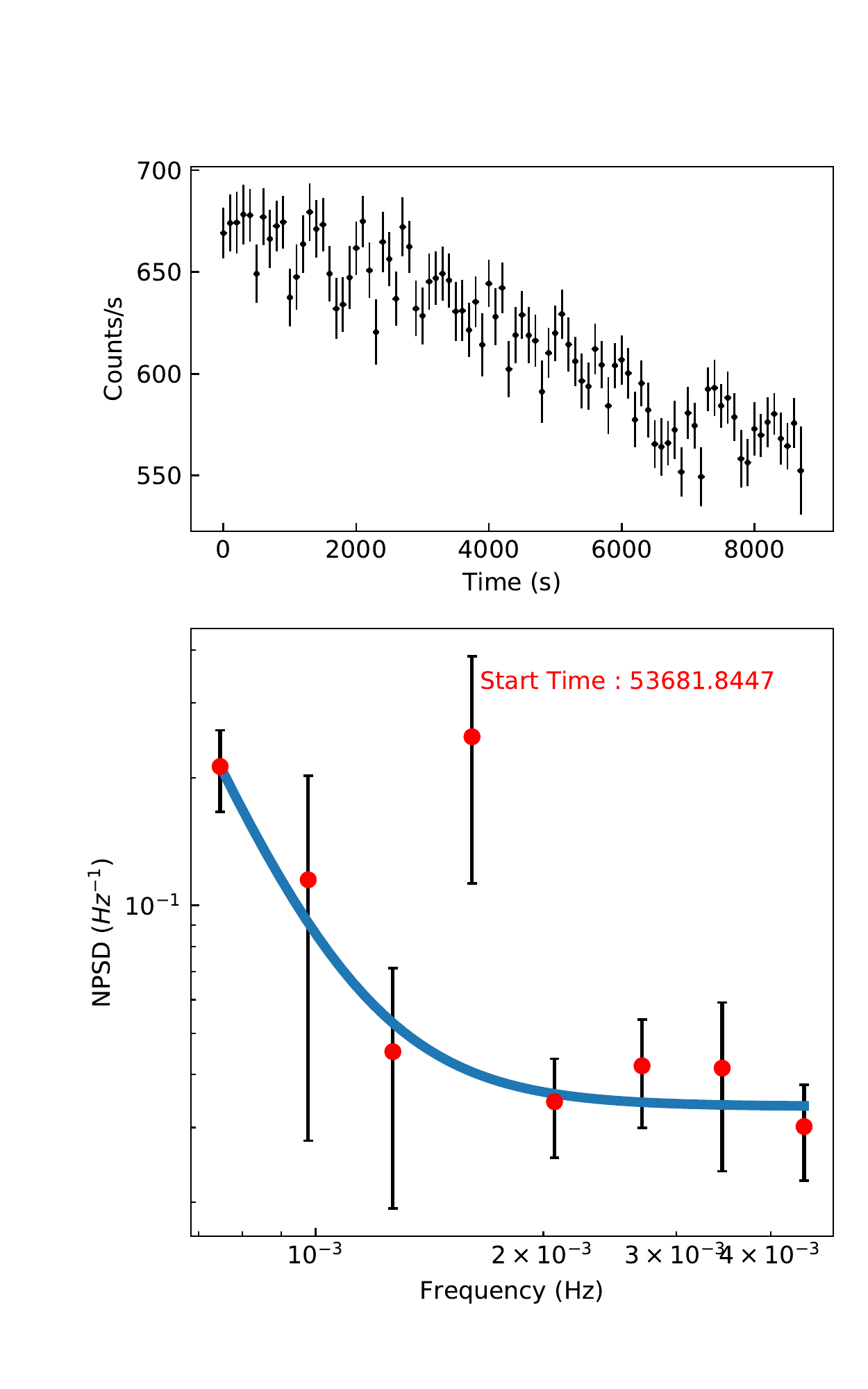}
	\includegraphics[scale=0.35]{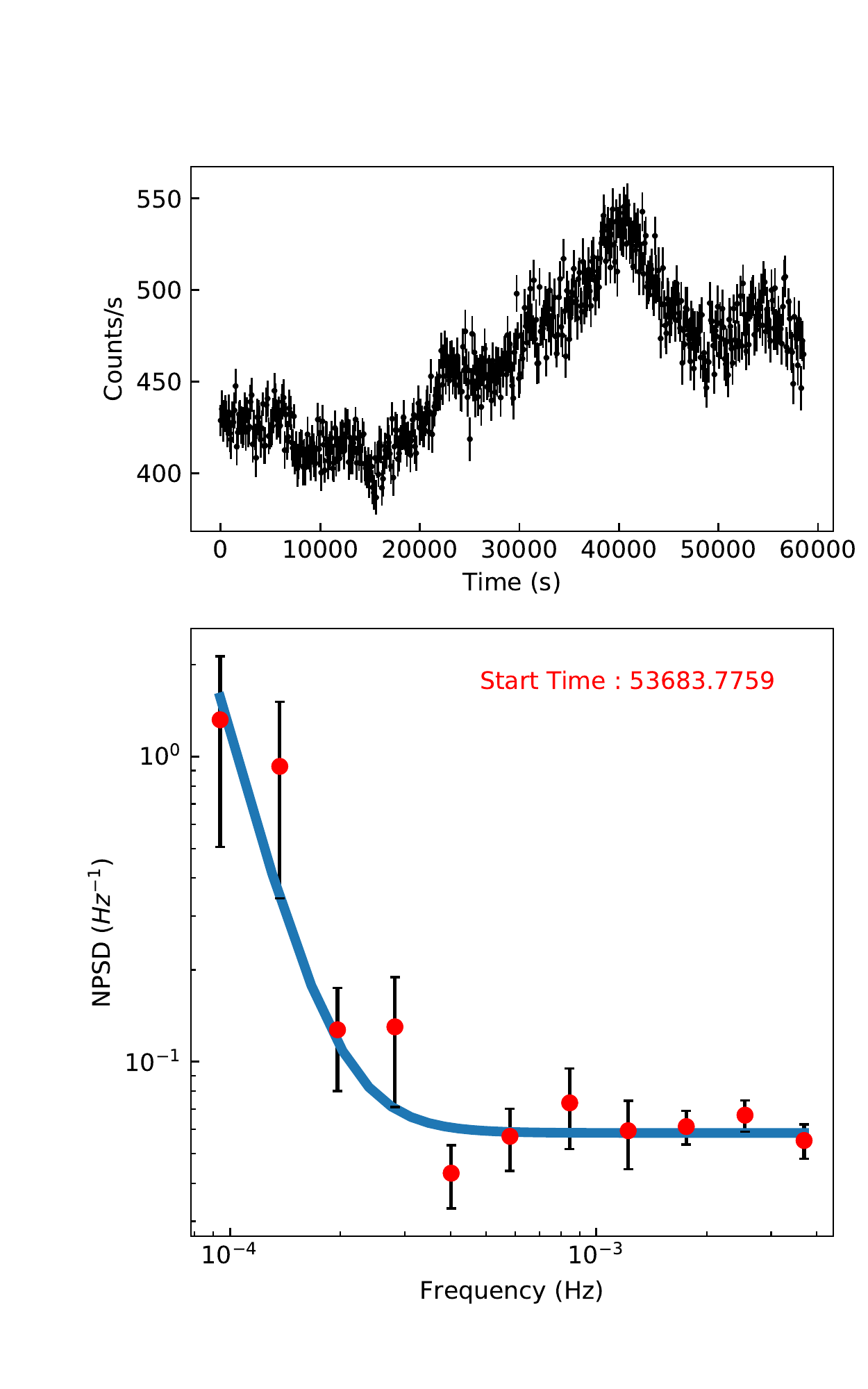}
	\includegraphics[scale=0.35]{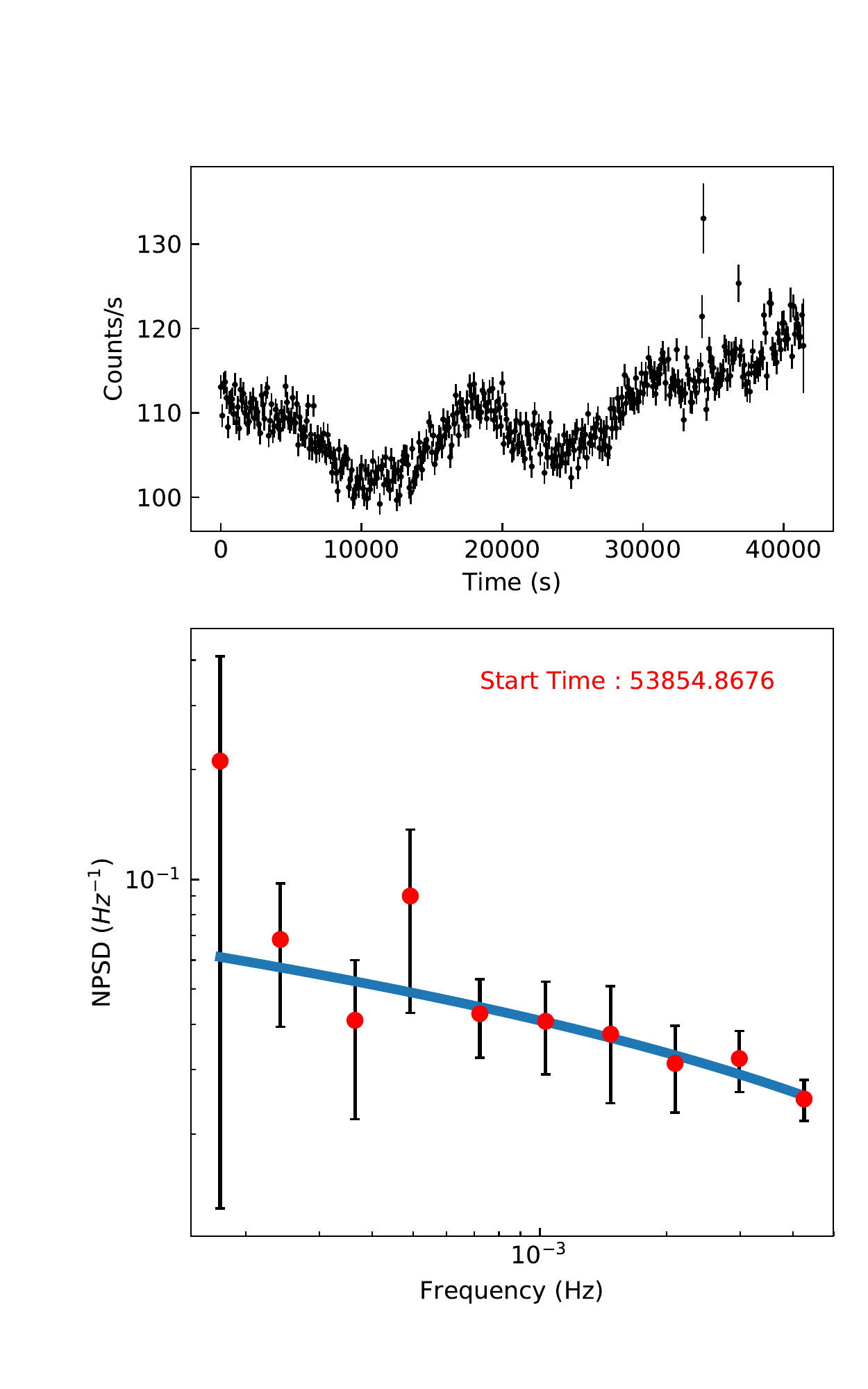}
	\includegraphics[scale=0.35]{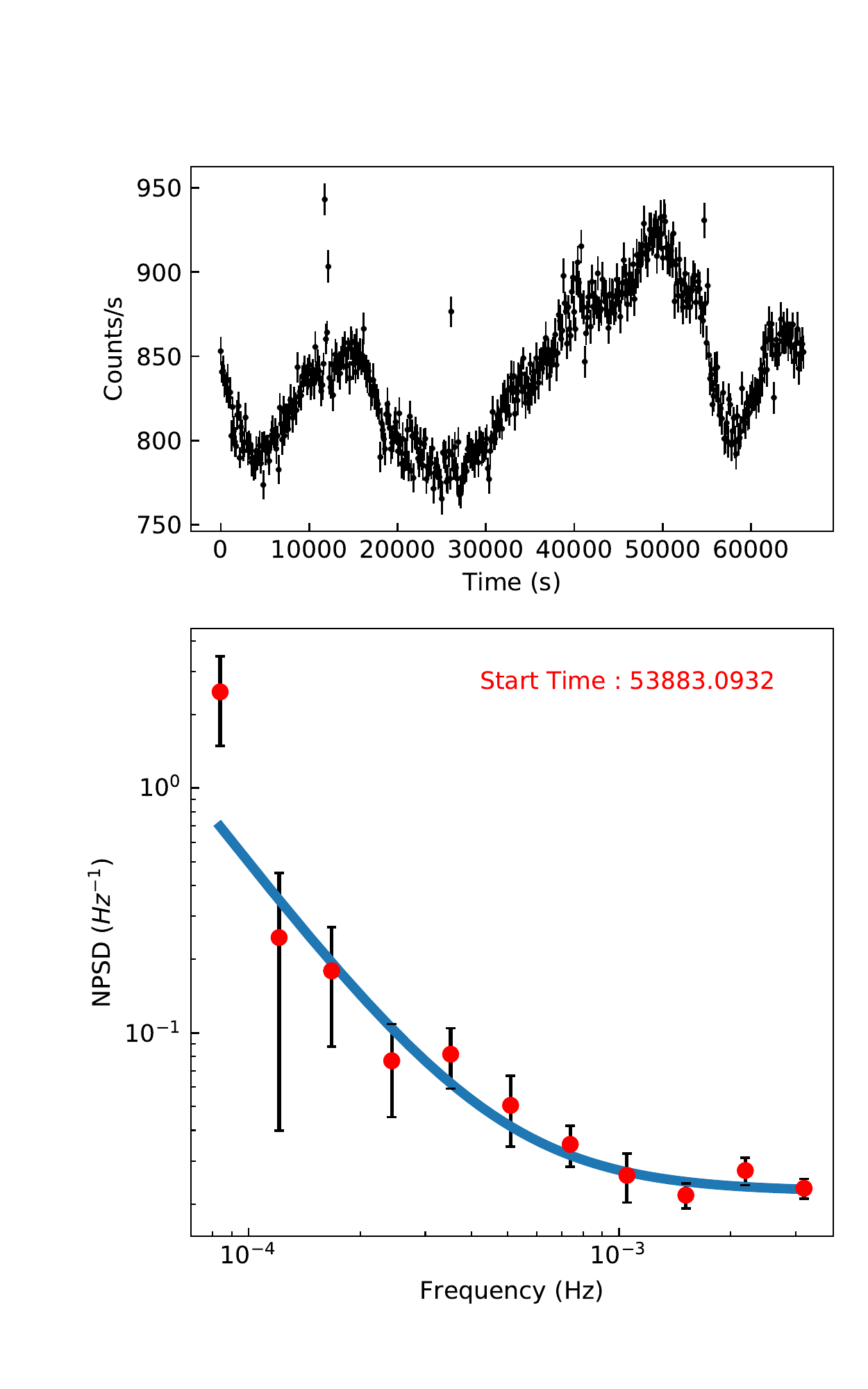}
	\caption{\em{Continued.}}
\end{figure}

\begin{figure}[!h]
	\figurenum{4}
	\includegraphics[scale=0.35]{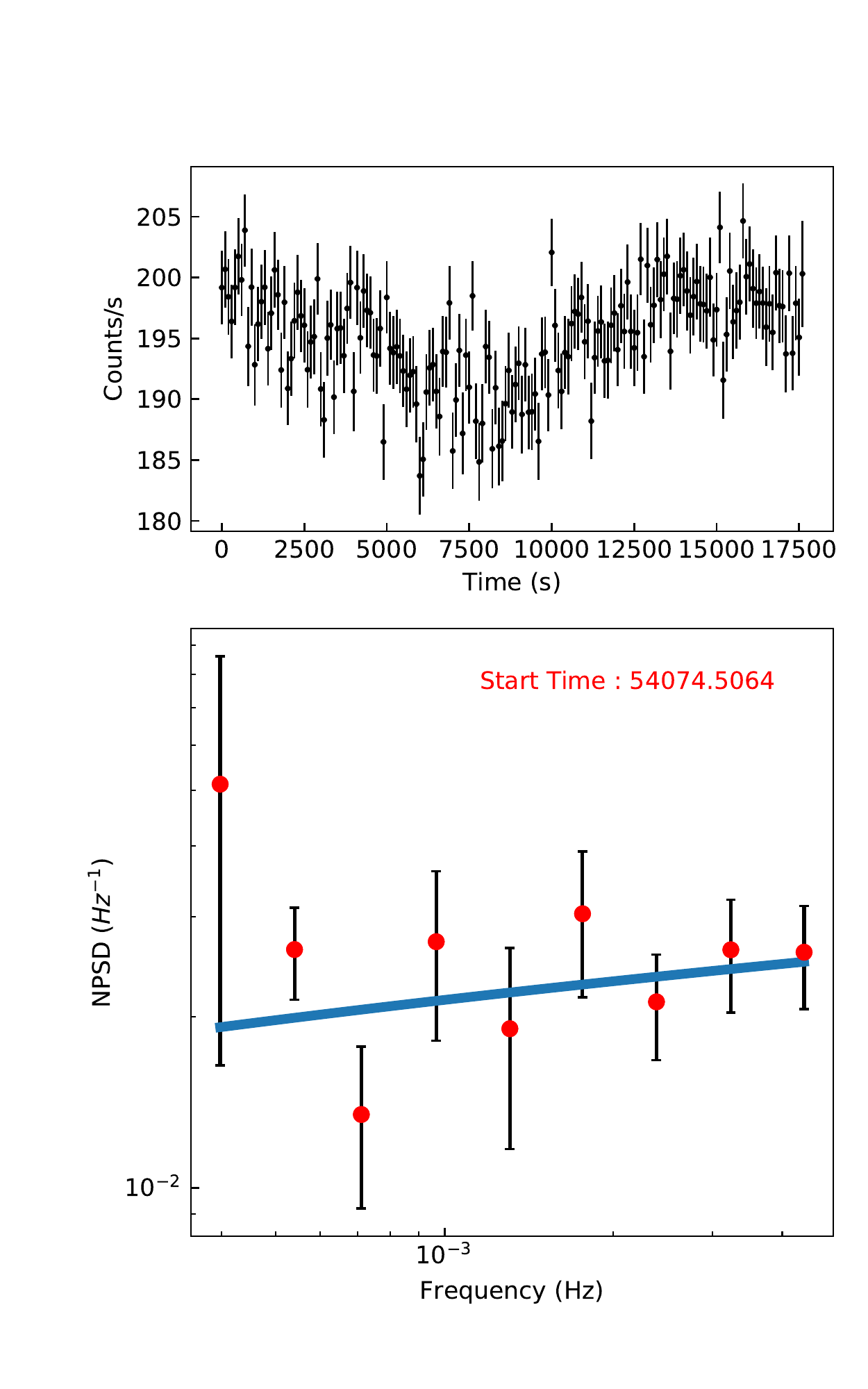}
	\includegraphics[scale=0.35]{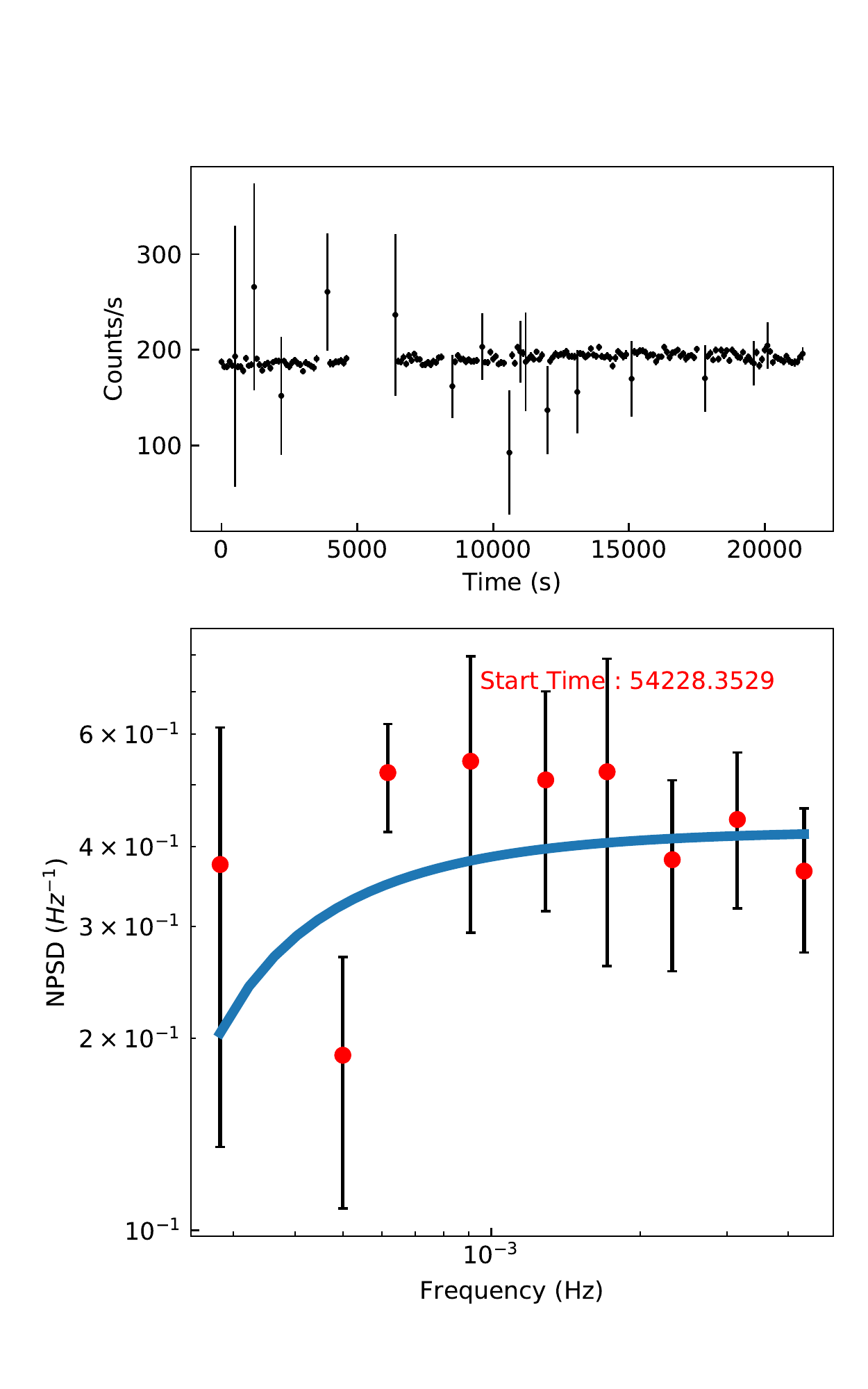}
	\includegraphics[scale=0.35]{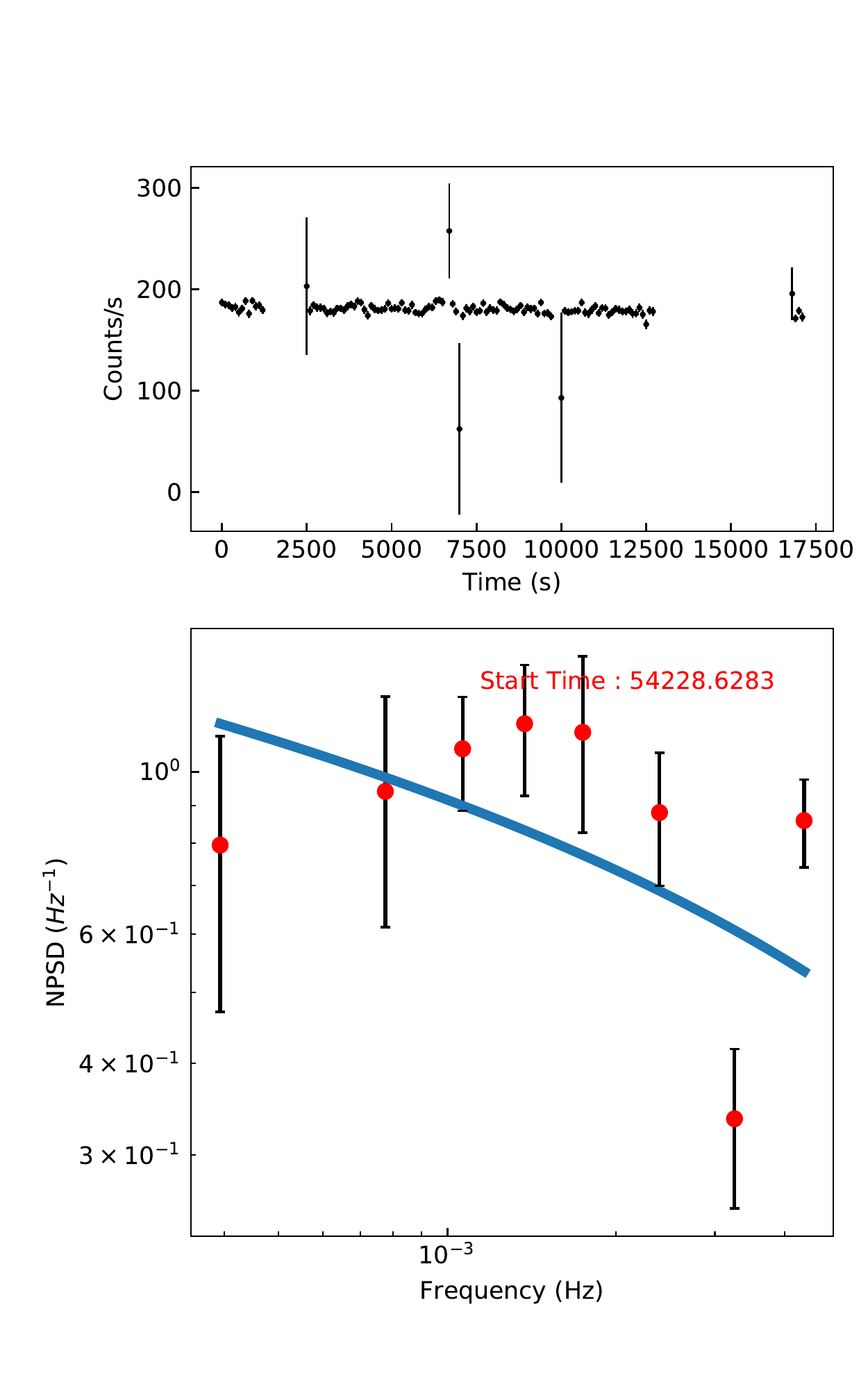}
	\includegraphics[scale=0.35]{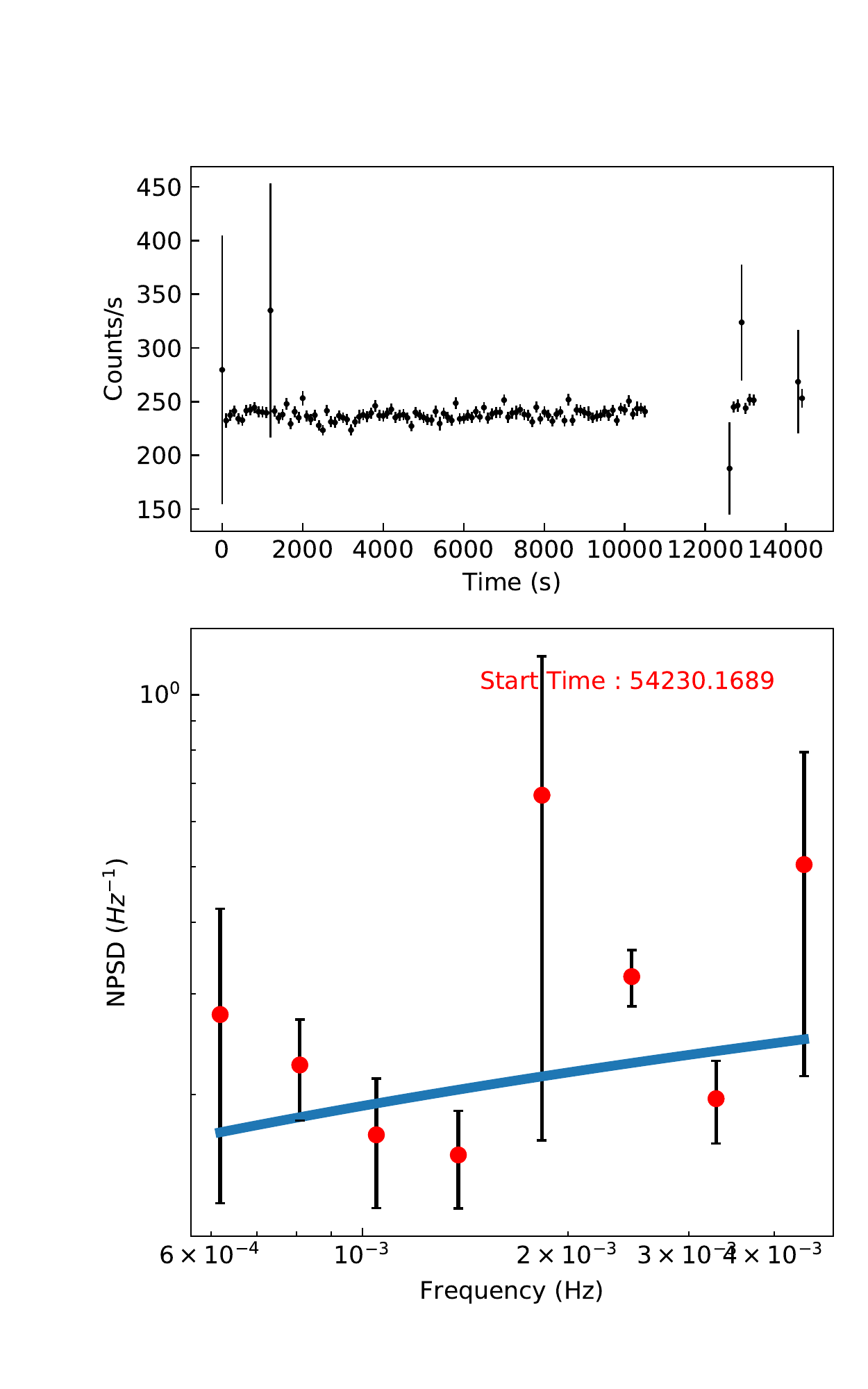}
	\includegraphics[scale=0.35]{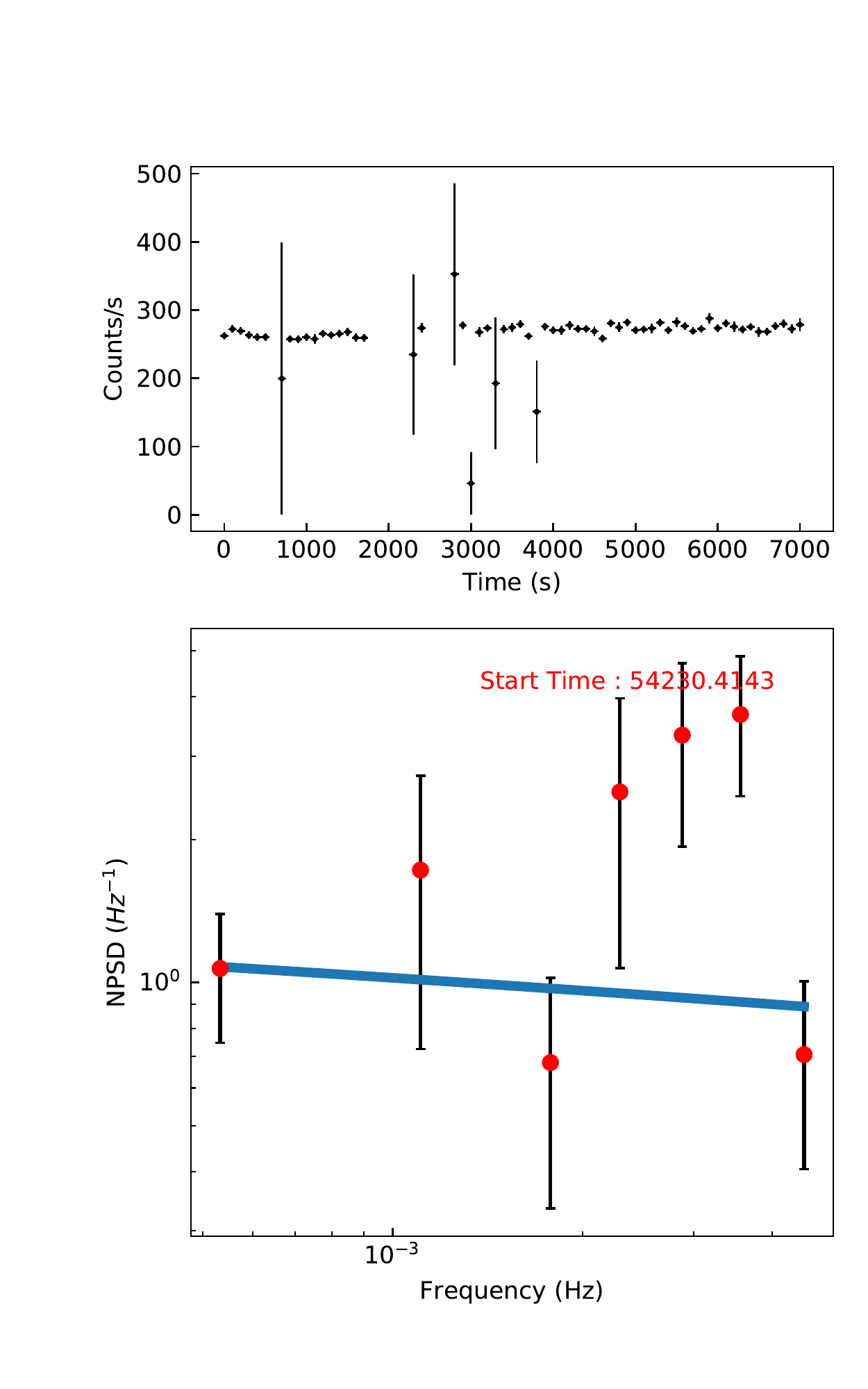}
	\includegraphics[scale=0.35]{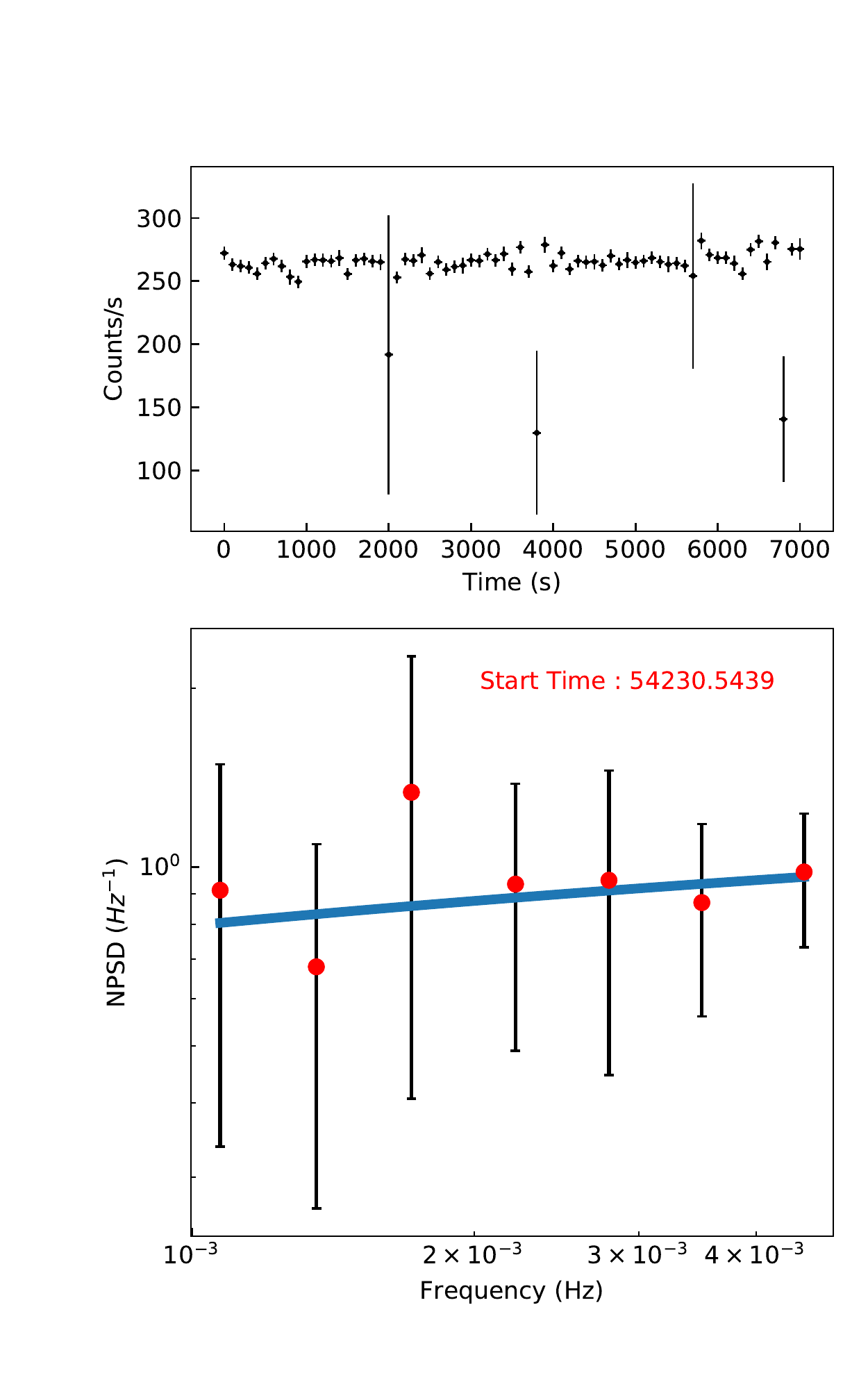}
	\includegraphics[scale=0.35]{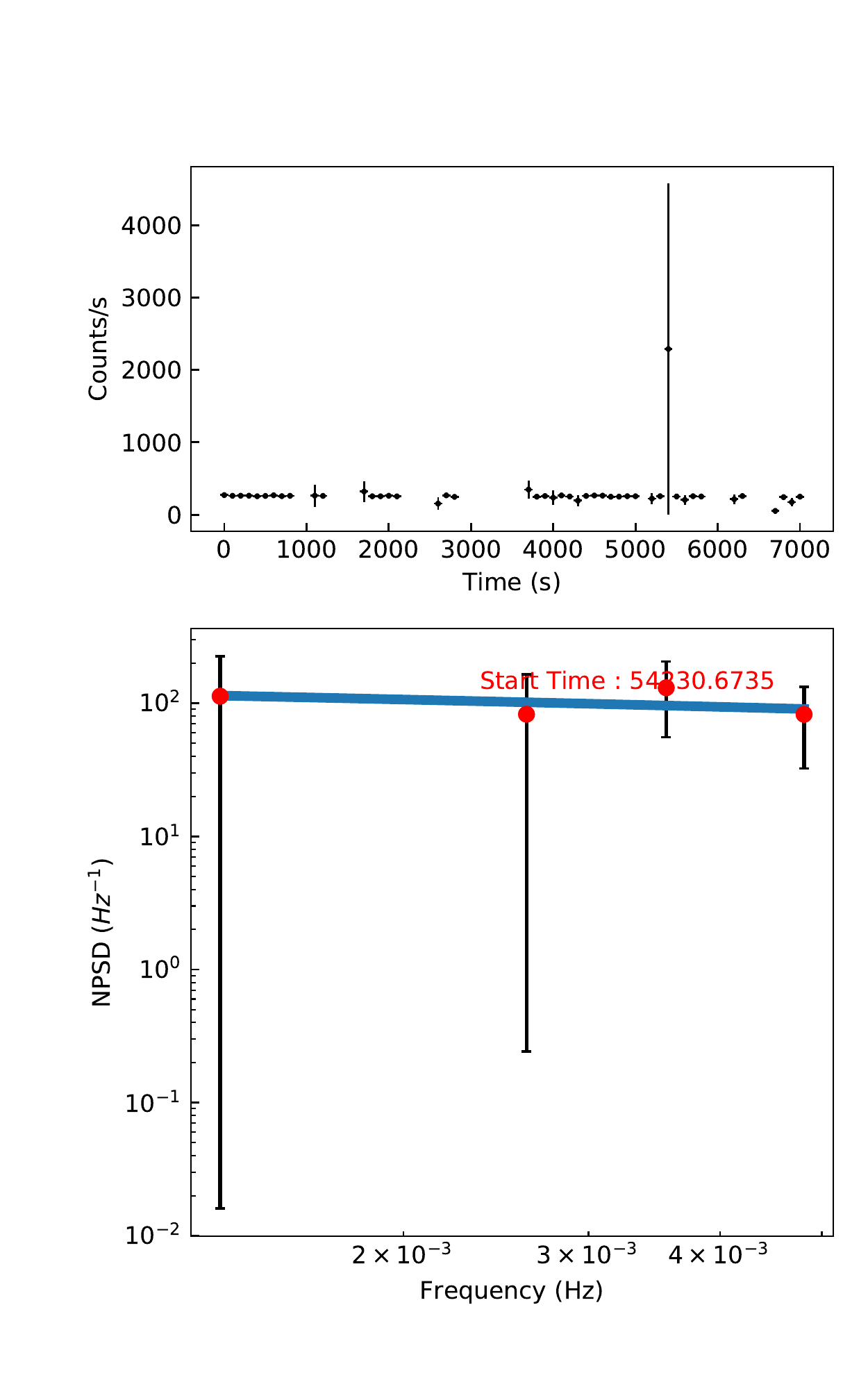}
	\includegraphics[scale=0.35]{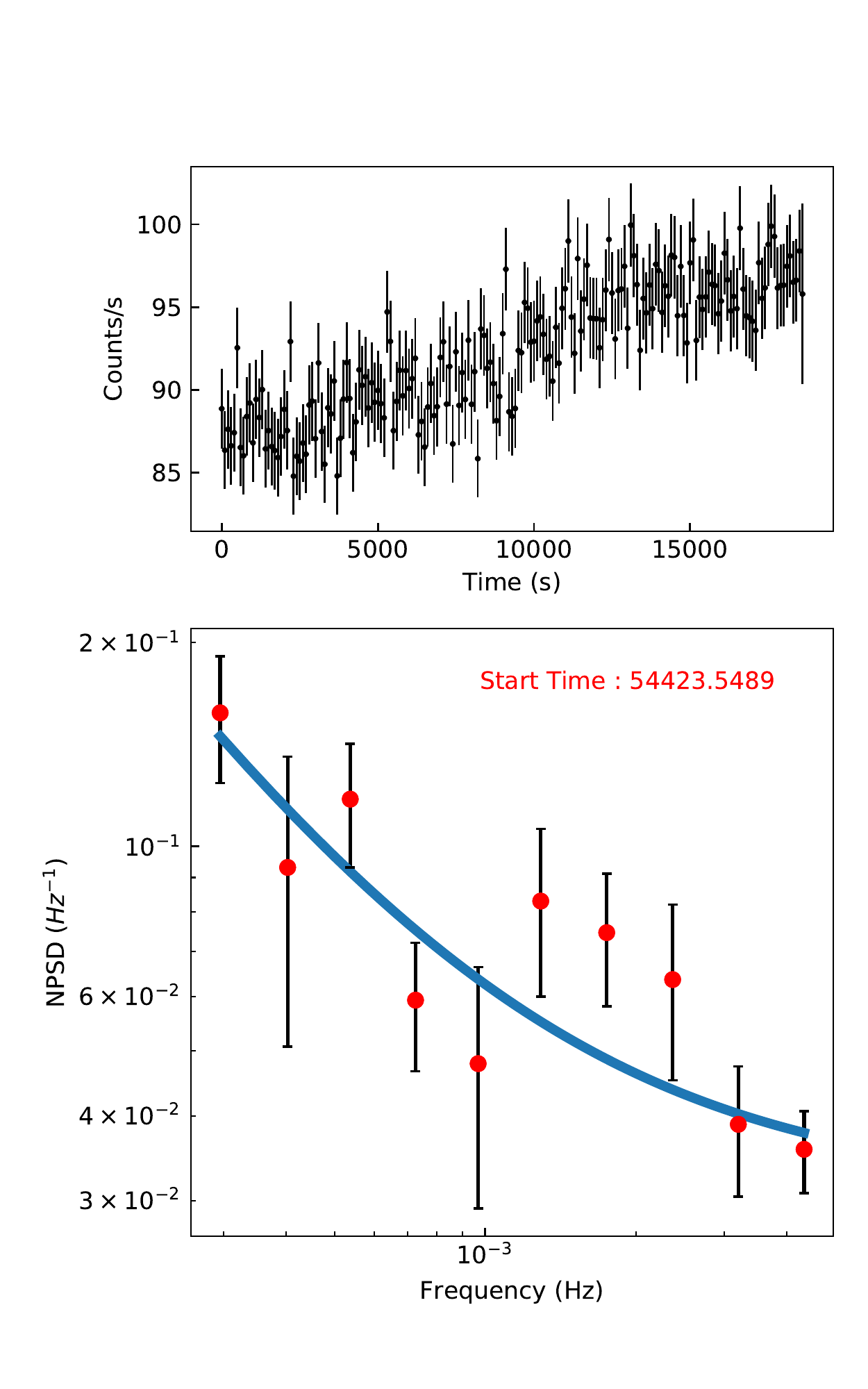}
	\includegraphics[scale=0.35]{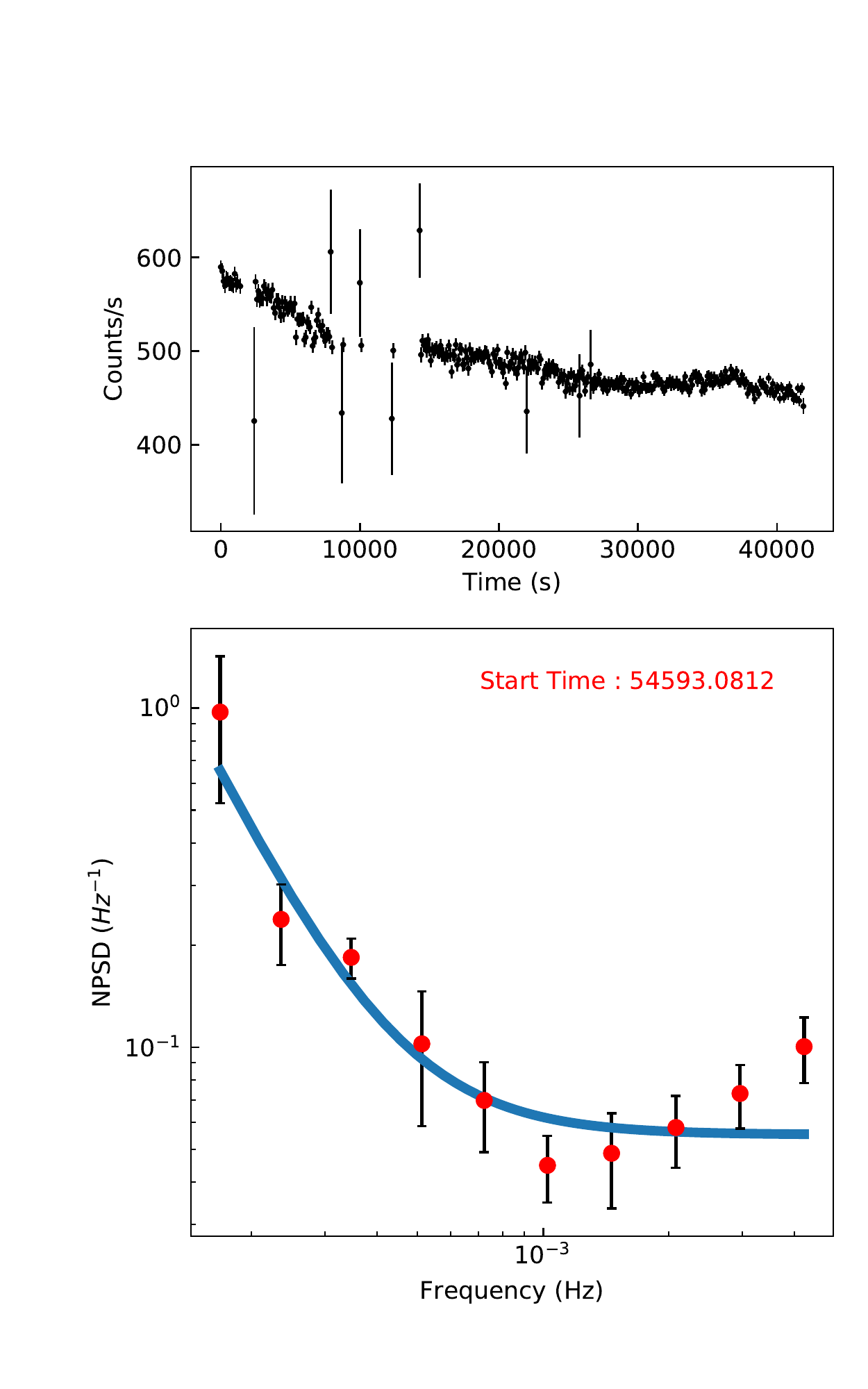}
	\includegraphics[scale=0.35]{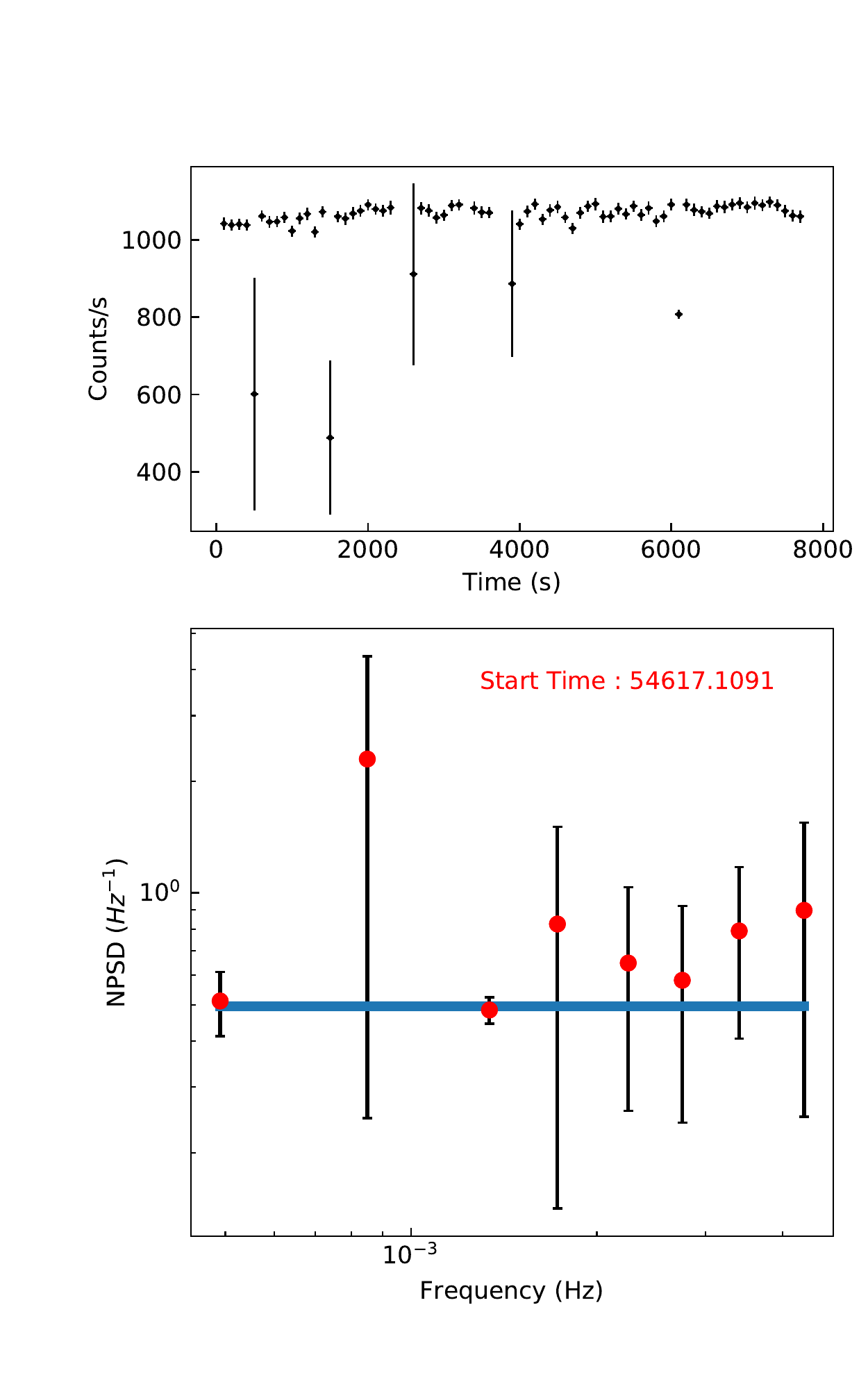}
	\includegraphics[scale=0.35]{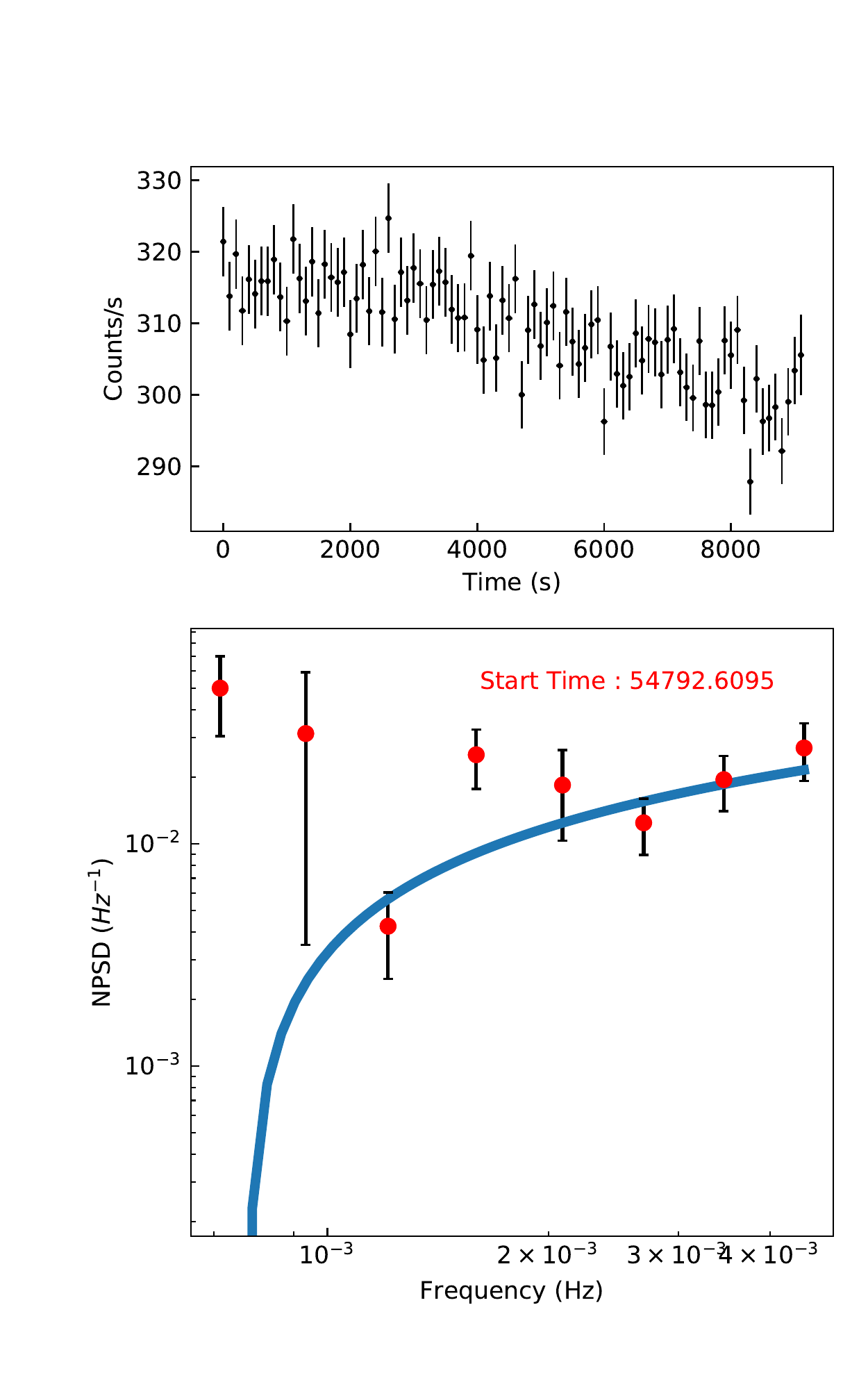}
	\includegraphics[scale=0.35]{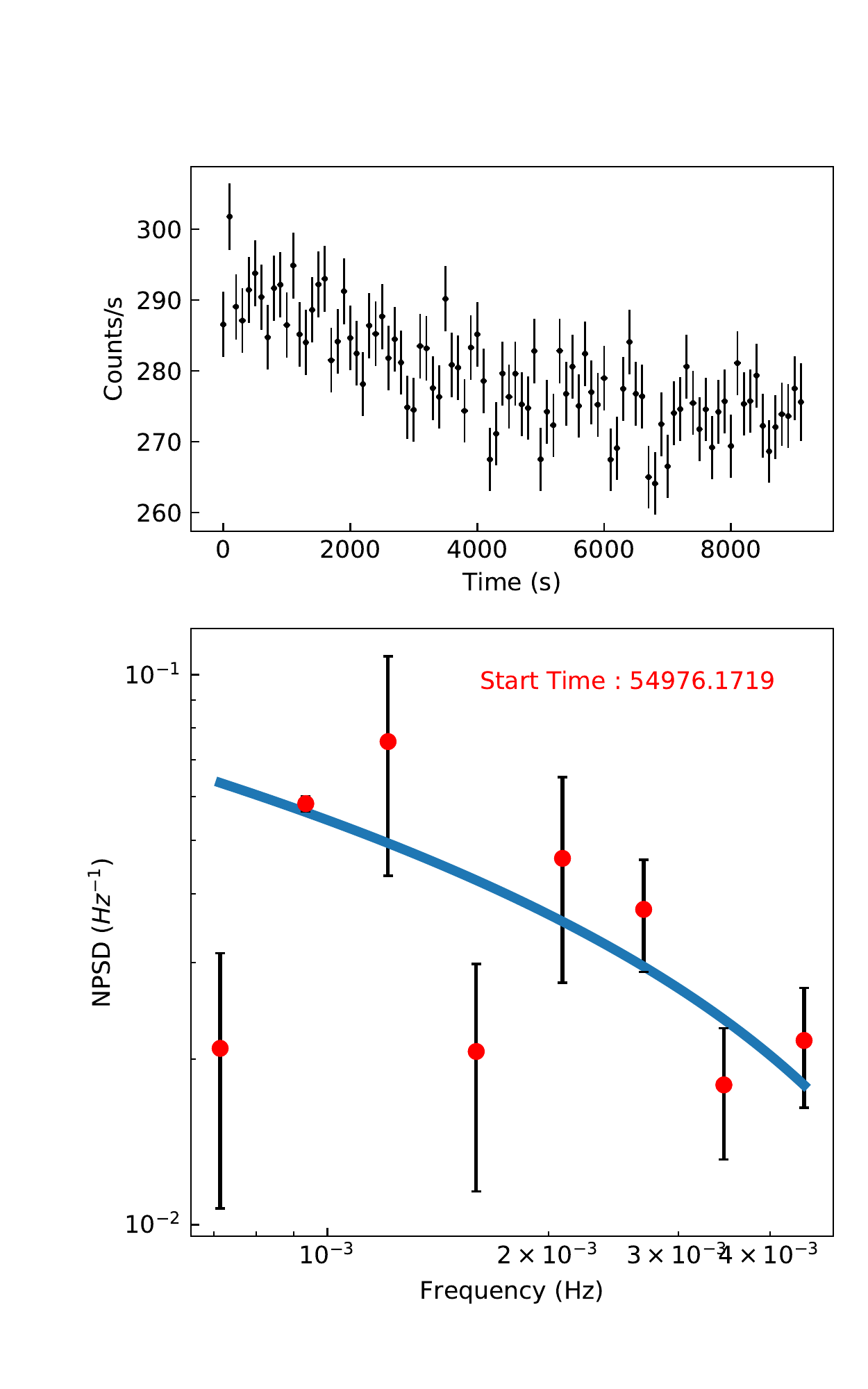}
	\caption{\em{Continued.}}
\end{figure}
\begin{figure}[!h]
	\figurenum{4}
	\includegraphics[scale=0.35]{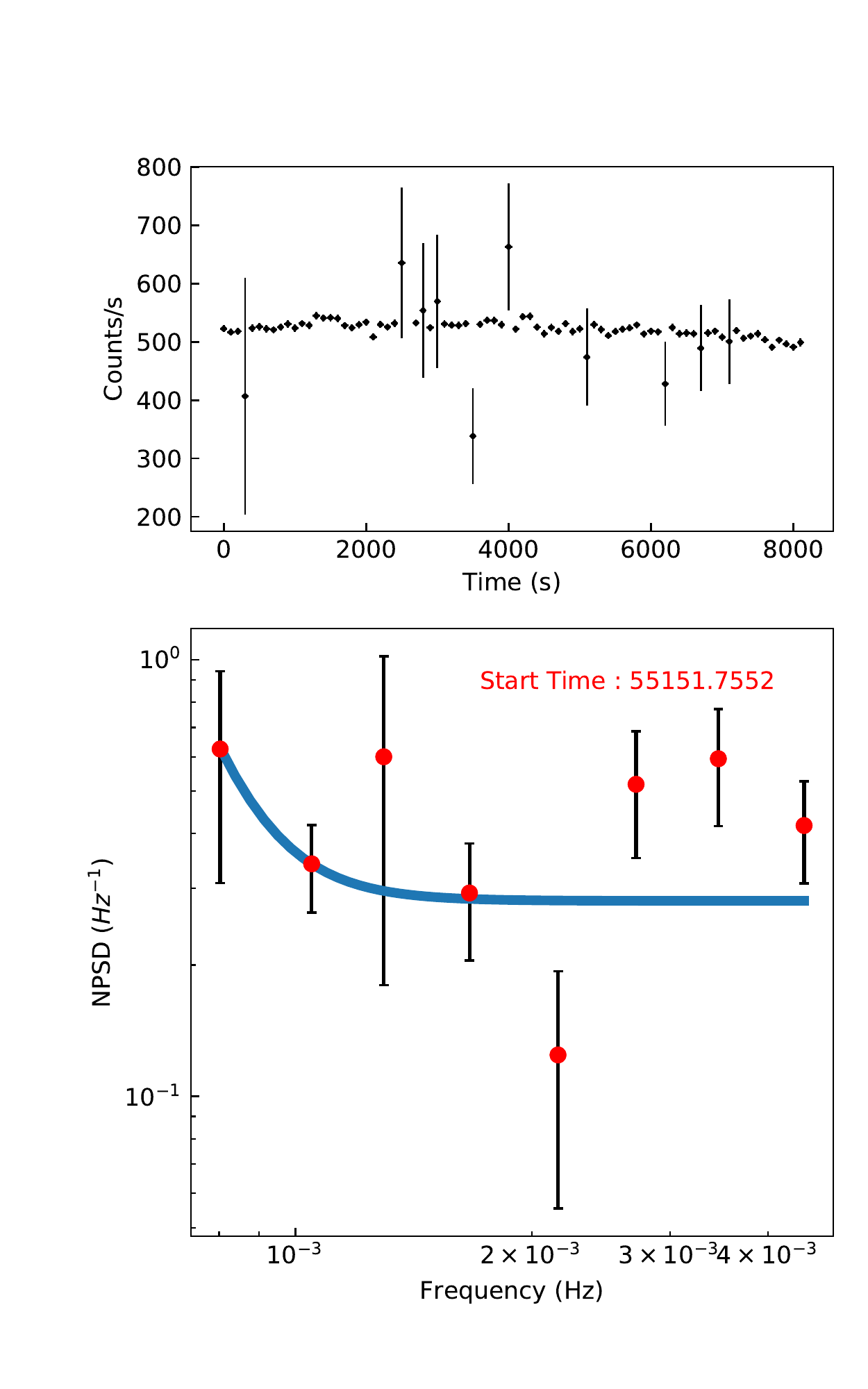}
	\includegraphics[scale=0.35]{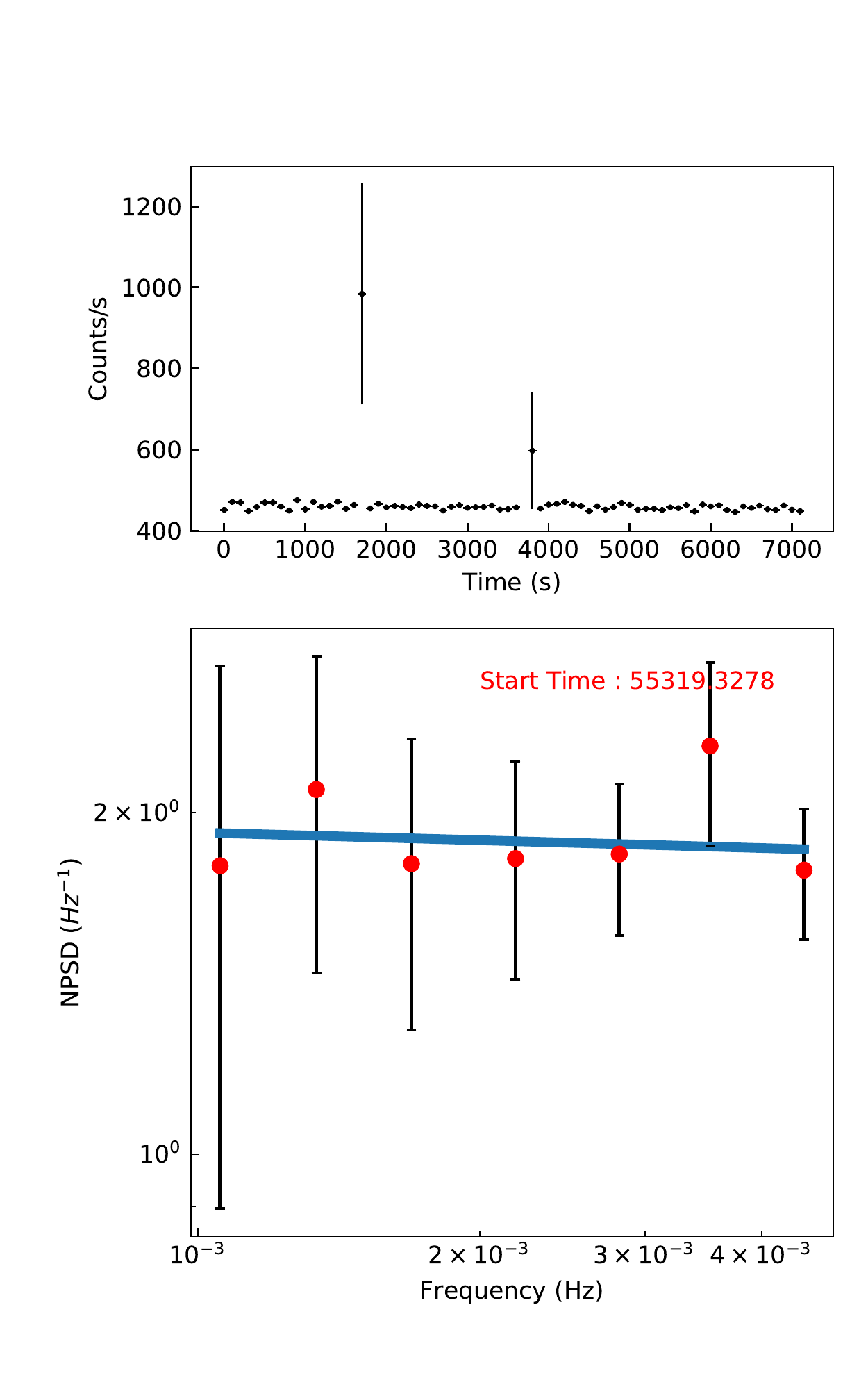}
	\includegraphics[scale=0.35]{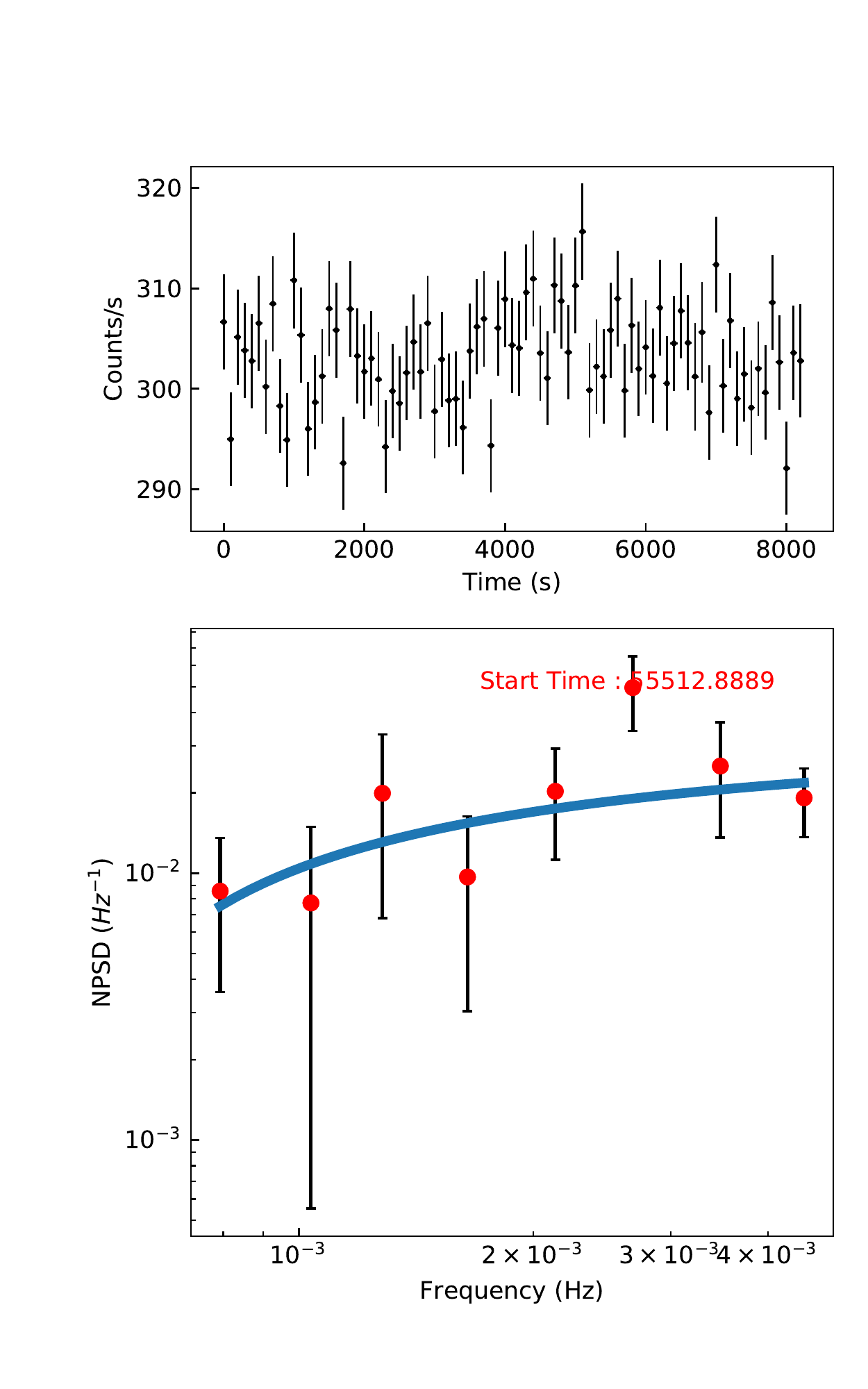}
	\includegraphics[scale=0.35]{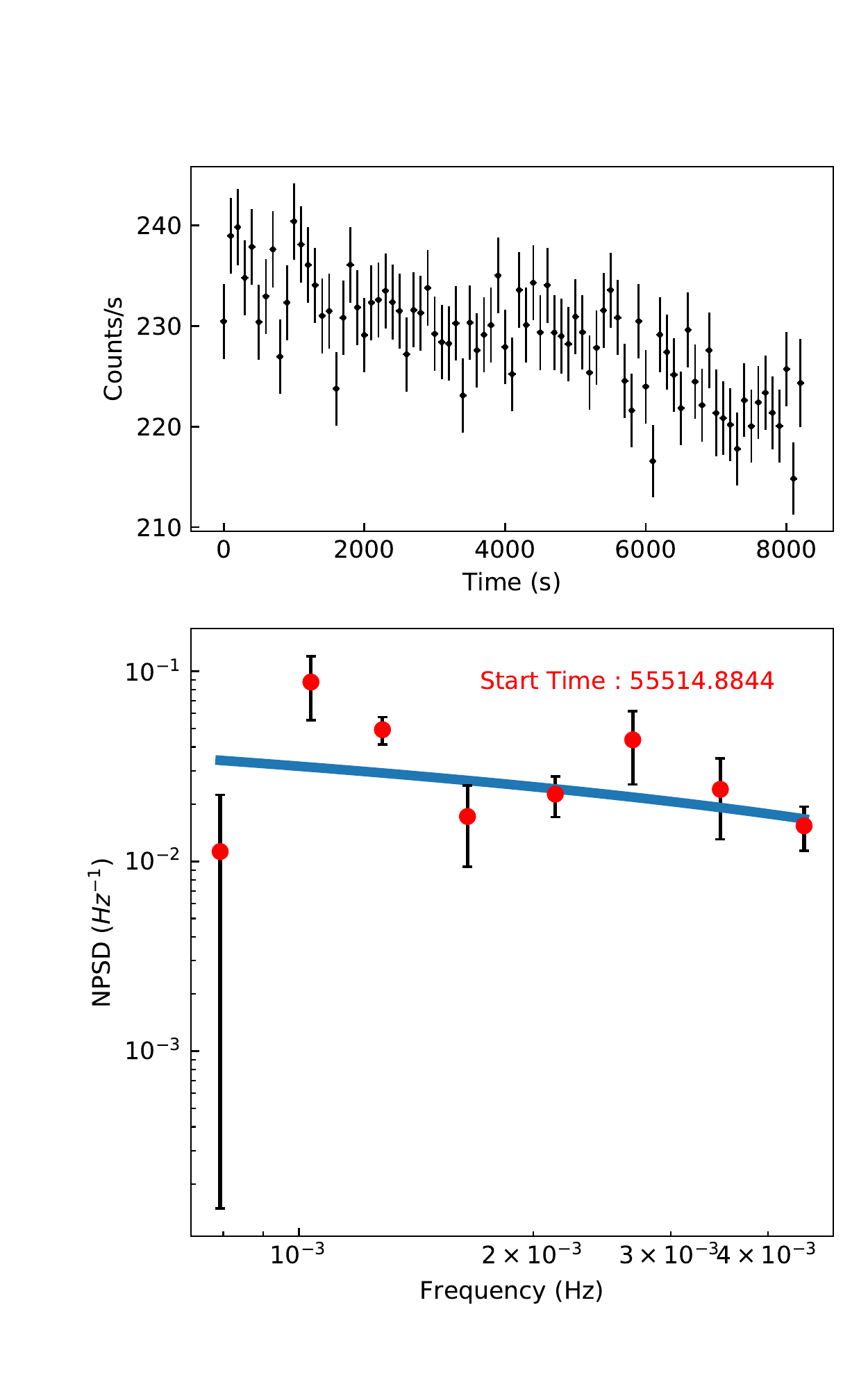}
	\includegraphics[scale=0.35]{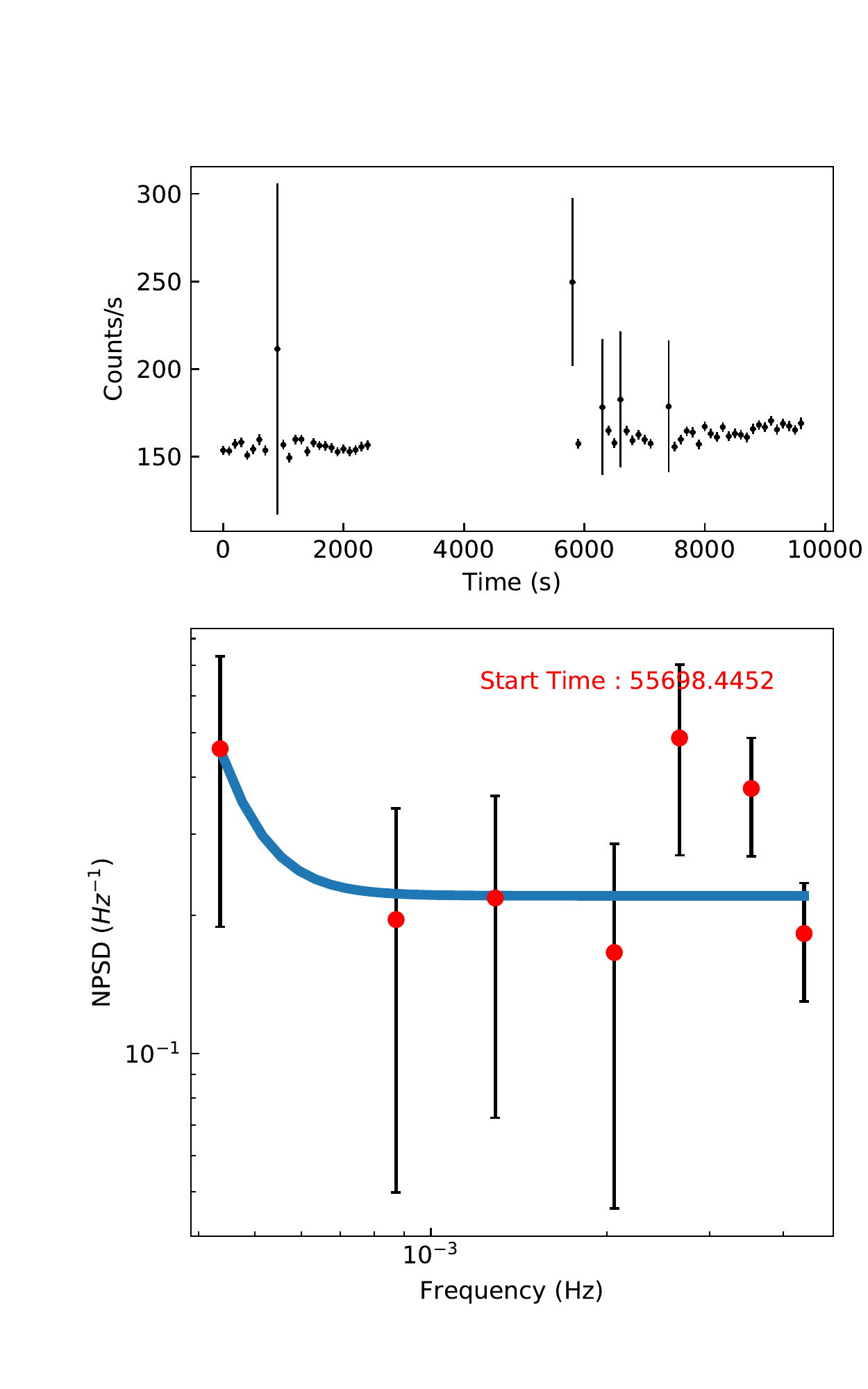}
	\includegraphics[scale=0.35]{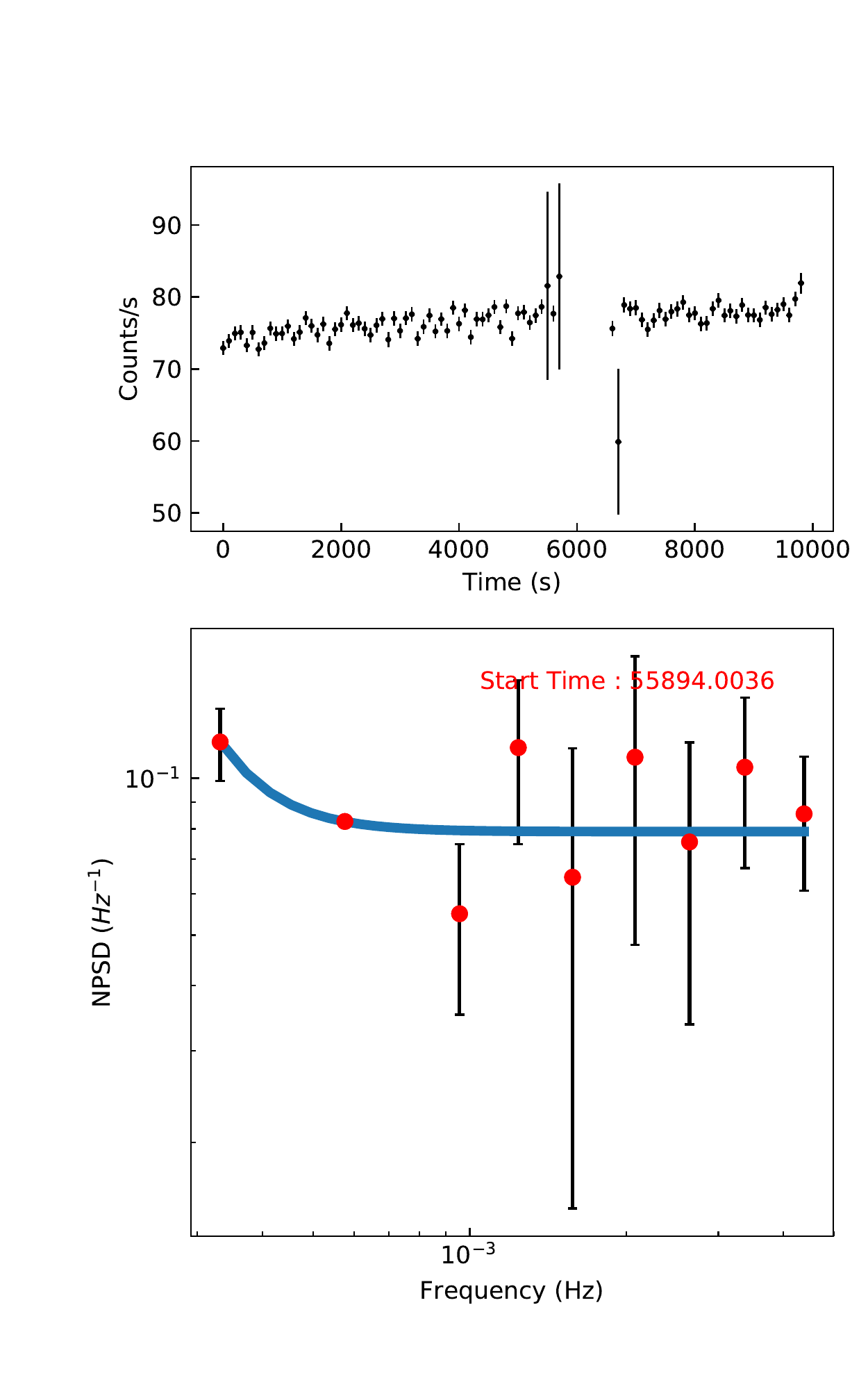}
	\includegraphics[scale=0.35]{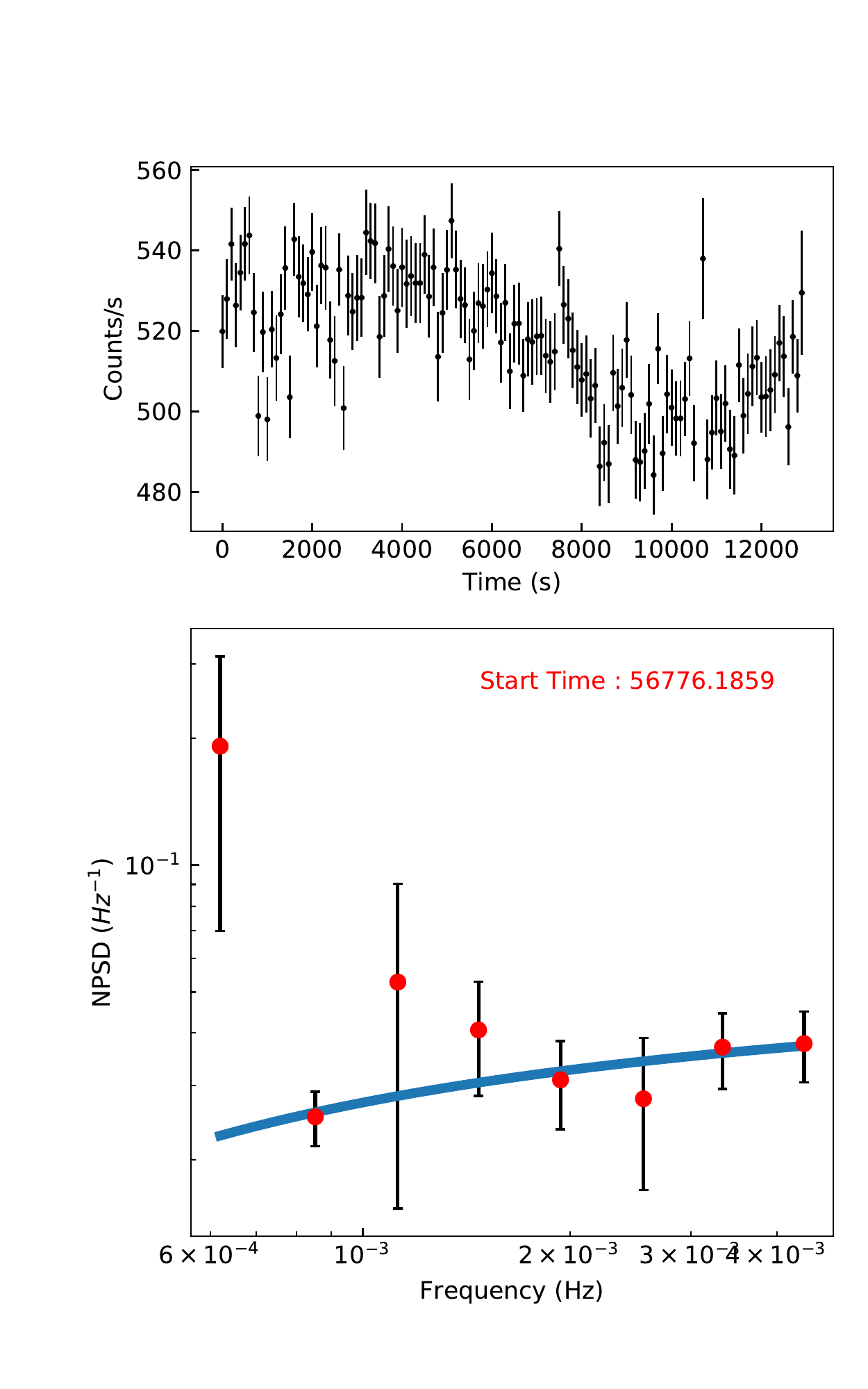}
	\includegraphics[scale=0.35]{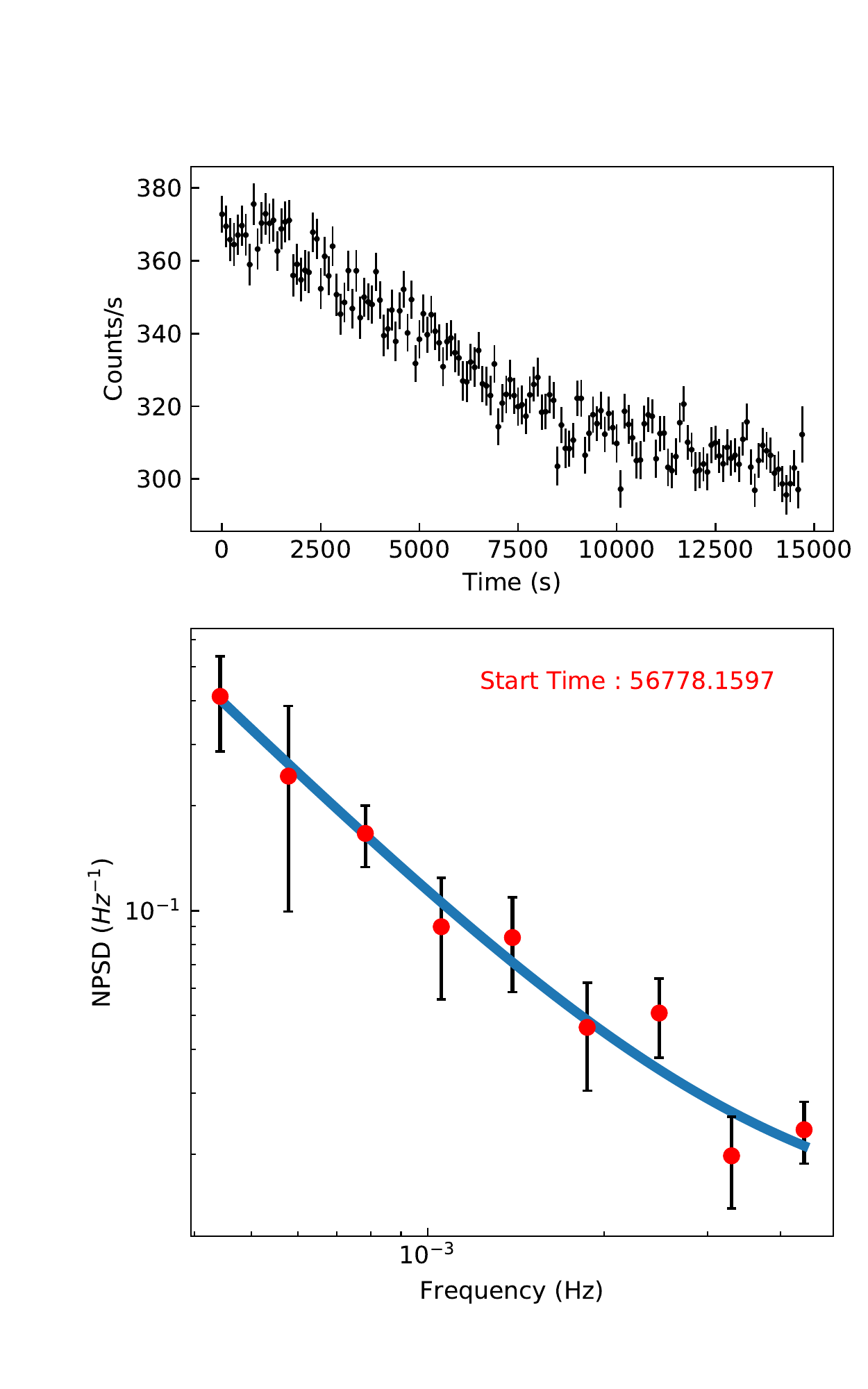}
	\includegraphics[scale=0.35]{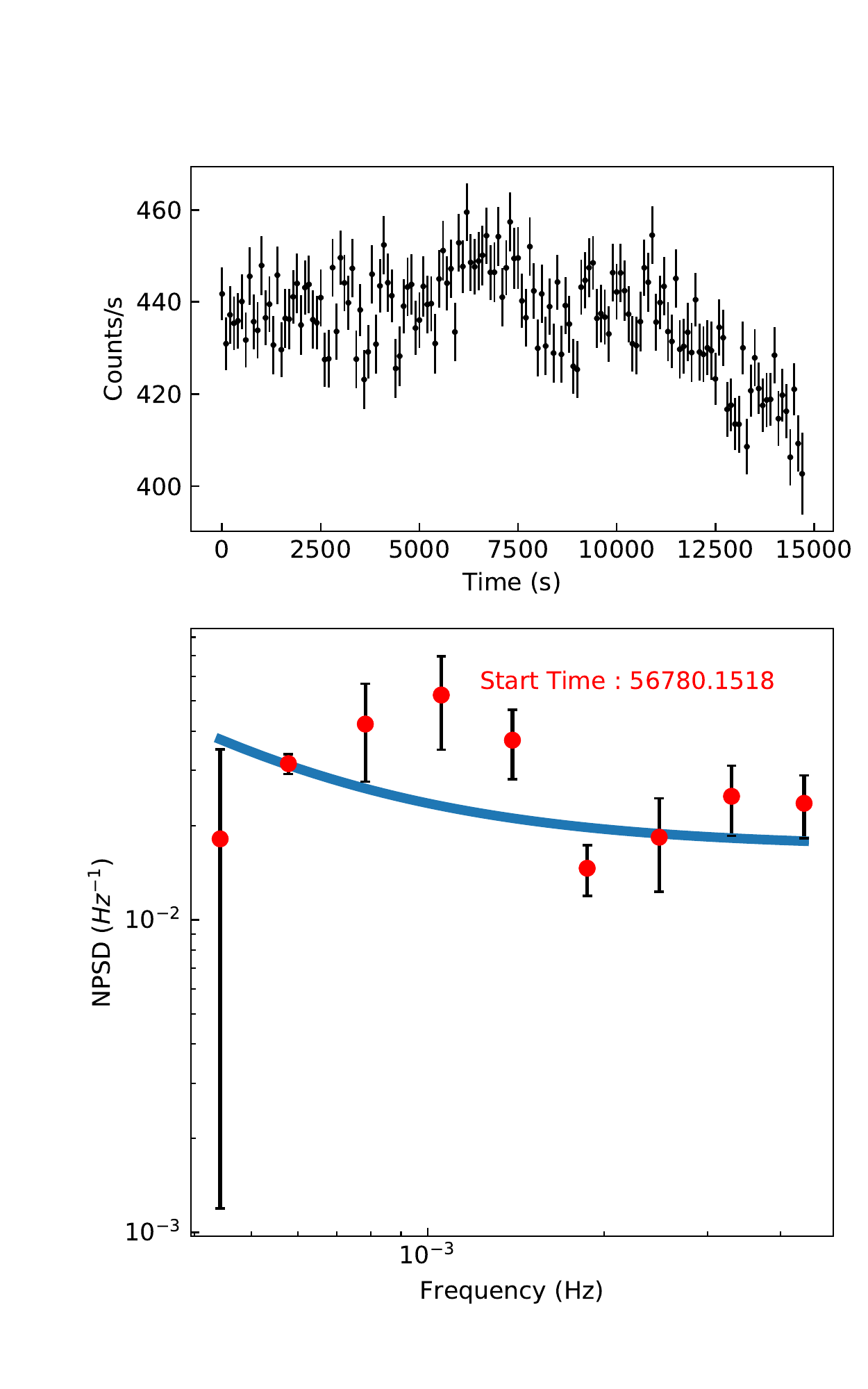}
	\includegraphics[scale=0.35]{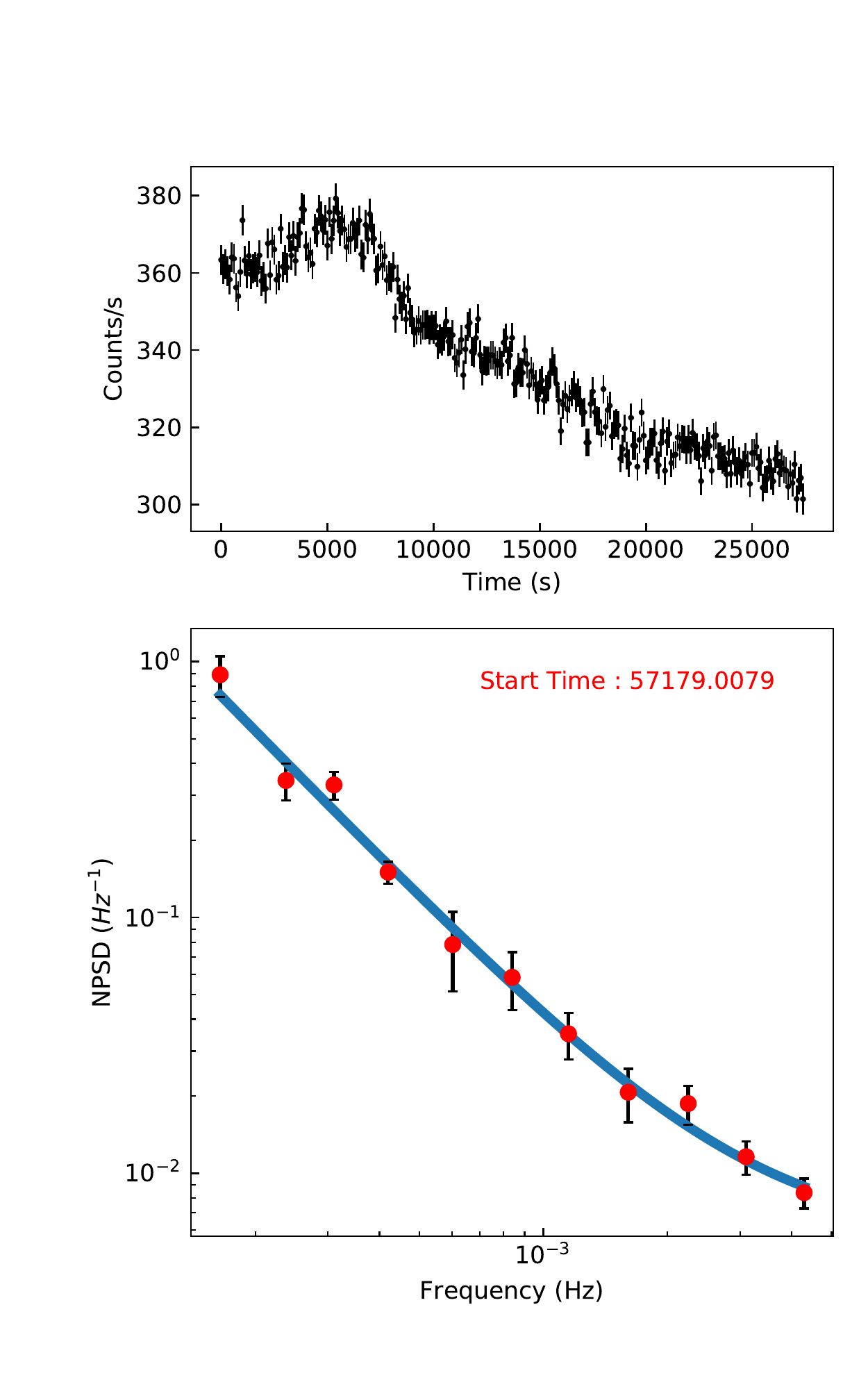}
	\includegraphics[scale=0.35]{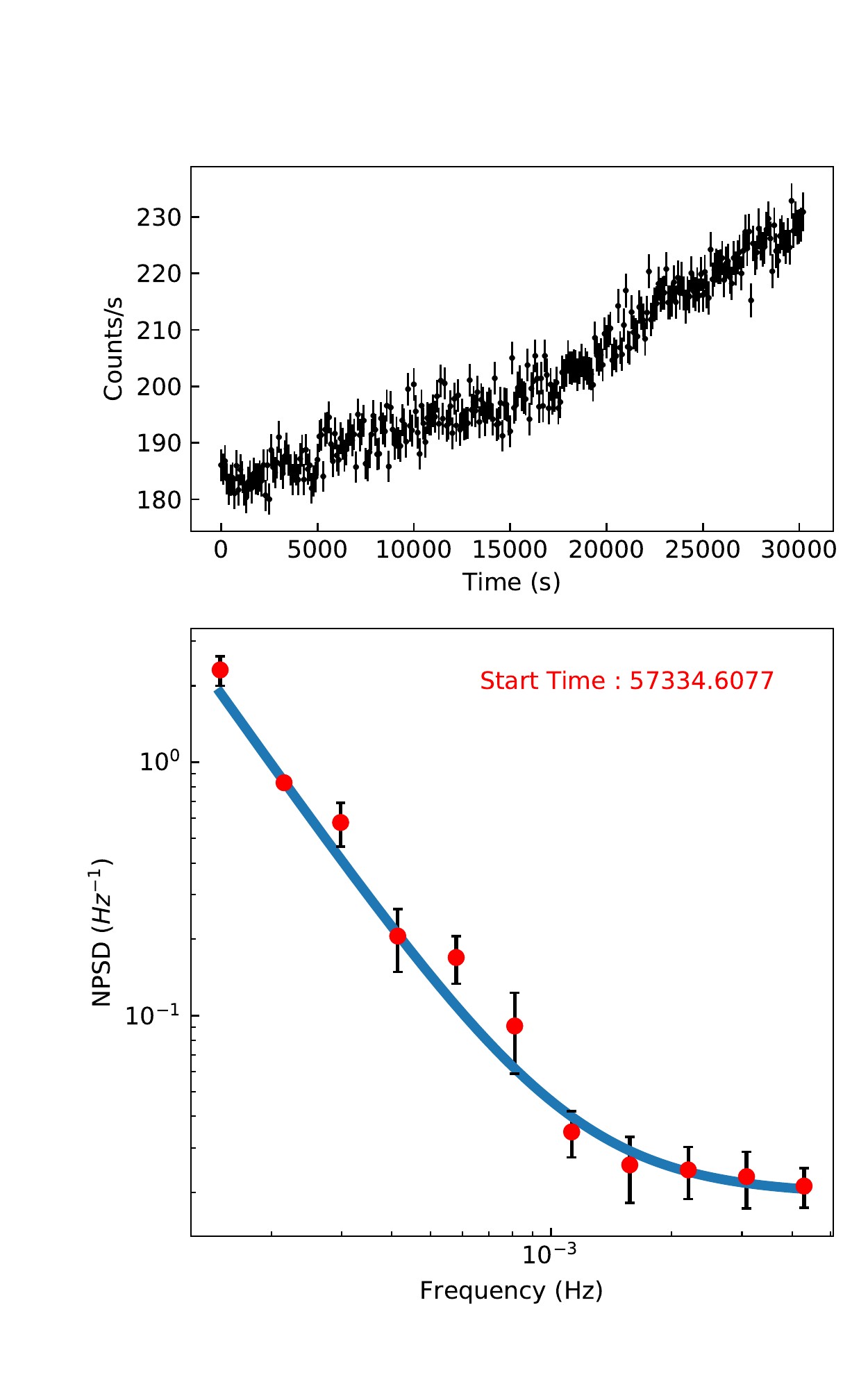}
	\includegraphics[scale=0.35]{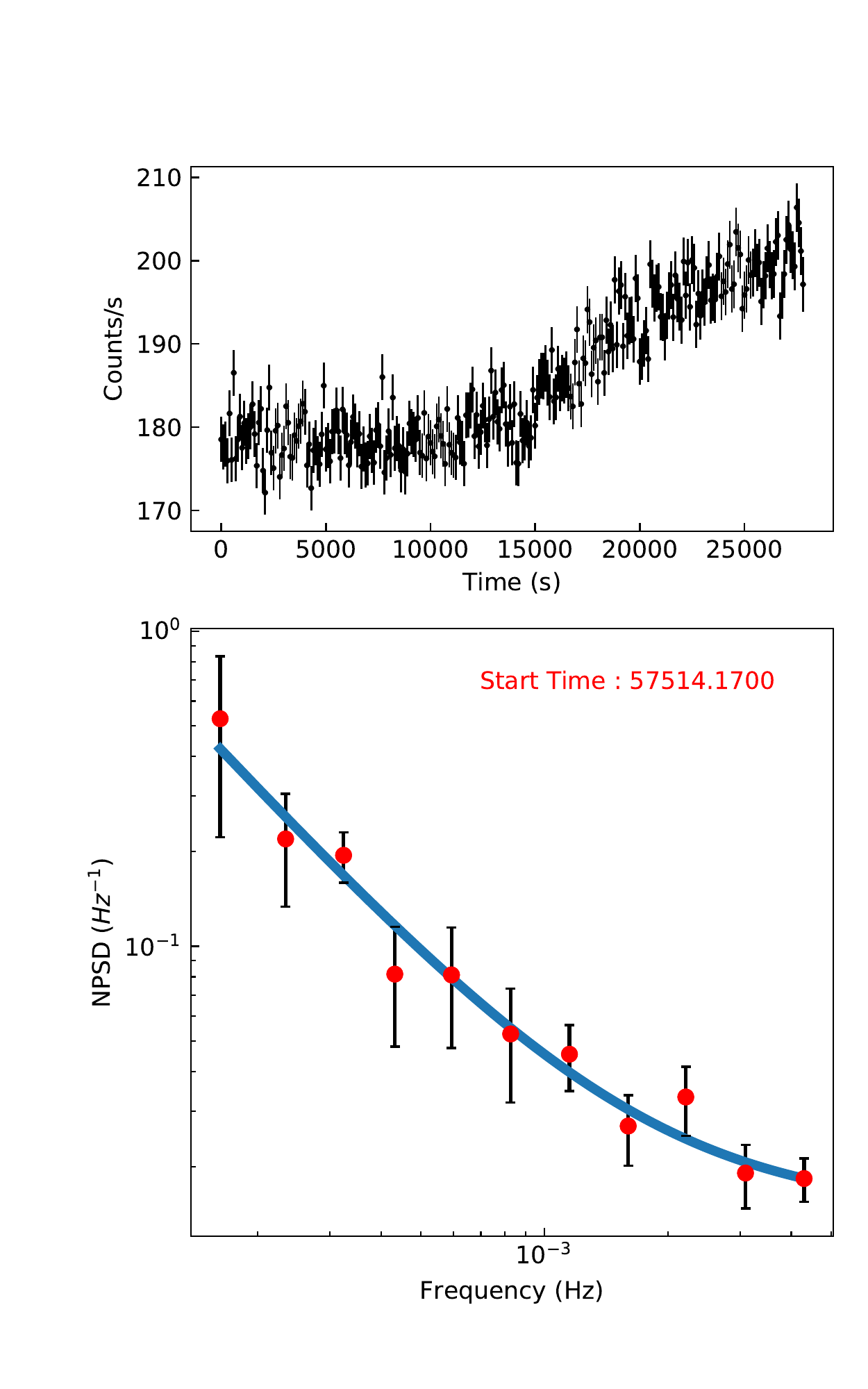}	
	\caption{\em{Continued.}}
\end{figure}
\begin{figure}[!h]
	\figurenum{4}
	\includegraphics[scale=0.35]{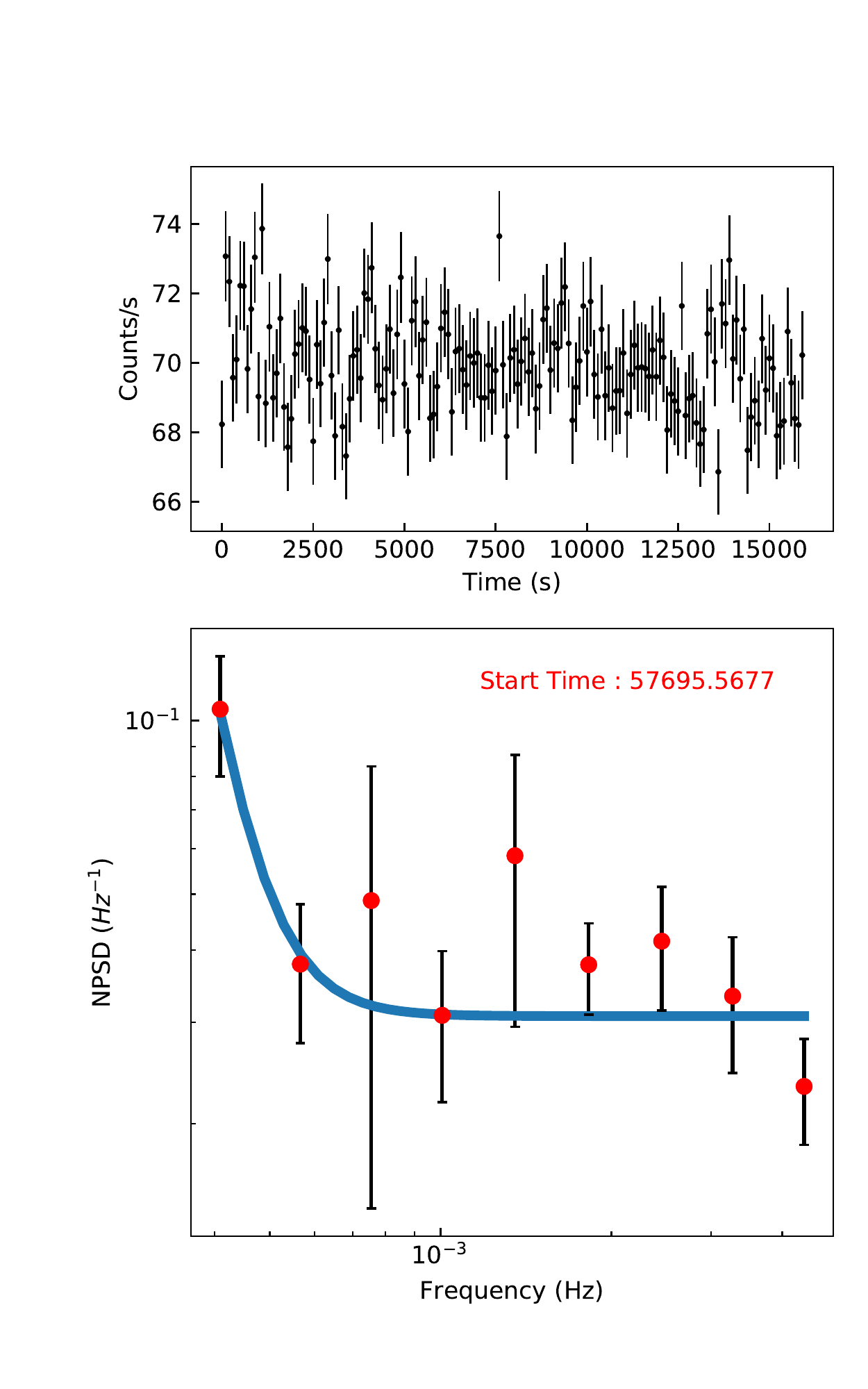}
	\includegraphics[scale=0.35]{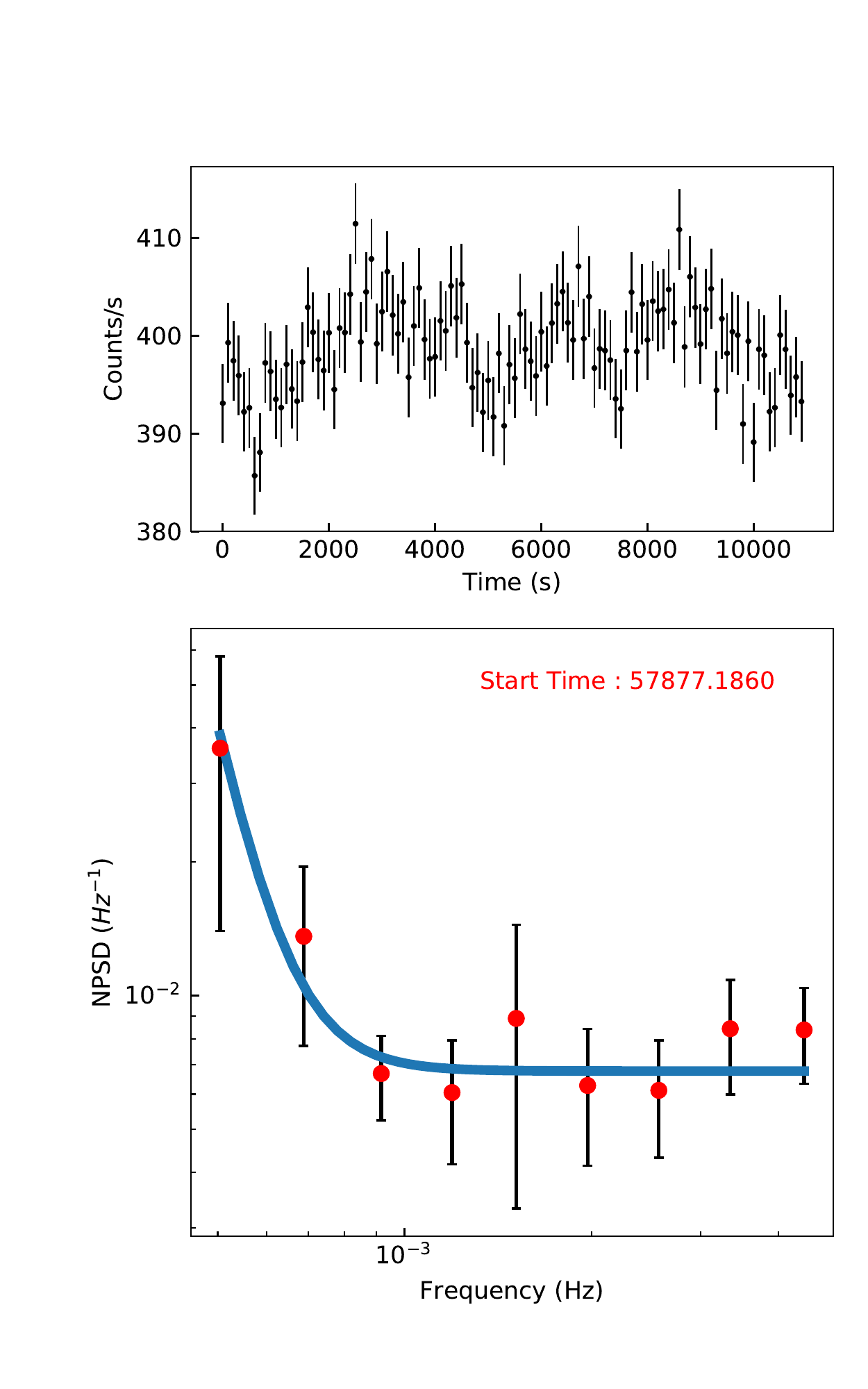}
	\caption{\em{Continued.}}
\end{figure}
%%%%%%%%%%%%%%%%%%%%%%%%%%
\clearpage
%%%%%%%%%%%%%%
\startlongtable
\begin{deluxetable*}{cccccc}
\tabletypesize{\small}
\tablenum{4}
\tablecaption{The NPSD best-fitting results and $F_{\rm{var}}$. (1) The start time, (2) the observation ID, (3) the power-law index, (4) the white noise power, (5) the reduced chi-square of the NPSD fitting, (6)the fraction variability amplitude of the light curve.  }
\label{tab:NPSDfvar}
\tablehead{\colhead{StartTime} & \colhead{Obs.ID} & \colhead{$\alpha$} & \colhead{$b$} & \colhead{$\chi^2_r$} & \colhead{$F_{\rm{var}}$} \\ 
\colhead{(MJD)} & \colhead{} & \colhead{} & \colhead{} & \colhead{} & \colhead{}  } 
%\decimals
\colnumbers
\startdata
51689.1624 & 0099280101 & 2.52 $\pm$ 0.42     & 0.0119    $\pm$ 0.0020      & 1.136      & 0.0609 $\pm$ 0.0004  \\
51689.4414 & 0099280101 & 1.34 $\pm$ 1.13     & 0.0657    $\pm$ 0.0231      & 0.427      & 0.1032 $\pm$ 0.0463  \\
51850.006  & 0099280201 & 2.42 $\pm$ 0.49     & 0.0224    $\pm$ 0.0031      & 1.381      & 0.0814 $\pm$ 0.0002  \\
51861.9324 & 0099280301 & 3.04 $\pm$ 0.30     & 0.0225    $\pm$ 0.0021      & 1.564      & 0.0320 $\pm$ 0.0063  \\
52037.3989 & 0136540101 & 2.68 $\pm$ 0.25     & 0.0325    $\pm$ 0.0017      & 0.441      & 0.0621 $\pm$ 0.0023  \\
52582.0308 & 0136540301 & 4.28 $\pm$ -     & 0.0355    $\pm$ -      & 0.552      & 0.0273 $\pm$ 0.0031  \\
52582.3202 & 0136540401 & 2.36 $\pm$ 0.30     & 0.0134    $\pm$ 0.0034      & 1.976      & 0.0433 $\pm$ 0.0003  \\
52592.8741 & 0136540801 & 5.32 $\pm$ -     & 0.0232    $\pm$ -      & 3.567      & 0.0761 $\pm$ 0.0251  \\
52609.9727 & 0136541001 & 3.16 $\pm$ 0.23     & 0.0142    $\pm$ 0.0011      & 0.998      & 0.0347 $\pm$ 0.0055  \\
52610.8351 & 0136541101 & 5.07 $\pm$ -     & 0.0486    $\pm$ -      & 1.452      & 0.0410 $\pm$ 0.0523  \\
52611.0031 & 0136541201 & 0.06 $\pm$ 2.64     & 0.1643    $\pm$ 6.3031      & 1.712      & 0.0080 $\pm$ 0.3564  \\
52957.6897 & 0150498701 & 2.11 $\pm$ 0.13     & 0.0677    $\pm$ 0.0058      & 0.621      & 0.0725 $\pm$ 0.5272  \\
52398.6795 & 0153950601 & 1.91 $\pm$ 0.06     & 0.0243    $\pm$ 0.0046      & 0.978      & 0.1242 $\pm$ 0.00002  \\
52399.1911 & 0153950701 & 0.01 $\pm$ 2.90     & 1.5690    $\pm$ 357.4908    & 7.162      & nan    $\pm$ nan     \\
53681.8447 & 0153951201 & 4.24 $\pm$ -     & 0.0336    $\pm$ 0.0041      & 0.662      & 0.0578 $\pm$ 0.1419  \\
53681.7058 & 0153951301 & 3.63 $\pm$ 12.24    & 0.9653    $\pm$ 0.2500      & 1.133      & 0.0389 $\pm$ 0.0164  \\
52791.557  & 0158970101 & 2.37 $\pm$ 0.36     & 0.0314    $\pm$ 0.0024      & 0.841      & 0.0635 $\pm$ 0.0001  \\
52792.0603 & 0158970201 & 0.01 $\pm$ 8.71     & -17.7281  $\pm$ 29382.3219  & 0.815      & nan    $\pm$ nan     \\
52797.897  & 0158970701 & 5.22 $\pm$ -     & 0.0468    $\pm$ -      & 4.091      & 0.0455 $\pm$ 0.0001  \\
53131.1251 & 0158971201 & 2.02 $\pm$ 0.82     & 0.1148    $\pm$ 0.0204      & 0.336      & 0.0690 $\pm$ 0.2558  \\
53683.7759 & 0158971301 & 4.42 $\pm$ -     & 0.0584    $\pm$ -      & 0.907      & 0.0786 $\pm$ 0.0082  \\
52983.8975 & 0162960101 & 3.80 $\pm$ -     & 0.0775    $\pm$ 0.0039      & 0.402      & 0.0325 $\pm$ 0.0928  \\
53854.8676 & 0302180101 & 0.05 $\pm$ 0.54     & -0.1917   $\pm$ 2.5311      & 0.312      & 0.0454 $\pm$ 0.0002  \\
53883.0932 & 0411080301 & 1.97 $\pm$ 0.43     & 0.0224    $\pm$ 0.0021      & 1.018      & 0.0468 $\pm$ 0.0037  \\
54074.5064 & 0411080701 & 0.00 $\pm$ 4.51     & 0.6957    $\pm$ 826.7401    & 1.183      & 0.0147 $\pm$ 0.0307  \\
54230.1689 & 0411081301 & 0.02 $\pm$ 6.58     & 2.3858    $\pm$ 768.9216    & 2.813      & nan    $\pm$ nan     \\
54230.4143 & 0411081401 & 0.01 $\pm$ 13.06    & -7.7254   $\pm$ 10649.9636  & 2.761      & nan    $\pm$ nan     \\
54230.5439 & 0411081501 & 0.12 $\pm$ 5.36     & 1.8319    $\pm$ 42.4040     & 0.126      & 0.0564 $\pm$ 4.4417  \\
54230.6735 & 0411081601 & 0.03 $\pm$ 12.65    & -518.7431 $\pm$ 252525.5077 & 0.292      & nan    $\pm$ nan     \\
54423.5489 & 0411081901 & 1.06 $\pm$ 0.58     & 0.0311    $\pm$ 0.0127      & 1.297      & 0.0330 $\pm$ 0.0104  \\
54617.1091 & 0411082701 & 0.77 $\pm$ 2115.67  & 0.4956    $\pm$ 0.6055      & 0.461      & 0.0749 $\pm$ 12.2353 \\
55151.7552 & 0411083201 & 6.54 $\pm$ -     & 0.2805    $\pm$ -      & 2.476      & nan    $\pm$ nan     \\
54593.0812 & 0502030101 & 2.50 $\pm$ 0.92     & 0.0552    $\pm$ 0.0101      & 1.744      & 0.0682 $\pm$ 0.0254  \\
54228.6283 & 0510610101 & 0.02 $\pm$ 2.42     & -14.6276  $\pm$ 2161.6860   & 5.009      & 0.0434 $\pm$ 2.7936  \\
54228.3529 & 0510610201 & 1.39 $\pm$ 2.79     & 0.4236    $\pm$ 0.1005      & 1.290      & nan    $\pm$ nan     \\
54792.6095 & 0560980101 & 0.01 $\pm$ 3.13     & 1.7160    $\pm$ 722.0796    & 2.956      & 0.0175 $\pm$ 0.0967  \\
54976.1719 & 0560983301 & 0.12 $\pm$ 1.95     & -0.1620   $\pm$ 3.0696      & 5.601      & 0.0216 $\pm$ 0.0635  \\
55319.3278 & 0656380101 & 0.01 $\pm$ 22.24    & -2.5869   $\pm$ 9925.8947   & 0.324      & 0.1127 $\pm$ 4.9136  \\
55512.8889 & 0656380801 & 0.60 $\pm$ 2.22     & 0.0296    $\pm$ 0.0502      & 1.126      & nan    $\pm$ nan     \\
55514.8844 & 0656381301 & 0.01 $\pm$ 3.52     & -1.0512   $\pm$ 399.0310    & 3.349      & 0.0174 $\pm$ 0.0731  \\
55698.4452 & 0658800101 & 6.71 $\pm$ -     & 0.2205    $\pm$ -      & 1.088      & nan    $\pm$ nan     \\
55894.0036 & 0658800801 & 4.32 $\pm$ -     & 0.0790    $\pm$ -      & 0.533      & 0.0123 $\pm$ 0.2107  \\
57179.0079 & 0658801301 & 1.66 $\pm$ 0.12     & 0.0056    $\pm$ 0.0016      & 0.890      & 0.0654 $\pm$ 0.0007  \\
57334.6077 & 0658801801 & 2.23 $\pm$ 0.18     & 0.0197    $\pm$ 0.0032      & 1.075      & 0.0677 $\pm$ 0.0006  \\
57514.17   & 0658802301 & 1.42 $\pm$ 0.20     & 0.0144    $\pm$ 0.0032      & 0.472      & 0.0444 $\pm$ 0.0016  \\
56776.1859 & 0670920301 & 0.36 $\pm$ 2.69     & 0.0513    $\pm$ 0.1450      & 0.693      & 0.0241 $\pm$ 0.2352  \\
56778.1597 & 0670920401 & 1.65 $\pm$ 0.32     & 0.0123    $\pm$ 0.0074      & 0.593      & 0.0676 $\pm$ 0.0051  \\
56780.1518 & 0670920501 & 1.41 $\pm$ 3.22     & 0.0170    $\pm$ 0.0113      & 2.390      & 0.0207 $\pm$ 0.0597  \\
57695.5677 & 0791780101 & 6.51 $\pm$ -     & 0.0307    $\pm$ -      & 0.981      & 0.0071 $\pm$ 0.0336  \\
57877.186  & 0791780601 & 6.92 $\pm$ -     & 0.0068    $\pm$ -      & 0.342      & 0.0057 $\pm$ 0.2461  \\
\hline
%all        & all        &      -          &           -            &            & 0.5490 $\pm$ 0.0012    \\
\enddata
%\tablecomments{The last line of this table is the $F_{\rm{var}}$ of all light curves of the PN data.}
\end{deluxetable*}
%%%%%%%%%%%%%%%

\end{document}